\documentclass[aps,prd,floatfix,preprintnumbers,altaffilletter,superscriptaddress,reprint,longbibliography,twocolumn,nofootinbib]{revtex4-2}
\usepackage{graphicx,url,amssymb,amsmath,rotating,color,units,wasysym,epsfig,multirow,epstopdf, stackengine}
\usepackage{amsmath}
\usepackage{cancel}
\usepackage{physics}
\usepackage{appendix}
\usepackage{amssymb}
\usepackage[colorlinks,urlcolor=blue,citecolor=blue,linkcolor=blue]{hyperref}
\usepackage[normalem]{ulem}
\usepackage{tabularx}
\newcolumntype{Y}{>{\centering\arraybackslash}X}

\usepackage{soul}

\usepackage[commentmarkup=uwave]{changes}
\definechangesauthor[name=Max, color=magenta]{MI}
\definechangesauthor[name=Will, color=cyan]{WF}
\definechangesauthor[name=Ethan, color=red]{EP}

\newcommand{\CIT}{\affiliation{Department of Physics, California Institute of Technology, Pasadena, California 91125, USA}}
\newcommand{\CITLab}{\affiliation{LIGO Laboratory, California Institute of Technology, Pasadena, California 91125, USA}}
\newcommand{\CCA}{\affiliation{Center for Computational Astrophysics, Flatiron Institute, 162 5th Ave, New York, New York 10010, USA}}
\newcommand{\StonyBrook}{\affiliation{Department of Physics and Astronomy, Stony Brook University, Stony Brook NY 11794, United States}}

\begin{document}

\title{Fortifying gravitational-wave tests of general relativity against astrophysical assumptions}

\author{Ethan Payne}
\email{epayne@caltech.edu}
\CIT
\CITLab

\author{Maximiliano Isi}
\email{misi@flatironinstitute.org}
\CCA

\author{Katerina Chatziioannou}
\email{kchatziioannou@caltech.edu}
\CIT
\CITLab

\author{Will M. Farr}
\email{wfarr@flatironinstitute.org}
\CCA
\StonyBrook

\begin{abstract}
Most tests of general relativity with gravitational-wave observations rely on inferring the degree to which a signal deviates from general relativity in conjunction with the astrophysical parameters of its source, such as the component masses and spins of a compact binary.
Due to features of the signal, measurements of these deviations are often highly correlated with the properties of astrophysical sources. 
As a consequence, prior assumptions about astrophysical parameters will generally affect the inferred magnitude of the deviations.
Incorporating information about the underlying astrophysical population is necessary to avoid biases in the inference of deviations from general relativity. 
Current tests assume that the astrophysical population follows an unrealistic fiducial prior chosen to ease sampling of the posterior---for example, a prior flat in component masses---which is is inconsistent with both astrophysical expectations and the distribution inferred from observations.
We propose a framework for fortifying tests of general relativity by simultaneously inferring the astrophysical population using a catalog of detections.
Although this method applies broadly, we demonstrate it concretely on massive graviton constraints and parameterized tests of deviations to the post-Newtonian phase coefficients. 
Using observations from LIGO-Virgo-KAGRA's third observing run, we show that concurrent inference of the astrophysical distribution strengthens constraints and improves overall consistency with general relativity. 
We provide updated constraints on deviations from the theory, finding that, upon modeling the astrophysical population, the 90\%-credible upper limit on the mass of the graviton improves by $25\%$ to $m_g \leq 9.6 \times 10^{-24}\, \mathrm{eV}/c^2$ and the inferred population-level post-Newtonian deviations move ${\sim} 0.4 \sigma$ closer to zero. 
\end{abstract}

\maketitle

\section{Motivation}~\label{sec:motivation}

Gravitational-wave observations from compact binary mergers have provided a unique laboratory to test Einstein's theory of gravity in the strong-field regime~\citep{GW150914_TGR, GW170104_TGR, GW170814_TGR, GW170817_TGR, GWTC1_TGR, GWTC2_TGR, GWTC3_TGR}. 
These individual detections by the Advanced LIGO~\cite{LIGO} and Advanced Virgo~\cite{Virgo} detectors allow for various tests---such as inspiral-merger-ringdown consistency~\citep{Ghosh2016qgn, Ghosh2017gfp}, parameterized inspiral deviations~\citep{Li2012PN, agathos2014TIGER, Mehta:2022pcn}, gravitational-wave dispersion~\citep{Will1997, Mirshekari2011}, birefringence \citep{Zhu:2023wci,Ng:2023jjt} and nontensorial polarizations~\citep{Eardley1973, Eardley1973b, Isi:2017fbj,  Pang2020pfz, Chatziioannou2021mij}, among many more; see Ref.~\citep{GWTC3_TGR} for recent results---to both target specific properties of general relativity (GR) as well as broadly explore its consistency with observations. 
Beyond analyzing events individually, the ensemble of detections can be analyzed collectively to study the possibility of deviations from GR at the population level~\cite{isi2019hier, saleem2022hier, GWTC2_TGR, GWTC3_TGR}. 
Hierarchical population tests rely on inferring the distribution of deviation parameters across all events and confirming that it is consistent with a globally vanishing deviation~\cite{isi2019hier, zimmerman2019combining,Isi:2022cii}.

In this study we explore the systematic impact of astrophysical population assumptions on these studies, show that they already come into play for current catalogs due to the increasing number of detections, and offer a solution under the framework of hierarchical population modeling.

In inferences about deviations from GR, there are strong likelihood-level correlations between the deviation parameters and the astrophysical parameters of the source, such as the masses and spins of compact binaries~\cite{GW150914_TGR, psaltis2021eht, wolfe2022hybrid}. 
Therefore, any inference of deviations from GR signals from black hole coalescences will be affected by assumptions about the distribution of binary black-hole masses and spins in the Universe---otherwise known as the astrophysical population distribution \cite{gwtc-3_pop}.
This is true at both the individual-event and catalog levels, regardless of the specific assumptions made in combining deviation parameters across events, whether the analysis is hierarchical or not.
Even when astrophysical parameters do not explicitly appear in the catalog-level test of GR, assumptions about these parameters are implicitly encoded in the individual-event deviation posteriors through the prior.
As the catalog of gravitational-wave observations grows and the precision of the measurements improves, these systematic effects become more important.

In presence of correlations between deviation and astrophysical parameters, we must simultaneously model the astrophysical population distribution in conjunction with testing GR.
By not explicitly doing so, as has been the case in previous tests of GR~\citep{GW150914_TGR, GW170104_TGR, GW170814_TGR, GW170817_TGR, GWTC1_TGR, GWTC2_TGR, isi2019hier, GWTC3_TGR}, the astrophysical population is typically implicitly assumed to be uniform in detector-frame masses and uniform in spin magnitude. 
This fiducial sampling prior is adopted to ensure broad coverage of the sampled parameter-space, and not to represent a realistic astrophysical population.
In reality, the primary-black hole mass population more closely follows a decreasing power-law with an excess of sources at ${\sim}35\,M_\odot$, and preferentially supports low spins~\cite{gwtc-2_pop, gwtc-3_pop}.
This mismatch can lead to biased inference regarding deviations from GR.
Simultaneously modeling the astrophysical and deviation distributions will not eliminate the influence of the former on the latter, but it will ensure that this interplay is informed by the data and not arbitrarily prescribed by analysis settings.

While this insight applies to all tests of GR, for concreteness we devote our attention to constraints on the mass of the graviton~\citep{Will1997, Mirshekari2011} and deviations in parameterized post-Newtonian (PN) coefficients~\cite{Yunes2009ke, Blanchet2014PN, arun2005PN, Li2012PN, agathos2014TIGER, chatziioannou2017PN}.
A massive graviton would affect the propagation of a gravitational wave over cosmological distances;
this leads to a frequency-dependent dephasing of the gravitational wave which is related to the mass of the graviton, $m_g$, and the propagated distance. 
The PN formalism describes the Fourier-domain phase of an inspiral signal under the stationary phase approximation through an expansion in the orbital velocity of the binary system;
each $k/2$ PN expansion order can then be modified by a deviation parameter, $\delta\varphi_k$, which vanishes in GR.
See App.~\ref{app:tests} for further details about both calculations. 
We focus on these tests as they target the signal inspiral phase, which also primarily informs astrophysical parameters such as masses and spins; we leave other tests~\citep{Ghosh2016qgn, Ghosh2017gfp, Will1997, Mirshekari2011, Eardley1973, Eardley1973b, Isi:2017fbj,  Pang2020pfz, Chatziioannou2021mij, GWTC1_TGR, GWTC2_TGR, GWTC3_TGR} to future work.

\begin{figure}
    \centering
    \includegraphics[width=\linewidth]{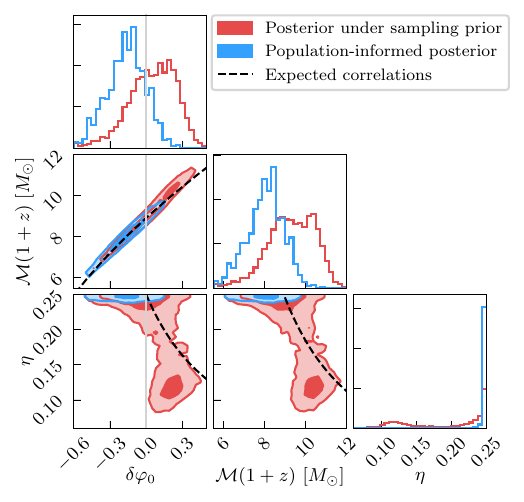}
    \caption{
    Posterior distributions for the 0PN deviation coefficient $\delta\varphi_0$, detector-frame chirp mass $\mathcal{M}(1+z)$, and symmetric mass ratio $\eta$ for the gravitational-wave event GW191216\_213338~\cite{GWTC2_TGR, gwtc2}, as inferred by a modified SEOBNRv4 waveform \cite{bohe2017, Cotesta2018, Cotesta2020,Brito2018, Mehta:2022pcn}. Posteriors are conditioned on two different astrophysical assumptions: the broad prior used during parameter estimation (red), and the astrophysical population inferred by the data using the model in Sec.~\ref{sec:model} (blue). The black dashed curves show the expected correlation (App.~\ref{app:expected}).
    Due to the correlations between astrophysical and deviation parameters, different astrophysical populations lead to different posteriors for $\delta\varphi_0$. 
    }
    \label{fig:correlated}
\end{figure}

As motivation, Fig.~\ref{fig:correlated} shows how inference on the 0PN coefficient of a real event (GW191216\_213338) depends on astrophysical assumptions.
This figure compares measurements with (blue) and without (red) a simultaneous measurement of the population of black hole masses and spins (see Sec.~\ref{sec:hier}).
The observed binary black-hole population shows a preference for systems with comparable masses;
as a consequence of the strong correlation between the 0PN deviation coefficient and the mass ratio of GW191216\_213338, this preference then ``pulls'' the system towards more equal masses and a more negative deviation coefficient.
This is a direct manifestation of the fact that tests of GR are contingent on our astrophysical assumptions.
Higher PN orders are expected to display similar correlations as in Fig.~\ref{fig:correlated} with these and other parameters. 
For example, spins are known to be correlated with the coupling constant of dynamical Chern-Simons gravity which modifies the phase at the 2PN order~\citep{Loutrel2018ydv, Perkins2021mhb, Loutrel2022tbk,Okounkova:2022grv}.
While we have constructed the posterior informed results here, it is more robust to simultaneously infer the astrophysical population while also testing GR. 
Fixing the prior to one astrophysical population realization or marginalizing over possible distributions from other analyses will not capture any correlated structure between the inferred deviation parameters and the astrophysical distributions. 
The above example serves only to illustrate the impact of the arbitrary choices previously made.

The remainder of the manuscript focuses on combining information from many observations to simultaneously infer the astrophysical population while testing GR; it is structured as follows. 
We first introduce our hierarchical analysis framework, as well as astrophysical and GR deviation models, in Sec.~\ref{sec:hier}.
We then demonstrate the impact of incorporating astrophysical information by constraining the graviton mass and inferring the PN deviation properties with an ensemble of gravitational-wave observations in Sec.~\ref{sec:app}. 
We analyze events from LIGO-Virgo-KAGRA (LVK)'s third observing run with individual-event results from Ref.~\citep{GWTC3_TGR} (the posterior samples are available in Ref.~\citep{gwtc3_tgr_release})  
 --- a subset of the events in GWTC-3 \cite{gwtc-3}.
The simultaneous modeling of the astrophysical population while testing GR tightens the graviton mass upper limit by $25\%$, and improves consistency with GR on the PN coefficients by ${\sim}0.4\sigma$, when using a modified {\tt\sc SEOBNRv4} waveform~\cite{bohe2017, Cotesta2018, Cotesta2020,Brito2018, Mehta:2022pcn}. 
Finally, we conclude in Sec.~\ref{sec:conc}, where we summarize the case for jointly modeling the astrophysical population when testing GR in order to avoid biases and hidden assumptions, and comment on how the same is true for gravitational-wave studies of cosmology or nuclear matter. 

\section{Population Analyses}~\label{sec:hier}

In this section, we introduce the fundamentals of inferring a population distribution from individual observations and discuss the population models we employ. 
We also outline the implementation and importance of observational selection effects in accounting for the events used within the analysis. 

\subsection{Preliminaries}~\label{subs:prelim}

We infer the astrophysical population distribution and deviations from GR~(see Refs.~\cite{Mandel2019, intro, Vitale2022} for a discussion of hierarchical inference in the context gravitational-wave astronomy).
This framework has already been extensively applied to tests of GR and astrophysical population inference separately~\cite{isi2019hier, saleem2022hier, GWTC2_TGR, GWTC3_TGR, gwtc-1_pop, gwtc-2_pop, gwtc-3_pop, Roulet_2021, Farr_2017, mass, spin, Callister_2021, Fishbach_2022, Biscoveanu_2022, Vitale_2017, Stevenson_2017, Miller2020, bbm, 2018_maya, Edelman_2022, Edelman_2023, golomb2022, Callister2023}.
Here we focus on combining both methods to jointly infer the astrophysical population while testing GR.

Our approach is based on a \textit{population likelihood}, $p(\{d\}|\Lambda)$, for the ensemble of observations, $\{d\}$, given population hyperparameters, $\Lambda = \{\Lambda_{\textrm{astro}},\, \Lambda_{\textrm{nGR}}\}$. 
We separate the hyperparameters into the parameters describing the astrophysical population distribution, $\Lambda_{\textrm{astro}}$, and parameters describing the deviation to GR, $\Lambda_{\textrm{nGR}}$. 
The hyperparameters encode the shape of the population distribution, $\pi(\theta|\Lambda)$, where $\theta$ are parameters of a single event;
we describe our population models in the following subsections.  
This hierarchical approach allows us to test GR while concurrently inferring the astrophysical population from the data. 
Given the likelihoods of individual events, $p(d_i|\theta_i)$, the population likelihood is
\begin{equation}~\label{eq:poplike}
    p(\{d\}|\Lambda) = \frac{1}{\xi(\Lambda)^N}\prod_{i=1}^N\int \dd \theta_i\,p(d_i|\theta_i)\, \pi(\theta_i|\Lambda)\,,
\end{equation}
where $d_i$ and $\theta_i$ are respectively the data and parameters for the $i$th event, and $\{d\}$ is the collection  of data for the ensemble of $N$ observations\footnote{Equation~\eqref{eq:poplike} assumes a prior on the rate of observations as $\pi(R)\propto 1/R$, which was analytically marginalized~\cite{2018_maya}.}. 
We address the technical aspects of the likelihood calculation in App.~\ref{app:marg}. 

In Eq.~\eqref{eq:poplike}, $\xi(\Lambda)$ is the detectable fraction of observations given a set of population hyperparameters and accounts for selection biases~\cite{Mandel2019}. It is defined as
\begin{equation} \label{eq:selection}
    \xi(\Lambda) = \int\dd\theta\, p_\textrm{det}(\theta)\,\pi(\theta|\Lambda)\,.
\end{equation}
Here $p_\textrm{det}(\theta)$ is the probability of detecting a binary black-hole system with parameters $\theta$. 
The selection factor in Eq.~\eqref{eq:selection} accounts for both the intrinsic selection bias of a gravitational-wave detector (e.g., heavier binaries are more detectable), as well as selection thresholds used when deciding which gravitational-wave events to analyze.
The detected fraction can also be framed as a ``normalizing factor,'' which relaxes the need for normalizable population distributions (so long as the integrals in Eqs.~\ref{eq:poplike} and~\ref{eq:selection} are finite)~\citep{farr2019}. 
This correction will become important in Sec.~\ref{subs:sel} when discussing the selection criteria for events to be included in the analysis. 

In theory, the selection factor should account for the effect of both astrophysical and deviation parameters. 
However, we ignore the latter here, the effect of which is subject of ongoing research~\citep{MageeInPrep}.
For the former, we compute the detectable fraction, $\xi(\Lambda)$, from a set of recovered injections,
\begin{equation}~\label{eq:sel}
    \xi(\Lambda) = \frac{1}{N_\textrm{inj}}\sum_{i=1}^{N_\textrm{rec}}\frac{\pi(\theta_i|\Lambda)}{\pi_\textrm{draw}(\theta_i)}\,,
\end{equation}
where $N_\textrm{inj}$ is the number of injected signals, $N_\textrm{rec}$ is the number of recovered signals, and $\pi_\textrm{draw}(\theta_i)$ is the distribution from which the injected signals were drawn (for more details see Refs.~\citep{intro, Mandel2019, Vitale2022, gwtc-1_pop, gwtc-2_pop, gwtc-3_pop}).  
The subset of injected signals that are recovered is determined by the particular thresholds used to determine which gravitational-wave observations to use within the hierarchical analysis.
To avoid biases, the criteria on the threshold for the detectable fraction calculation must match that of the observed signals. 
We address the specifics of the relevant criteria for our analysis in Sec.~\ref{subs:sel}. 

Finally, Eq.~\eqref{eq:poplike} explicitly shows the need for jointly modeling the astrophysical population when testing GR. 
While the astrophysical population may be separable from the deviation distribution so that $\pi(\theta|\Lambda) = \pi(\theta_{\textrm{astro}}|\Lambda_{\textrm{astro}})\,\pi(\theta_{\textrm{nGR}}|\Lambda_{\textrm{nGR}})$, this factorization cannot be undertaken for individual event likelihoods, as the deviations are often correlated with astrophysics (see Fig.~\ref{fig:correlated}), i.e. $p(\{d_i\}|\theta) \neq p(\{d\}|\theta_{\textrm{nGR}})\,p(\{d\}|\theta_{\textrm{astro}})$. 
Therefore, the integrals of Eq.~\eqref{eq:poplike} do not separate and tests of GR cannot be undertaken in isolation from the astrophysics. 

From the hyperposterior distribution on the population parameters, we can construct the individual event population-informed posteriors following Refs.~\citep{Moore2021xhn, farr_pop_informed, callister_pop_informed} (and references therein).
Such distributions represent our best inference about the properties of a given event in the context of the entire catalog of observed signals.
These calculations are subtle as they avoid ``double-counting'' the gravitational-wave events which also used to infer the population distribution. 

\subsection{Population models}~\label{sec:model}

In this subsection, we outline the population models for both the GR deviations and the astrophysical population.  
While many astrophysical population models have been proposed~\cite{gwtc-1_pop, gwtc-2_pop, gwtc-3_pop, Roulet_2021, Farr_2017, mass, spin, Callister_2021, Fishbach_2022, Biscoveanu_2022, Vitale_2017, Stevenson_2017, Miller2020, bbm, 2018_maya, Edelman_2022, Edelman_2023, golomb2022, Callister2023} as a product of the increasing number of observations~\cite{gwtc-3, gwtc-3_pop}, in this work we restrict ourselves to standard parameterized models motivated by previous analyses. 

\subsubsection{GR deviation population models}

There are two typical approaches to combining posteriors on GR deviation parameters obtained from different gravitational-wave observations, each stemming from different assumptions behind the deviations (see, e.g., discussions in\cite{GWTC2_TGR,GWTC3_TGR}). 
The first, more general approach is to assume that the population describing deviations from GR is, to the lowest order, a Gaussian distribution with a mean, $\mu$, and standard deviation, $\sigma$~\cite{zimmerman2019combining, isi2019hier}. 
In the limit that all observations are consistent with GR, $(\mu,\sigma)\to(0,0)$ and the inferred distribution approaches a Dirac delta function at the origin.
Since a Gaussian distribution encapsulates the lowest order moments of more complicated distributions, given enough events any deviation from a delta function at the origin will be identified as a violation of GR, even if the exact shape of the deviation distribution is not captured by a Gaussian \cite{isi2019hier,Isi:2022cii}. 
This approach is now routinely applied to post-Newtonian deviations tests, inspiral-merger-ringdown consistency tests and ringdown analyses~\cite{isi2019hier, GWTC2_TGR, GWTC3_TGR}, but it can be naturally extended to any analysis that recovers GR in the limit of some vanishing parameter.
This method provides a null test in cases where the exact nature of the deviation is unknown.

The second approach assumes all observations share the same value of the deviation parameter~\cite{GWTC1_TGR, Li2012PN, agathos2014TIGER, Pozzo2011, Meidam2014, Ghosh2016, Ghosh2018, Meidam2018, Brito2018, Mehta:2022pcn}. 
This is the limiting case of the aforementioned Gaussian model when $\sigma \to 0$.
This model (in the absence of astrophysical information) is equivalent to simply multiplying the marginal likelihoods of the deviation parameter obtained from the individual events. 
The assumption of a shared parameter is only suitable in the context of specific theories or models, in which case the expected degree of deviation for each event can be predicted exactly as a function system specific parameters (e.g., BH masses and spins) and universal, theory-specific parameters (e.g., coupling constants), the second of which can be measured jointly from a catalog of detections by multiplying likelihoods. In practice, the lack of complete waveform models beyond GR means that this approach has so far only been well-suited for measurements such as the mass of the graviton, and features of the propagation of gravitational waves whose observational signatures are independent of specific source properties by construction~\cite{GWTC1_TGR, GWTC2_TGR, GWTC3_TGR}.

\subsubsection{Astrophysical population models}

Following Refs.~\citep{gwtc-1_pop, gwtc-2_pop, gwtc-3_pop}, we model the primary black-hole mass ($m_1$) distribution as a power-law whose slope is given by an index $\alpha$, with a sharp cut-off governed by the minimum mass, $m_\textrm{min}$, and a higher-mass Gaussian peak,
\begin{align}
    \pi(m_1|\Lambda) = (1&-f_{\rm peak})\,\mathcal{P}[\alpha, m_{\rm min}](m_1) + \nonumber\\
    &f_{\rm peak}\,\mathcal{N}[\mu_{\rm peak},\sigma_{\rm peak}^2](m_1)\,.
\end{align}
Here, $f_{\rm peak}$ is the fraction of binaries in the Gaussian peak, the powerlaw is given by
\begin{equation}
    \mathcal{P}[\alpha, m_{\rm min}](m_1) \propto \begin{cases}
        m_1^{-\alpha},&m_1 \geq m_\textrm{min}\\
        0,&m_1 < m_\textrm{min}\,,
    \end{cases}
\end{equation}
and $\mathcal{N}[\mu,\sigma^2](x)$ is the probability density function for a Gaussian with mean $\mu$ and variance $\sigma^2$.
We fix $m_\textrm{min} = 5\,M_\odot$ for simplicity.
Unlike other studies~\citep{mass, gwtc-2_pop, gwtc-3_pop}, we do not infer much structure in the Gaussian peak as higher mass features become unresolvable when looking at the light binary systems that provide constraints of PN coefficients (see Sec.~\ref{subs:sel}).

We parameterize the distribution of mass ratios, $q\equiv m_2/m_1$, as a conditional power-law, with index $\beta$, and a sharp cut-off imposed by $m_{\rm min}$, such that
\begin{equation}
    \pi(q|m_1; \Lambda) \propto \begin{cases}
        q^{\beta},& 1 \geq q \geq m_\textrm{min}/m_1\\
        0,& q \leq m_\textrm{min}/m_1\,.
    \end{cases}
\end{equation}
Here $\beta$ can take any value without leading to a singularity due to the lower bound on the mass ratio. 

We adopt a truncated Gaussian population model for the component spins with a mean, $\mu_\chi$, and standard deviation, $\sigma_\chi$, bounded between zero and one, assuming both spins are drawn independently from the same population distribution. 
This differs from standard Beta distribution utilized in many recent analyses~\citep{Wysocki2019, spin, gwtc-1_pop, gwtc-2_pop, gwtc-3_pop}. as it allows for non-zero support at the edges of the spin-magnitude domain~\citep{Callister2022}. 
Furthermore, adopting a Gaussian model allows for efficient computation of the population likelihood via analytic integration (see App.~\ref{app:marg}). 
For individual-event analyses where the spins are assumed to be aligned with the orbital angular momentum (as is the case for posteriors using a modified {\tt \sc SEOBNRv4} waveform~\citep{bohe2017, Cotesta2018, Cotesta2020, Brito2018, Mehta:2022pcn}), this model treats the measured spin along the orbital angular momentum as the total spin magnitude. 

For analyses where the individual event inferences also possess information about the spin-precession degrees of freedom, we adopt a model for the spin tilts, $\cos\theta_{1/2}$, whereby the population is 
parameterized as a mixture of isotropically distributed and preferentially aligned spins~\citep{spin},
\begin{align}
    \pi(\cos\theta_1,& \cos\theta_2|\Lambda) = \frac{f_\textrm{iso}}{4} + (1-f_\textrm{iso})\times\nonumber\\
    &\mathcal{N}[1, \sigma_\theta^2](\cos\theta_1)\,\mathcal{N}[1, \sigma_\theta^2](\cos\theta_2)\,,
\end{align}
where $f_\textrm{iso}$ is the mixing fraction, and $\sigma_\theta$ is the standard deviation of the preferentially aligned Gaussian component.
This model is only relevant for analyses with precessing spins. 
In this manuscript, this includes the massive graviton constraints (Sec.~\ref{subs:same}), and PN deviation tests with the {\sc IMRPhenomPv2}~\citep{arun2005PN, khan2016PN,chatziioannou2017PN} waveform (App.~\ref{app:imr}).

Finally, we also adopt a power-law model for the merger-rate density as a function of redshift~\cite{2018_maya},
\begin{equation}
    \pi(z|\Lambda) \propto \frac{1}{1+z}\dv{V_c}{z}() (1+z)^{\lambda}\,,
\end{equation}
where ${\dd V_c}/{\dd z}$ denotes the evolution of the comoving volume with redshift, and $\lambda$ is the power-law index.
When $\lambda=0$, the binary black-hole population is uniformly distributed within the source-frame comoving volume. 

\subsection{Selection criteria and observations}~\label{subs:sel}

We limit ourselves to binary black-hole observations made during LIGO-Virgo-KAGRA's third observing run~\citep{gwtc-3} with false-alarm-rates of less than $10^{-3}$ per year\footnote{For comparison, the population analyses presented Ref.~\citep{gwtc-3_pop} used a false alarm rate threshold of $1$ per year. A more stringent false-alarm-rate threshold is often adopted when testing GR to avoid contaminating from false detections.}. 
This mirrors the selection criteria chosen for the tests of GR within Refs.~\citep{GWTC1_TGR, GWTC2_TGR, GWTC3_TGR}, and therefore we do need not reanalyze any individual gravitational-wave observations~\citep{gwtc2_tgr_release, gwtc3_tgr_release}.
The events that pass these criteria are listed in Table IV of Ref.~\citep{GWTC2_TGR} and Table V of Ref.~\citep{GWTC3_TGR}. 
In future studies, the false-alarm-rate threshold could be raised to increase the number of included gravitational-wave events. 
This would likely improve inference of the astrophysical population and GR deviation constraints due to the larger catalog of observations. 
In our analyses, we exclude GW190814~\citep{gw190814} as it is an outlier from the binary black-hole population~\citep{gwtc-2_pop} and GW200115\_042309 since it is a black hole-neutron star merger~\citep{gw200115}.
It is straightforward to extend this analysis to additionally incorporate binary neutron star and neutron star-black hole mergers by adopting a mixture model of the different source classifications (see Ref.~\citep{gwtc-3_pop} for one example).
We then use all events except GW200316\_215756\footnote{GW200316\_215756 was excluded from propagation tests within Ref.~\citep{GWTC3_TGR} due to poor sampling convergence.} when inferring the mass of the graviton, mirroring the analysis in Ref.~\citep{GWTC3_TGR}. 
When constraining the PN deviation coefficients, we include the additional criterion that signal-to-noise ratios (SNRs) during the binaries' inspiral must be greater than 6, again mirroring previous analyses~\citep{GWTC2_TGR, GWTC3_TGR}.

We use posteriors for the graviton's mass inferred using a modified {\tt \sc IMRPhenomPv2}~\citep{arun2005PN, khan2016PN,chatziioannou2017PN} waveform, whereas we use both modified {\tt \sc SEOBNRv4}~\citep{bohe2017, Cotesta2018, Cotesta2020, Brito2018, Mehta:2022pcn} (for results in Sec.~\ref{subs:pn}) and modified {\tt \sc IMRPhenomPv2}~\citep{arun2005PN, khan2016PN,chatziioannou2017PN,agathos2014TIGER,Li2012PN, Meidam2014, Meidam2018} (for results in App.~\ref{app:imr})\footnote{Single-event results with {\tt \sc IMRPhenomPv2} were only produced during the first half of the third observing run~\citep{GWTC2_TGR,GWTC3_TGR}.} waveform models when inferring the PN deviations. 
We summarize these events and their relevant properties in Tab.~\ref{tab:events}. 
We do not include gravitational-wave events from the first and second LIGO-Virgo observing runs, as a semi-analytic approximation was used to estimate the sensitivity of the detector network during that time~\citep{GWTC1_TGR}. 
This approximation does not compute a false-alarm rate and therefore cannot be unambiguously incorporated into this methodology. 

\begin{table}
    \centering
    \renewcommand*{\arraystretch}{1.4}
    \caption{Observations from the LIGO-Virgo-KAGRA's third observing run that pass our selection criteria~\citep{gwtc-2, gwtc-3, GWTC2_TGR, GWTC3_TGR}.
    The different columns outline the gravitational-wave event, the detector-frame chirp mass, the total and inspiral \textit{maximum a posteriori} SNRs ($\rho_\mathrm{tot}$ and $\rho_\mathrm{insp}$ respectively), and whether it was included in the graviton constraint calculation ($m_g$) or the post-Newtonian deviation tests (PN). 
    Horizontal lines split events from the two halves of the third observing period.
    While we use all events marked under ``PN'' in Sec.~\ref{subs:pn}, we are limited to the first half of observing run when using {\tt \sc IMRPhenomPv2} posterior samples in App.~\ref{app:imr}.}
    \begin{tabularx}{\linewidth}{l|Y|Y|Y|c|c}
         \multicolumn{1}{c|}{Event} & $(1+z)\mathcal{M}$ [$M_\odot$] & $\rho_\textrm{tot}$ & $\rho_\textrm{insp}$ & $m_g$ & PN \\\hline
         GW190408\_181802 & $23.7^{+1.4}_{-1.7}$ & 15.0 & 8.3 & \checkmark & \checkmark \\
         GW190412 & $30.1^{+4.7}_{-5.1}$ & 19.1 & 15.1 & \checkmark & \checkmark \\
         GW190421\_213856 & $46.6^{+6.6}_{-6.0}$ &  10.4 & 2.9  & \checkmark & - \\
         GW190503\_185404 & $38.6^{+5.3}_{-6.0}$ &  13.7 & 4.3 & \checkmark & - \\
         GW190512\_180714 & $18.6^{+0.9}_{-0.8}$ & 12.8 & 10.5 & \checkmark & \checkmark \\
         GW190513\_205428 & $29.5^{+5.6}_{-2.5}$ & 13.3 & 5.1 & \checkmark & - \\
         GW190517\_055101 & $35.9^{+4.0}_{-3.4}$ & 11.1 & 3.4 & \checkmark & - \\
         GW190519\_153544 & $65.1^{+7.7}_{-10.3}$ & 15.0 & 0.0 & \checkmark & - \\
         GW190521\_074359 & $39.8^{+2.2}_{-3.0}$& 25.4 & 9.7 & \checkmark & \checkmark \\
         GW190602\_175927 & $72.9^{+10.8}_{-13.7}$ & 13.1 & 0.0 & \checkmark & - \\
         GW190630\_185205 & $29.4^{+1.6}_{-1.5}$& 16.3 & 8.1 & \checkmark & \checkmark \\
         GW170706\_222641 & $75.1^{+11.0}_{-17.5}$ & 12.7 & 0.0 & \checkmark & - \\
         GW190707\_093326 & $9.89^{+0.1}_{-0.09}$& 13.4 & 12.2 & \checkmark & \checkmark \\
         GW190708\_232457 & $15.5^{+0.3}_{-0.2}$& 13.7 & 11.1 & \checkmark & \checkmark \\
         GW190720\_000836 & $10.4^{+0.2}_{-0.1}$& 10.5 & 9.2 & \checkmark & \checkmark \\
         GW170727\_060333 & $44.7^{+5.3}_{-5.7}$ & 12.3 & 2.0 & \checkmark & - \\
         GW190728\_064510 & $10.1^{+0.09}_{-0.08}$& 12.6 & 11.4 & \checkmark & \checkmark \\
         GW190828\_063405 & $34.5^{+2.9}_{-2.8}$& 16.2 & 6.0 & \checkmark & \checkmark \\
         GW190828\_065509 & $17.4^{+0.6}_{-0.7}$& 9.9 & 6.3 & \checkmark & \checkmark \\
         GW190910\_112807 & $43.9^{+4.6}_{-3.6}$ & 14.4 & 3.3 & \checkmark & - \\
         GW190915\_235702 & $33.1^{+3.3}_{-3.9}$ & 13.1 & 3.7 & \checkmark & - \\
         GW190924\_021846 & $6.44^{+0.04}_{-0.03}$& 12.2 & 11.8 & \checkmark & \checkmark\\\hline\hline
         GW191129\_134029 & $8.49^{+0.06}_{-0.05}$& 14.1 & 12.8 & \checkmark & \checkmark \\
         GW191204\_171526 & $9.70^{+0.05}_{-0.05}$& 18.0 & 16.3 & \checkmark & \checkmark \\
         GW191215\_223052 & $24.9^{+1.5}_{-1.4}$ & 10.6 & 5.5 & \checkmark & - \\
         GW191216\_213338 & $8.94^{+0.05}_{-0.05}$& 17.9 & 15.6 & \checkmark &\checkmark  \\
         GW191222\_033537 & $51.0^{+7.2}_{-6.5}$ & 13.1 & 3.1 & \checkmark & - \\
         GW200129\_065458 & $32.1^{+1.8}_{-2.6}$& 25.7 & 10.4 & \checkmark & \checkmark \\
         GW200202\_154313 & $8.15^{+0.05}_{-0.05}$& 11.1 & 10.5 & \checkmark & \checkmark \\
         GW200208\_130117 & $38.8^{+5.2}_{-4.8}$ & 9.9 & 3.0 & \checkmark & - \\
         GW200219\_094415 & $43.7^{+6.3}_{-6.2}$ & 11.2 & 2.8 & \checkmark & - \\
         GW200224\_222234 & $40.9^{+3.5}_{-3.8}$ & 19.4 & 4.7 & \checkmark & - \\
         GW200225\_060421 & $17.7^{+1.0}_{-2.0}$& 12.9 & 6.8 & \checkmark & \checkmark \\
         GW200311\_115853 & $32.7^{+2.7}_{-2.8}$& 17.5 & 6.5 & \checkmark & \checkmark \\
         GW200316\_215756 & $10.7^{+0.1}_{-0.1}$& 11.5 & 10.7 & - & \checkmark \\\hline
    \end{tabularx}
    \label{tab:events}
\end{table}

\begin{figure}
    \centering
    \includegraphics[width=\linewidth]{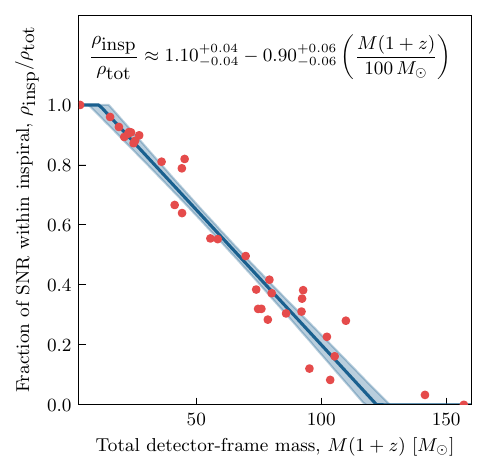}
    \caption{Ratio between the network \textit{maximum a posteriori} gravitational-wave inspiral and the total SNRs as a function of detector-frame total mass, $M(1+z) \equiv (m_1 + m_2)(1+z)$, for all gravitational-wave observations in the LIGO-Virgo-KAGRA third observing run~\citep{gwtc-2, gwtc-3, GWTC2_TGR, GWTC3_TGR} with a false-alarm rate less than $10^{-3}/\textrm{yr}$. 
    The solid blue line is the median best-fit line to the observations, with the band representing the 90\%-credible uncertainty. 
    While computing this fit, we also estimate the uncertainty in the individual data points. 
    We use this fit to compute the inspiral SNR for the injections used to estimate the detection probability, $p_\textrm{det}(\theta)$, as described in Sec.~\ref{subs:sel}.
    }
    \label{fig:snr_ratio}
\end{figure}

As described in Sec.~\ref{sec:hier}, selection effects are estimated through an injection campaign.
While we know the total network SNR of the individual injections, part of our selection criteria is based on the inspiral network SNR. 
We approximate the inspiral SNR from the total SNR by constructing a linear fit to their ratio as a function of detector-frame total mass (Fig.~\ref{fig:snr_ratio}). 
This fit is constructed by inferring the slope and offset of the line, as well as the uncertainty on the data points.
We assume identical uncertainties on all SNR ratios, and marginalize over this parameter to fit the line.
We validate this approximation by computing the detection probability $p_\textrm{det}(\theta)$ with different draws of the linear fit. 
We find that different realizations of the approximation do not change the detection probability, and so we consider this approximation to be sufficiently accurate for our purposes.
Future injection campaigns may also opt to compute the inspiral SNR directly. 

\section{Results}~\label{sec:app}

In this section we simultaneously infer the astrophysical population while testing GR and quantify the impact of fixing the population distribution to the sampling prior. 
Throughout, we use the nomenclature ``fixed'' and ``inferred'' to refer to whether the analysis uses the fixed sampling prior or infers the distribution from data, respectively. 
We implement the analyses using {\sc NumPyro}~\citep{phan2019composable, bingham2019pyro} and {\sc JAX}~\citep{jax2018github}, leveraging {\sc AstroPy}~\citep{astropy:2013, astropy:2018, astropy:2022} and {\sc SciPy}~\citep{2020SciPy-NMeth} for additional calculations, and {\sc matplotlib}~\citep{Hunter:2007}, {\sc arViz}~\citep{arviz_2019} and {\sc corner}~\citep{corner} for plotting purposes. 
The code for the hierarchical tests is available in Ref.~\citep{code_release}. 

\subsection{Massive graviton constraints}~\label{subs:same}

We begin by demonstrating that astrophysical assumptions are crucial even in the simplest scenarios, where a global deviation parameter is shared across events.
This is the case for the mass of the graviton, $m_g$ \citep{Will1997, Mirshekari2011} (see App.~\ref{app:mg}), for which we produce an updated upper limit by simultaneously inferring the astrophysical distribution.

We combine results from individual-event likelihoods under the assumption of a shared deviation parameter as described in Sec.~\ref{sec:model}. 
In practice, we compute this as the limit of a vanishing standard deviation of the hierarchical analysis described in Sec.~\ref{sec:hier}.
For technical reasons, we assume a uniform prior distribution on $\log_{10}(m_g)$ when combining observations, which differs from Refs.~\citep{GWTC1_TGR, GWTC2_TGR, GWTC3_TGR} which applied a uniform prior prior on $m_g$ itself;
this is to avoid poor convergence when reweighting between individual-event posterior distributions. 
In the end, we reweight the shared graviton mass inference to a uniform prior to report upper limits on $m_g$.
We compare this to results obtained assuming the sampling prior for the astrophysical parameters. 

\begin{figure}
    \centering
    \includegraphics[width=\linewidth]{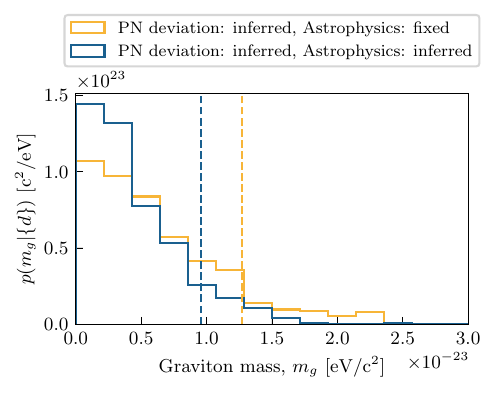}
    \caption{Marginal one-dimensional posterior distributions for the mass of a massive graviton.
    In practice, we compute the shared value of graviton mass by assuming a shared deviation parameter $\log_{10}(m_g c^2/\textrm{eV})$ then reweighting to a uniform graviton mass prior.
    The dashed lines correspond to the 90\% upper limits from the two analyses.
    We compare the result when astrophysical information is not included, equivalent to multiplying individual event likelihood functions (yellow), to also modeling the astrophysical population (dark blue). The result shifts towards smaller values of $m_g$ if simultaneously modelling the astrophysical population and the graviton's mass.
    }
    \label{fig:shared}
\end{figure}

The one-dimensional marginal distributions of the shared mass of the graviton are shown in Fig.~\ref{fig:shared}. 
The inclusion of astrophysical information changes the inferred distributions of the graviton's mass increasing support for $m_g = 0$. 
When using the sampling prior for the astrophysical population (and thereby assuming the incorrect distribution), the graviton's mass is constrained to be $m_g \leq 1.3\times10^{-23}\,\textrm{eV}/c^2$ at the 90\% level\footnote{This constraint differs from the 90\% upper limit of $1.27\times10^{-23}$ $\textrm{eV}/c^2$ calculated in Ref.~\citep{GWTC3_TGR}, which is determined by additionally incorporating observations from the first and second LIGO-Virgo-KAGRA observing periods~\citep{gwtc1, GWTC1_TGR}. We do not include these observations due to the ambiguity in the detector network sensitivity during these periods.};
however, upon inferring the astrophysical population the graviton's mass becomes more constrained, with $m_g \leq 9.6\times10^{-24}\,\textrm{eV}/c^2$ at the 90\% credible level.
Under the expectation that GR is correct and $m_g=0$, a reduced constraint is generically expected as we have included the correct information regarding the astrophysical population. 
This highlights the effect of unreasonable astrophysical assumptions, which are inconsistent with the observed population, on tests of GR.

\subsection{Hierarchical post-Newtonian deviation constraints from {\tt \sc SEOBNRv4}}~\label{subs:pn}

\begin{figure*}
    \centering
    \includegraphics[width=\linewidth]{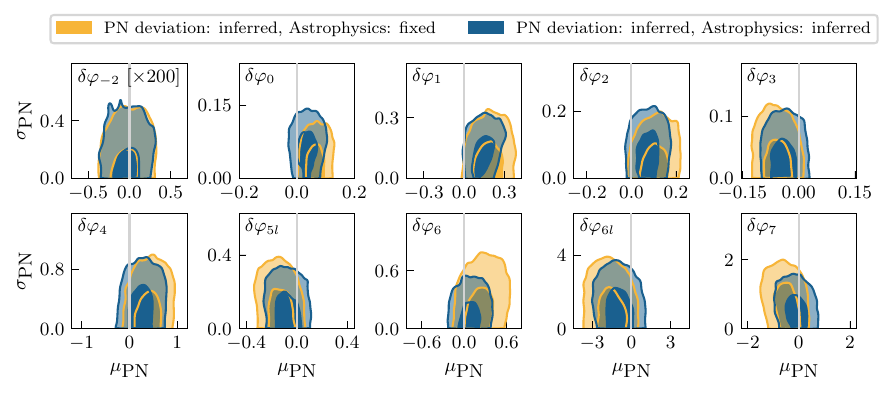}
    \caption{Two-dimensional marginal posterior distributions for the hyperparameters of the Gaussian PN deviation distribution informed by the 20 events in the third LIGO-Virgo-KAGRA observing run passing the selection criteria, analysed with a modified {\tt \sc SEOBNRv4}~\citep{bohe2017, Cotesta2018, Cotesta2020, Brito2018, Mehta:2022pcn} waveform. 
    The contours indicate the 50\% and 90\% credible regions.
    Each panel corresponds to a separate analysis where the coefficient varied was at a different PN order. 
    The analysis was undertaken with an implicitly assumed, astrophysically-unrealistic population (yellow), and a model which simultaneously infers the astrophysical population model (dark blue).
    Modelling both the astrophysical population and the PN deviation population systematically shifts the inferred mean, $\mu_\textrm{PN}$, closer to zero.
    }
    \label{fig:tgr_mu_sigma_seob}
\end{figure*}

\begin{figure}
    \centering
    \includegraphics[width=\linewidth]{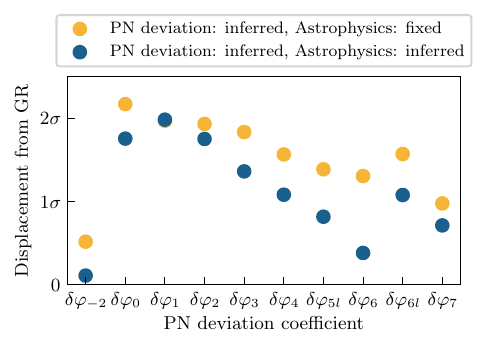}
    \caption{Displacement of the deviation parameter distribution  from GR for each PN deviation coefficient. The displacement corresponds to the credible levels at which the hyperparameter values corresponding to GR, $(\mu_\textrm{PN}, \sigma_\textrm{PN}) = (0,0)$, reside for two different models as shown in Fig.~\ref{fig:tgr_mu_sigma_seob}. 
    This quantity is indicative of the relative position of the posterior to the GR value. 
    Incorporating the astrophysical population as well as the hierarchical model for the PN deviation leads to an inferred result more consistent with GR for most cases.}
    \label{fig:quant_seob}
\end{figure}

We repeat the population analysis, this time measuring the hierarchical PN deviation distribution with a mean, $\mu_\textrm{PN}$, and standard deviation, $\sigma_\textrm{PN}$, for all PN orders.
This is corresponds to ten separate analyses where only one PN deviation coefficient is allowed to vary. 
To compare with the default approach (which implicitly assumes a flat-in-detector-frame mass, uniform mass ratio, uniform spin-magnitude aligned spin, and comoving volume redshift distributions), we also fit the GR deviation in isolation under the assumption of the (astrophysically unrealistic) sampling prior~\cite{GWTC2_TGR, GWTC3_TGR}.

Figure~\ref{fig:tgr_mu_sigma_seob} shows the two-dimensional posterior distribution of the deviation hyperparameters for $-1$ through to 3.5 PN orders. 
The standard results implicitly using the sampling prior are shown in yellow, while the results from the simultaneous modeling of the astrophysical and deviation populations are shown in dark blue.
When concurrently modeling the astrophysical distribution, in all PN deviation parameters the inferred mean resides closer to zero, i.e., the expected value from GR, while there is no clear trend in $\sigma_\textrm{PN}$.
Overall, $(\mu_\textrm{PN}, \sigma_\textrm{PN}) = (0,0)$ is retained with greater significance for almost all PN orders. 

We quantify this improvement by comparing the two-dimensional credible level\footnote{This ``displacement'' is the quantile, $\mathcal{Q}_{\rm GR}$, reported in Refs.~\citep{GWTC2_TGR, GWTC3_TGR} as $(\textrm{displacement})^2= -2\ln(1-\mathcal{Q}_{\rm GR})\, \sigma^2$. The quantile is computed by integrating over all regions of the hyperposterior distribution which are at a higher probability than $(\mu_\textrm{PN}, \sigma_\textrm{PN}) = (0,0)$. We report values in terms of the standard deviation in two dimensions, $1\sigma$ and $2\sigma$ correspond to ${\sim}39.3\%$ and ${\sim}86.5\%$ credibility, respectively.} at which the expected GR value, $(\mu_\textrm{PN}, \sigma_\textrm{PN}) = (0,0)$, resides in Fig.~\ref{fig:quant_seob}.
A lower value for the credible region implies that the value of hyperparameters expected from GR resides closer to the bulk of the distribution. 
In all but one PN order, jointly inferring the astrophysical and PN deviation distributions moves the inferred distribution to be more consistent with GR. 
For the 0.5PN deviation coefficient, $\delta\varphi_1$, there is little change in the credible level at which GR is recovered. 
Generally, inference of the astrophysical population allows our inferences of GR deviations to be more consistent with GR, with an average improvement of $0.4\sigma$. 

To shed further light on the interaction between the GR and astrophysics parameters, we focus on two specific deviation parameters. 
In particular, we draw attention to the 3PN coefficient (which shows the largest tightening of the supported hyperparameter space in Fig.~\ref{fig:tgr_mu_sigma_seob}) and the 0PN coefficient (where the PN deviation is most inconsistent with GR in Fig.~\ref{fig:quant_seob}). 

\subsubsection{Example: 3PN deviation coefficient, $\delta\varphi_6$}

\begin{figure*}[p!]
    \centering
    \includegraphics[width=\linewidth]{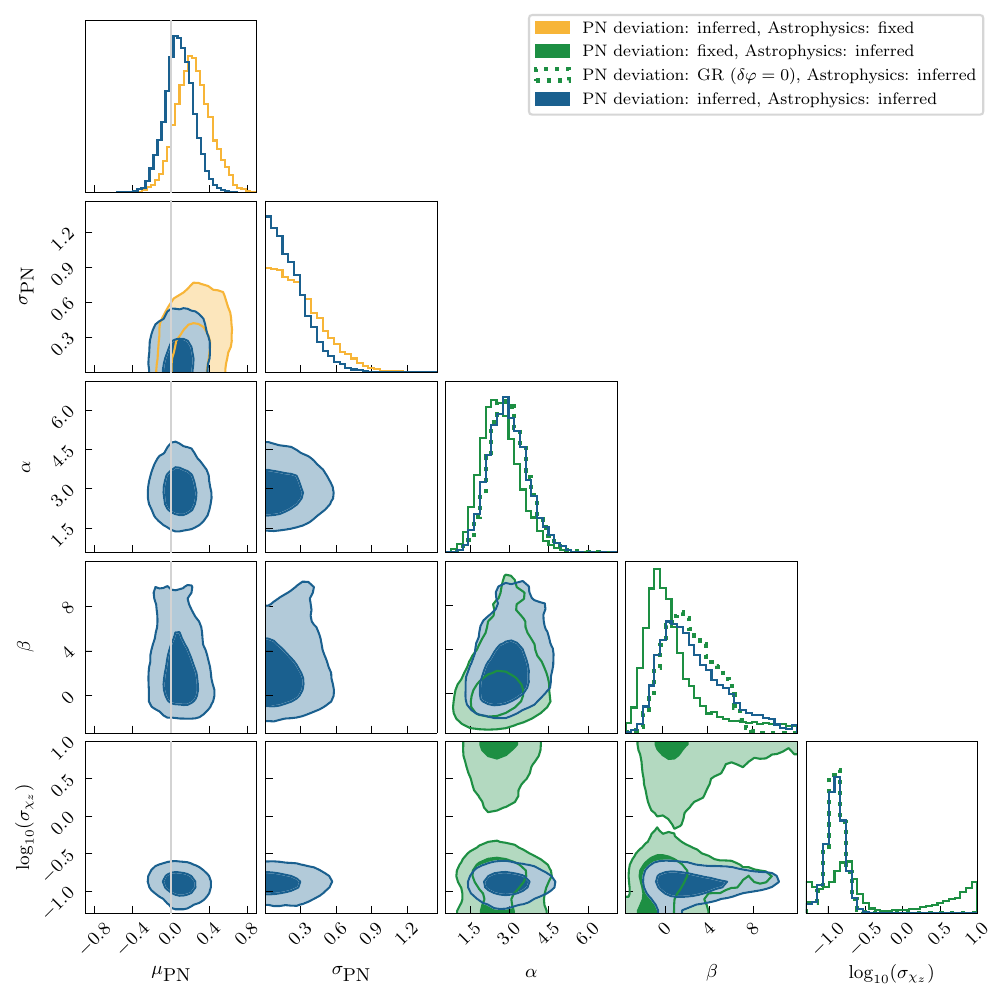}
    \caption{Marginal one- and two-dimensional posterior distributions for the $\delta\varphi_6$ PN deviation and a subset of astrophysical population hyperparameters. 
    Contours correspond to the 50\% and 90\% credible regions. 
    Results from four analyses are shown --- population inference using the PN deviation population only with the ``default'' sampling prior astrophysical population (yellow), astrophysical population only (green), astrophysical population under the assumption that GR is correct (dashed green), and the joint analysis inferring the post-Newtonian deviation and astrophysical populations simultaneously (dark blue). 
    No strong correlations exist between either the mean or standard deviation of the deviation Gaussian and astrophysical population parameters. 
    The starkest difference is that inferring the population when the PN deviation population is ignored leads to broad spin magnitude populations. 
    }
    \label{fig:full_corner_dchi6}
\end{figure*}

To understand the origin of the improved measurement for $\delta\varphi_6$ when modeling astrophysics in Fig.~\ref{fig:tgr_mu_sigma_seob}, we show an expanded corner plot in Fig.~\ref{fig:full_corner_dchi6} with an additional subset of the hyperparameter posterior distributions. 
The top left corner reproduces the corresponding panel in Fig.~\ref{fig:tgr_mu_sigma_seob}, wherein the yellow posterior distribution is obtained under the assumption of the astrophysical population given by the sampling priors, while the dark blue is obtained by simultaneously inferring the astrophysical-population and the GR deviation parameters.  

Additionally, we use the same set of individual-event posterior samples to \emph{separately} infer the astrophysical population independently of the PN deviation parameters, which amounts to assuming a uniform distribution of deviations across events (solid green). 
This differs from standard astrophysical population inference, which assumes that GR is correct \textit{a priori} and thus starts from individual-event posteriors conditioned on $\delta\varphi = 0$~\citep{gwtc-1_pop, gwtc-2_pop, gwtc-3_pop}.
Finally, we also compute the astrophysical population under the assumption that GR is correct, $(\mu_\textrm{PN}, \sigma_\textrm{PN}) = (0,0)$ (dashed green).
The result assuming GR is correct is computed by fixing $(\mu_\textrm{PN}, \sigma_\textrm{PN}) \rightarrow (0,0)$ to ensure equivalent samples are used between analyses, and is consistent with the usual population inference modulo model choices at the individual-event and population levels~\citep{gwtc-1_pop, gwtc-2_pop, gwtc-3_pop}. 

From the two-dimensional marginal distributions, the most apparent feature is that inferring the astrophysical population under the assumption of a broad uniform GR deviation population
(shown in solid green) leads to inferences consistent with broad spin populations (large $\sigma_{\chi_z}$) and populations favoring uneven mass ratios ($\beta < 0$). 
This can be straightforwardly explained by the presence of correlated structure between $\delta\varphi_6$, mass ratio, and the component spins at the individual-event level. 

\begin{figure}
    \centering
    \includegraphics[width=\linewidth]{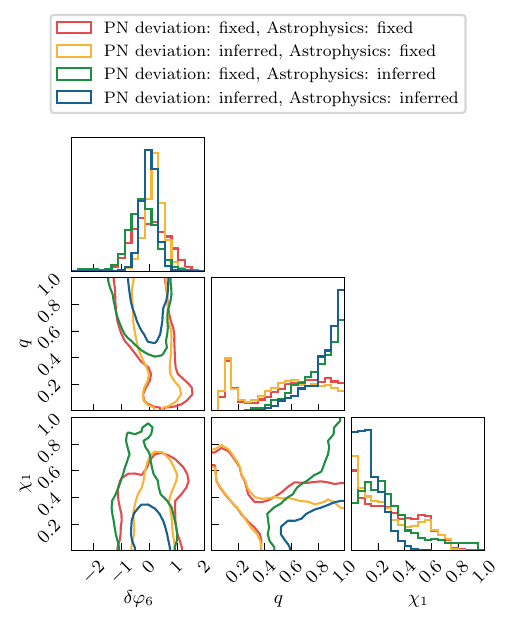}
    \caption{One- and two-dimensional posterior distributions for the 3PN deviation parameter, the mass ratio, and the primary black-hole spin for GW191216\_213338 under four different assumptions: broad sampling priors (red), informed by the GR deviation population analysis (yellow), informed by the astrophysical population (green), informed by the joint inference of PN deviation and astrophysical populations (dark blue). 
    Contours indicate the 90\% credible region. 
    Evidence for both a low mass ratio and larger primary spins is strongly contingent upon the astrophysical assumptions. 
    Broad priors such as those used while sampling the posterior distribution have significant support for lower mass ratios. 
    Inclusion of information from both the deviation population and the astrophysics leads to an inferred result with both low primary spin and high mass ratio. }
    \label{fig:indiv_event}
\end{figure}

To demonstrate this, Fig.~\ref{fig:indiv_event} shows four different posteriors for GW191216\_213338 under different priors. 
The four distributions shown are the posterior obtained with the sampling priors (red), the one informed by the GR deviation population only analysis (yellow), the one informed by the astrophysical population only analysis (green), and the one informed by the jointly-inferred GR deviation and astrophysical populations (blue).
The posteriors which involve information from inferred populations are computed following Ref.~\citep{Moore2021xhn}, and do not double-count the data from GW191216\_213338, as discussed in Sec.~\ref{subs:prelim}.

Under the sampling astrophysical prior, posteriors exhibit a low-$q$, high-$\chi_1$ mode. 
Since the inferred astrophysical population is inconsistent with low mass ratios and high spin magnitudes, the astrophysical-population-informed posteriors have reduced support for unequal masses (compare the red contour to the green one). 
Additionally incorporating the GR deviation information (blue), the population-informed posterior further reduces support for high-spinning systems. 
The similarity of the results under the sampling prior (red) with those in which only the GR deviation population is inferred (yellow), suggests that inferring small GR deviations is on its own not enough to significantly affect the inference of the astrophysical parameters in this case. 

The tightening of the $\sigma_{\chi_z}$ hyperposterior distribution (i.e., inferring a more narrow spin population) when jointly inferring the GR deviation and astrophysical populations is precisely what we observe at the population level in Fig.~\ref{fig:full_corner_dchi6} comparing the dark blue and green contours. 
Additionally, when enforcing that $\delta\varphi_6=0$ for all events (dashed green), we no longer recover support for broad spin populations.
Interestingly, the astrophysical population inferred jointly with the GR deviation population is very similar to the result obtained when fixing $\delta\varphi_6=0$. 
This illustrates that, if we allow the model to infer that the scale of GR deviations is small, we will recover similar inferences overall as if we had fixed $\delta\varphi=0$ \textit{a priori}: we are learning \emph{both} that spins are small \emph{and} that any GR deviation must be small at this PN order. 
Conversely, an assumption of a broad GR deviation population leads to unrealistic astrophysical populations to account for the far-fetched astrophysical systems such analyses allow. 
We can also use this example to understand why inferring the deviation population in the absence of astrophysical modelling leads to a different deviation population with a larger inferred mean.
Figure~\ref{fig:indiv_event} shows that $q$ and $\delta\varphi_6$ are correlated at the individual-event level, and therefore a broader $q$ distribution will lend more support to the higher values of $\delta\varphi_6$. 
This correlation then systematically pulls the mean of the PN deviation distribution to higher values. 

\subsubsection{Example: 0PN deviation coefficient, $\delta\varphi_0$}~\label{ssub:dphi0}

We now turn to $\delta\varphi_0$, for which the standard analysis with a fixed astrophysical prior finds the least consistency with GR, at the $2.2\sigma$ credible level (yellow circle for $\delta\varphi_0$ in Fig.~\ref{fig:quant_seob}), driven by a displacement away from $\mu_\textrm{PN}=0$ (Fig.~\ref{fig:tgr_mu_sigma_seob}). 
Since this parameter is strongly correlated with the chirp mass and mass ratio (Fig.~\ref{fig:correlated}), we expect improvements when jointly modeling the astrophysical and deviation distributions; indeed that is the case, with GR recovered at the $1.6\sigma$ level (blue circle in Fig.~\ref{fig:quant_seob}). This analysis infers a $\sigma_\textrm{PN}$ distribution that peaks slightly away from zero.

\begin{figure*}[p!]
    \centering
    \includegraphics[width=\linewidth]{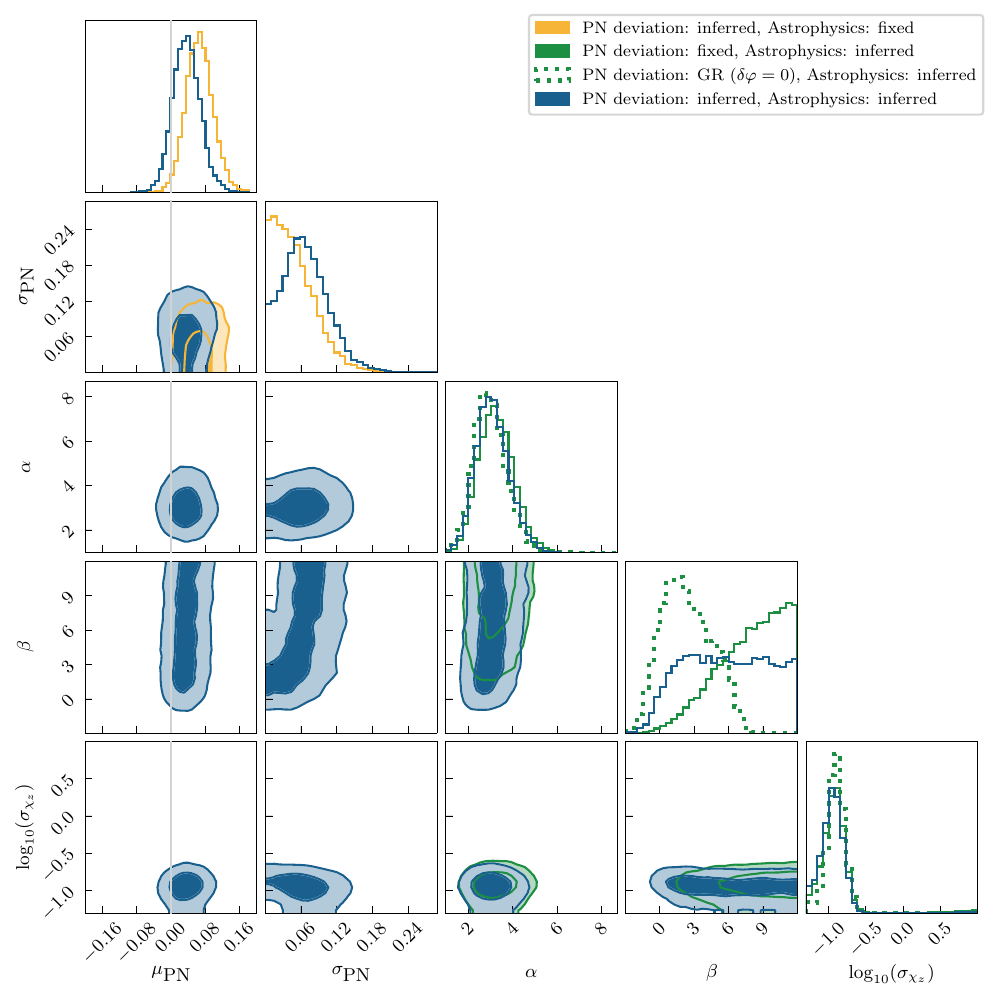}
    \caption{Similar to Fig.~\ref{fig:full_corner_dchi6}, one- and two-dimensional posterior distributions for the $\delta\varphi_0$ deviation and a subset of astrophysical population hyperparameters. 
    A strong correlation is found between the width of the inferred post-Newtonian deviation population and the index of the mass ratio power-law when jointly inferring the deviation and astrophysical population models. 
    There is also a less pronounced correlation between the deviation and spin population standard deviations. 
    In the absence of modelling the astrophysical population, the inferred PN population is pulled to a higher mean with a reduced width. 
    }
    \label{fig:full_corner_dchi0}
\end{figure*}

We can understand this behavior with Fig.~\ref{fig:full_corner_dchi0}, where we plot a subset of the two-dimensional marginal population posterior distributions in the same color scheme as Fig.~\ref{fig:full_corner_dchi6}. 
The structure of the PN deviation distribution is directly correlated with the mass ratio power-law index, $\beta$: steeper power-laws correspond to more variance in the GR deviation (larger $\beta$, larger $\sigma_{\rm PN}$). 
This is also manifested in the fact that when the PN deviation is assumed to be uniformly distributed (solid green), the astrophysical inference prefers steeper mass ratio power-laws (larger $\beta$), and that the analysis with deviations fixed to zero (dashed green) leads to a shallower slope ($\beta \lesssim 6$). 
There is also a correlation between $\sigma_\textrm{PN}$ and the width of the spin distribution, $\sigma_{\chi_z}$, by which a narrower spin distribution demands for a greater spread in deviation parameters within the population.

Such correlations highlight precisely why we need to account for the astrophysical population when testing GR. 
By assuming a particular, fixed model for the astrophysical population, the hyperparameter correlations will not be captured in the marginal posterior for the GR deviation population.
The analysis assuming the sampling prior for the astrophysical population (yellow), infers a value of $\sigma_\textrm{PN}$ which peaks at zero. 
Among other hyperparameters, the sampling prior corresponds to a uniform ($\beta=0$) mass-ratio distribution. 
Fixing the astrophysical population in such a way will lead to the hyperparameter posterior peaking at $\sigma_\textrm{PN}=0$, as seen in Fig.~\ref{fig:full_corner_dchi0}. 

\section{Conclusions}~\label{sec:conc}

In this study, we have shown the importance of modeling the astrophysical population when testing GR with gravitational waves. 
Current tests do not explicitly model the astrophysical population, and therefore implicitly treat the prior used for sampling the posterior distribution as the assumed astrophysical population.
Due to the presence of correlations between many GR deviations and astrophysical parameters, inappropriate astrophysical population choices will bias the test of GR.
Like other sources of systematics, including waveform modeling \citep{Moore2021, Hu2022bji, Saini:2022igm, Bhat:2022amc}, the severity of this bias increases with the number of detections. We have shown that the effect of this bias is already being felt in the present catalog.
This issue can only be fully addressed by simultaneously modelling both the astrophysical population in addition to the GR deviations.

We demonstrate the effect of inappropriate astrophysical models using constraints of the graviton's mass and tests of PN deviations as concrete examples. 
We show that jointly modeling the astrophysical population distribution while testing GR leads to results more consistent with GR.
Furthermore, for some deviations at various PN orders there are correlations between hyperparameters governing the astrophysical and deviation populations. 
The impact of the astrophysical distribution is not just important for these parameters and these hierarchical models: any test of GR should accurately account for the astrophysical population. 
In fact, this problem is not unique to tests of GR--- attempts to infer cosmological properties~\citep{LIGOScientific2021aug} or the equation of state of dense nuclear matter~\citep{Wysocki:2020myz} are also impacted by these same considerations. 

We can generically understand the impact of folding in the astrophysical population as follows. 
The standard sampling prior is chosen to broadly cover the parameter range of interest, and not to accurately represent the true astrophysical population. 
The actual population distribution will then typically provide support on a more narrow region of parameter space than the sampling prior. 
As a result, population-informed posteriors will not only avoid systematic biases but will also provide more stringent constraints on GR due to the additional information from the associated narrower population. 

\begin{figure*}
    \centering
    \includegraphics[width=\linewidth]{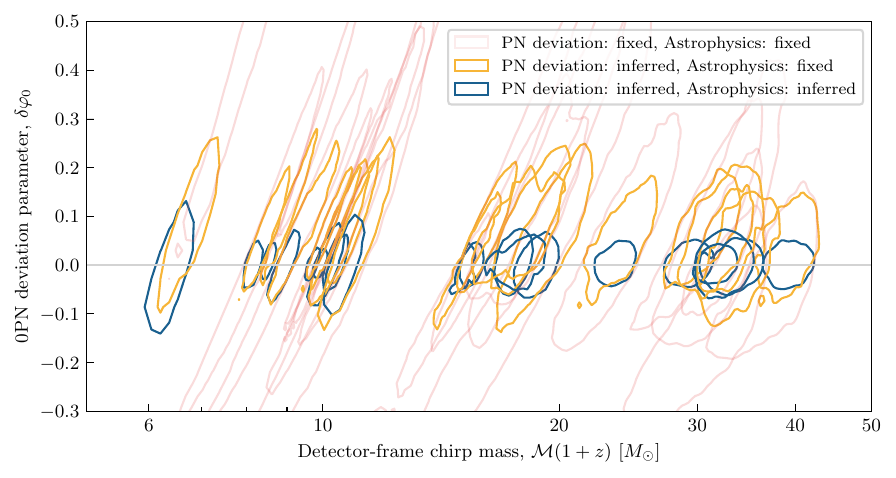}
    \caption{Marginal two-dimensional posterior distributions for the 0PN deviation coefficient and the detector-frame chirp mass for the events analyzed under the broad prior assumptions (light red), informed PN deviation population only (yellow), and informed by the jointly inferred deviation and astrophysical populations (dark blue). 
    Contours indicate the 90\% credible regions. 
    This result demonstrates that as additional information is incorporated into the population distribution, more stringent constraints on the deviation parameters are placed on an individual event level. 
    In the case demonstrated here, this pulls the inferred value towards $\delta\varphi=0$ for all events.}
    \label{fig:dphi0_correlations}
\end{figure*}

This posterior shrinkage is illustrated in Fig.~\ref{fig:dphi0_correlations}, which shows the 0PN deviation parameter and detector frame chirp mass for the 20 events considered in our study (Table \ref{tab:events}). 
The three sets of distributions correspond to the posteriors under different priors: fixed sampling priors (light red), fixed astrophysical prior and an inferred PN deviation population (yellow), as well as the case where both PN-deviation and astrophysics distributions are inferred (blue).
As more information about the GR deviation distribution is included, the inferred posterior of 0PN deviation parameter and the detector-frame chirp mass is more constrained. 
The posteriors are then constrained further still as additional information regarding the astrophysical population is included. 

There are a number of directions in which to extend our work. The first would be to account for selection effects on the hyperparameters of the GR deviation distribution; this is to be addressed in upcoming work~\citep{MageeInPrep}. 
Additionally, here we have assumed a strongly parameterized model for the astrophysical population, with a power law and a Gaussian peak.
This model is currently flexible enough given the number of events, with the primary mass Gaussian peak not impacting the inferred PN deviations with the selection of events considered.
As the number of events used with these tests increases, and subtle features in the astrophysical population reveal themselves, we will likely need more flexible models~\citep{Edelman_2022, Edelman_2023, golomb2022, Callister2023} to further avoid biases from misspecified population models~\citep{wmf, Gelman, Payne2023}.
Furthermore, in the case of PN coefficients, one would ideally constrain all orders simultaneously, in addition to the astrophysical parameters \citep{GW150914_TGR, Gupta:2020lxa, Datta:2020vcj, Shoom:2021mdj,  Perkins:2022fhr, Datta:2022izc}.

Concurrently modeling the astrophysical population when testing GR is inevitable. 
Models that do not include a parameterized astrophysical population are implicitly assuming the sampling prior as the fixed population model. 
Such an assumption may induce systematic biases, cause false detections of GR violations, or incorrectly claim a stronger confirmation of GR than is warranted by the data.
Moreover, even when accounting for the astrophysical population, correlations between GR deviation and astrophysical hyperparameters suggest that a true deviation could be absorbed into an unphysical inferred astrophysical population, a case that can only be noticed in studying the hyperposterior relating astrophysical to deviation parameters. 
Hierarchically modeling the astrophysical population while testing GR provides the solution to the implicit bias of assuming a fixed astrophysical population, and allows us to explore correlations between astrophysical parameters and deviations from GR, with fewer hidden assumptions. 

\section{Acknowledgements}

We thank Jacob Golomb and Alan Weinstein for insightful discussions, and Carl-Johan Haster for useful comments on the manuscript. 
Computing resources were provided by the Flatiron Institute. The Flatiron Institute is funded by the Simons Foundation.
EP was supported by NSF Grant PHY-1764464.
KC was supported by NSF Grant PHY-2110111.
This material is based upon work supported by NSF's LIGO Laboratory which is a major facility fully funded by the National Science Foundation.
This research has made use of data, software and/or web tools obtained from the Gravitational Wave Open Science Center (https://www.gw-openscience.org), a service of LIGO Laboratory, the LIGO Scientific Collaboration and the Virgo Collaboration.
Virgo is funded by the French Centre National de Recherche Scientifique (CNRS), the Italian Istituto Nazionale della Fisica Nucleare (INFN) and the Dutch Nikhef, with contributions by Polish and Hungarian institutes.
The authors are grateful for computational resources provided by the LIGO Laboratory and supported by National Science Foundation Grants PHY-0757058 and PHY-0823459.
This manuscript carries LIGO Document Number \#P2300292.

\appendix

\section{Formulation of parameterized tests of general relativity}~\label{app:tests}

In this appendix we outline the calculations required to constrain the graviton's mass (App.~\ref{app:mg}) and infer the PN deviation parameters (App.~\ref{app:pn}). 

\subsection{Massive graviton measurements}~\label{app:mg}

The impact of a massive graviton on the propagation of gravitational waves has been studied in Refs.~\citep{Will1997, Mirshekari2011} and references therein. 
A graviton with mass $m_g$ modifies the dispersion relation of the gravitational wave.
In a cosmological background, $g_{\mu\nu}$, 
\begin{equation}
    g_{\mu\nu}p^{\mu}p^{\nu} = -m_g^2
\end{equation}
where $p^{\mu}$ is the 4-momentum of the graviton. 
This leads to a dephasing of the gravitational wave, $\delta\Phi(f)$, that scales with the distance over which the signal propagates,
\begin{equation}
    \delta\Phi(f) = -\frac{\pi(1+z)D_L^2m_g^2c^3}{D_0h^2}f^{-1}\,,
\end{equation}
where $D_L$ is the luminosity distance, $h$ is Planck's constant, and 
\begin{equation}
    D_0 = \frac{c(1+z)}{H_0}\int_0^z\dd z'\,\frac{(1+z')^{-2}}{\sqrt{\Omega_m(1+z')^3 + \Omega_\Lambda}}\,.
\end{equation}
Here, $H_0 = 67.9\,\textrm{km s}^{-1}\, {\rm Mpc}^{-1}$ is the Hubble constant, and $\Omega_m = 0.3065$ and $\Omega_\Lambda = 0.6935$ are the matter and dark energy density parameters, respectively, adopting the values used in previous analyses~\citep{gwtc-3, GWTC3_TGR, Planck:2015fie}. 

\subsection{Post-Newtonian deviation tests}~\label{app:pn}

Current parameterized PN tests are constructed by single-parameter modifications to the post-Newtonian description of the inspiral gravitational-wave phase in the frequency domain. This is given by~\cite{sathyaprakash1991PN, arun2005PN}
\begin{align}~\label{eq:pn}
\Phi(f) &=2\pi f t_c - \phi_c - \frac{\pi}{4} +\frac{3}{128}\times\nonumber\\&\sum_{k=0}^7 \frac{1}{\eta^{k/5}}\Big(\varphi_k + \varphi_{k,l}\ln\tilde{f}\Big)\tilde{f}^{(k-5)/3}\,.
\end{align}
Here, $\Phi(f)$ is the frequency-domain gravitational-wave phase under the stationary-phase approximation, $\tilde{f} = \pi G\mathcal{M}(1+z)f/c^3$, where $\mathcal{M}(1+z)$ is the redshifted chirp mass, $\mathcal{M} = (m_1m_2)^{3/5}/(m_1+m_2)^{1/5}$ is the source-frame chirp mass, $\eta = m_1m_2/M^2$ is the symmetric mass ratio, $t_c$ and $\phi_c$ are the coalescence time and phase of the binary; finally, $k$ indexes the $k/2$ PN order, and $\varphi_k$ and $\varphi_{k,l}$ are the PN coefficients.
The logarithmic coefficients, $\varphi_{k,l}$ only enter at 2.5 and 3.5 PN orders and otherwise vanish~\citep{Blanchet1993tails, Blanchet1994tails}. 
In GR, the coefficients are functions of the intrinsic parameters of the binary, their masses and spins. 
From this prescription, modifications to GR are incorporated by modifying~\cite{Yunes2009ke, Li2012PN, agathos2014TIGER}
\begin{equation}
\varphi_k\rightarrow \left(1+\delta\varphi_k\right)\varphi_k\,,
\end{equation}
except for $k$'s for which $\varphi_k = 0$ in GR ($k=-2, 1$); in these cases, the modification is $\varphi_k \to \delta \varphi_k$, and $\delta \varphi_k$ is an absolute deviation~\cite{Will2014confront}.

In practice, modifications to \textsc{IMRPhenomPv2}~\citep{arun2005PN, khan2016PN,chatziioannou2017PN,agathos2014TIGER,Li2012PN, Meidam2014, Meidam2018} and \textsc{SEOBNRv4}~\citep{bohe2017, Cotesta2018, Cotesta2020, Brito2018, Mehta:2022pcn} waveforms are computed differently, then the latter is transformed to the former.
For the modified \textsc{SEOBNRv4} waveform, the deviation is applied as above~\citep{Mehta:2022pcn}.
While, \textsc{IMRPhenomPv2} is modified to only apply the deviation is onto the nonspinning portion of the PN coefficient~\citep{agathos2014TIGER,Li2012PN}. 
We translate all inferred deviation parameters to the \textsc{IMRPhenomPv2} deviation parameter $\delta\varphi^{\textrm{IMR}}_k$ for consistency, 
\begin{equation}
    \delta\varphi^{\textrm{IMR}}_k = \delta\varphi_k\frac{\varphi_k}{\varphi^{\textrm{NS}}_k}\,,
\end{equation}
where $\varphi^{\textrm{NS}}_k$ is the nonspinning value of the PN coefficient --- calculated by setting the spins to zero for a particular set of compact binary masses. 
Additionally, care needs to be taken when translating to a uniform prior on $\delta\varphi^{\textrm{IMR}}_k$, as the appropriate Jacobian, 
\begin{equation}
    \dv{\delta\varphi^{\textrm{IMR}}_k}{\delta\varphi_k} = \frac{\varphi_k}{\varphi^{\textrm{NS}}_k}\,,
\end{equation}
is necessary. 
If the original prior is uniform on $\delta\varphi_k$, then the $\delta\varphi^{\textrm{IMR}}_k$ must be weighted by the Jacobian to be effectively translated to another uniform prior.

\section{Computing expected parameter correlations}~\label{app:expected}

Correlations between GR deviation and astrophysical parameters can be analytically approximated by identifying regions of the parameter space that lead to a similar frequency evolution~\citep{psaltis2021eht} and signal duration.
The dominant correlation is the one between the detector-frame chirp mass, $\mathcal{M}(1+z)$, and the symmetric mass ratio, $\eta$. 
The duration of a gravitational-wave signal is related to the detector-frame chirp mass and some fiducial cut-off frequency~\citep{Cutler1994},
\begin{equation}
    T \propto \mathcal{M}^{5/3}(1+z)^{5/3}f_\textrm{cut}^{-8/3}\,. 
\end{equation}
If we relate the final frequency to the innermost stable orbit or any cut-off which scales inversely with the binary's total mass, then $T \propto \eta^{-8/5}\mathcal{M}^{13/3}(1+z)^{13/3}$. 
A constant duration then imlies
\begin{equation}
    \mathcal{M}(1+z) \propto \eta^{-24/65}\,.
\end{equation}
Here we have ignored both the contributions of a spin-induced ``hang-up'' effect~\citep{Campanelli2006uy} and GR deviations. 

Correlations between astrophysical parameters and GR deviations can then be computed at lowest order~\citep{psaltis2021eht} by enforcing that the second-order derivative of the phase evolution as a function of frequency be constant. 
As an example, for the correlation in Fig.~\ref{fig:correlated}, we compare the phase evolution when $\delta\varphi_0 = 0$ and when varying $\delta\varphi_0$ at the leading PN order, resulting in
\begin{equation}
   \mathcal{M}_0^{-5/3}(1+z_0)^{-5/3} \sim (1+\delta\varphi)\mathcal{M}^{-5/3}(1+z)^{-5/3}\,.
\end{equation}
Here $\mathcal{M}_0$ and $z_0$ are the values of the chirp mass and redshift when there is no deviation. 
We find the 0PN deviation coefficient to only be directly correlated with the detector frame chirp mass,
\begin{equation}
    \delta\varphi_0 \sim \Bigg(\frac{\mathcal{M}(1+z)}{\mathcal{M}_0(1+z_0)}\Bigg)^{5/3} - 1\,. 
\end{equation}
This calculation can be repeated for higher PN orders as well, however care needs to be taken as lower PN orders need to be retained when computing higher PN deviation coefficient correlations. 

\section{Population likelihood approximation}~\label{app:marg}

In practice, we carry out single-event parameter estimation with a fiducial sampling prior, $\pi(\theta)$, before the hierarchical population analysis. 
We therefore do not possess representations of the individual event likelihoods, $p(d | \theta)$, but rather samples drawn from the fiducial posterior distribution $p(\theta | d) \propto p(d|\theta)\, \pi(\theta)$.
Therefore, it is common to instead reformulate the integral within Eq.~\eqref{eq:poplike} as an average over samples drawn from each event's posterior distribution~\cite{Mandel2019, intro, Vitale2022}, 
\begin{equation}
    p(\{d\}|\Lambda) \propto \frac{1}{\xi(\Lambda)^N}\prod_{i=1}^N\frac{1}{M_i}\sum_{k=1}^{M_i}\frac{\pi(\theta_{i,k}|\Lambda)}{\pi(\theta_{i,k})}\,,~\label{eq:monte}
\end{equation}
where $M_i$ is the number of posterior samples for the $i$th event.
It is possible for this Monte Carlo integration to not converge---particularly if the population distribution $\pi(\theta|\Lambda)$ is narrower than posterior distributions for individual events~\cite{farr2019, intro, bbm, Callister2022, essick2022, talbot2023}.
This is particularly important in our scenario, since the inferred population of deviations from GR is typically narrower than marginal measurements from many individual events. 
This leads to a dearth of samples within the inferred GR deviation population, which subsequently leads to unreliable Monte Carlo integration in Eq.~\eqref{eq:monte}.

To address this issue, we use Gaussian kernel density estimates to represent the individual-event posteriors in a number of parameters, and simplify the calculation analytically by leveraging Gaussian population models.
Dividing the parameters into the subset described by the Gaussian population distributions, $\theta^\textrm{G}$, and the non-Gaussian distributions, $\theta^\textrm{NG}$, we can analytically integrate over the former without resorting to Eq.~\eqref{eq:monte}.
The Gaussian population parameters are the GR deviation parameter and the binary-hole spin magnitudes, whereas the black-hole primary mass and mass ratio, redshift, and spin tilts (for the analysis in App.~\ref{app:imr}) are included in the non-Gaussian set of parameters.
For the kernel density estimation, we determine the corresponding covariance matrix for each individual event's distribution using Scott's rule~\citep{scott1979optimal}, 
\begin{equation}
    \Sigma_{BW,i} \approx \frac{\Sigma_i}{n_{\textrm{eff},i}^{2/(d+4)}}\,,
\end{equation}
where $\Sigma_i$ is the weighted covariance matrix of the parameters being estimated, $d$ is the number of Gaussian dimensions, and $n_\textrm{eff}$ is the effective number of samples~\citep{Kish, Elvira},
\begin{equation}
    n_{\textrm{eff},i} = \frac{\Big(\sum_{k=1}^{M_i}w(\theta^G_{i,k})\Big)^2}{\sum_{k=1}^{M_i}w(\theta^G_{i,k})^2}\,,
\end{equation}
with the weights, $w(\theta^G_{i,k}) = 1/\pi(\theta^G_{i,k})$.

Since the integrand in the $\theta^\textrm{G}$-space is a product of Gaussian distributions, the resulting integral is also a Gaussian~\citep{Hogg2020}. 
This leads to the straightforward expression for the likelihood function
\begin{align}
p(\{d\}|\Lambda) \propto &\frac{1}{\xi(\Lambda)^N}\prod_{i=1}^N\frac{1}{M_i}\sum_{k=1}^{M_i}\frac{\pi(\theta^\textrm{NG}_{i,k}|\Lambda)}{\pi(\theta_{i,k})}\,\times\nonumber\\
&\mathcal{N}[\mu(\Lambda), \Sigma_{BW} + \Sigma(\Lambda)](\theta^\textrm{G}_{i,k})\,,
\end{align}
where $\mu(\Lambda) = (\mu, \mu_\chi, \mu_\chi)$ and $\Sigma(\Lambda) = \textrm{diag}(\sigma^2, \sigma^2_\chi, \sigma^2_\chi)$, though more complicated structure can be imposed on the population model. 
Since this integral is computed analytically, we empirically find improved convergence. 

\section{Constraints from {\tt \sc IMRPhenomPv2}}~\label{app:imr}

\begin{figure*}
    \centering
    \includegraphics[width=\linewidth]{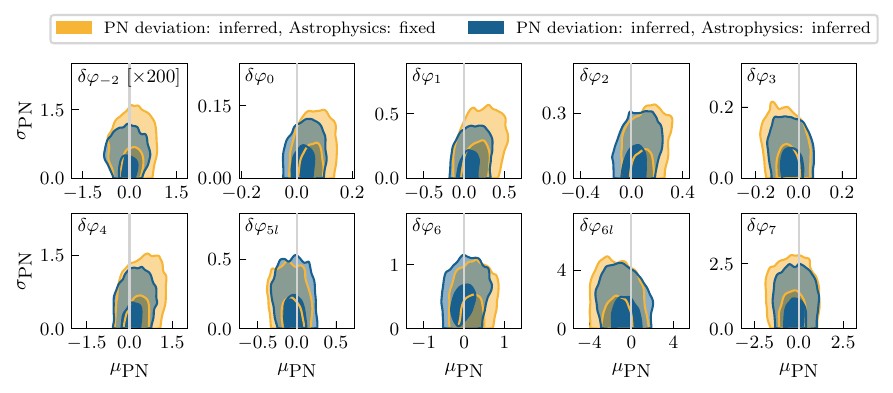}
    \caption{Same figure as Fig.~\ref{fig:tgr_mu_sigma_seob} but using 12 events from the first half of the third LIGO-Virgo-KAGRA observing run, with individual event posterior distributions constructed with {\tt \sc IMRPhenomPv2}. 
    We generally observe similar structure to the results with {\tt \sc SEOBNRv4}, although parameters are less constrained---likely due to fewer observations incorporated.}
    \label{fig:tgr_mu_sigma_phenom}
\end{figure*}

\begin{figure}
    \centering
    \includegraphics[width=\linewidth]{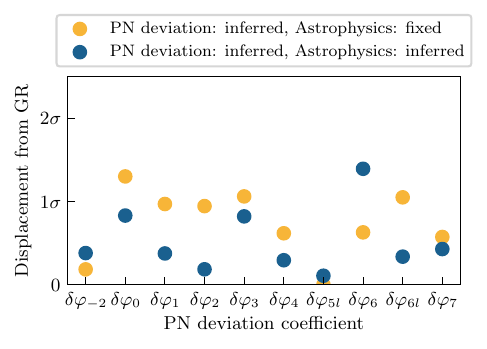}
    \caption{Same as Fig.~\ref{fig:quant_seob}, for the results from the {\tt \sc IMRPhenomPv2} analysis. 
    As seen throughout the manuscript, inclusion of the astrophysical population model in general leads to improved consistency with GR. 
    Furthermore, the posterior distributions sit closer to GR for {\tt \sc IMRPhenomPv2} than \textsc{SEOBNRv4}, likely as a result of analyzing fewer events.}
    \label{fig:phenom_quant}
\end{figure}

While we have focused on results from {\tt \sc SEOBNRv4}~\citep{bohe2017, Cotesta2018, Cotesta2020, Brito2018, Mehta:2022pcn}, these analyses do not include precessing degrees of freedom. 
However, evidence for precession has been found at the population level within gravitational-wave observations~\cite{gwtc-2_pop, gwtc-3_pop}.
Therefore, to explore if there are any major changes when incorporating precession effects, we use the 12 events from the first half of the third observing run analysed with {\tt \sc IMRPhenomPv2}~\citep{arun2005PN, khan2016PN,chatziioannou2017PN,agathos2014TIGER,Li2012PN, Meidam2014, Meidam2018} which meet our selection criteria~\citep{GWTC2_TGR}. 
There are no equivalent results from the second half of the third observing run~\citep{GWTC3_TGR}.
We show the summary of the marginal two-dimensional posterior distribution for the Gaussian population hyperparameters with and without the inclusion of astrophysical information in Fig.~\ref{fig:tgr_mu_sigma_phenom}. 
Generally, these results are less constrained due to the smaller number of events, though we still witness a similar shift in the means of the Gaussian populations as in Fig.~\ref{fig:tgr_mu_sigma_seob}. 
We also summarize the quantiles at which the expectation from GR presides in Fig.~\ref{fig:phenom_quant}.
Generally, the {\tt \sc IMRPhenomPv2} results are more consistent with GR than the equivalent {\tt \sc SEOBNRv4} results presented in Sec.~\ref{subs:pn}.
This could be a product of this waveform model incorporating precession, or simply that fewer events were analyzed, leading to a decrease in precision.

\bibliography{refs}

\begin{thebibliography}{122}%
\makeatletter
\providecommand \@ifxundefined [1]{%
 \@ifx{#1\undefined}
}%
\providecommand \@ifnum [1]{%
 \ifnum #1\expandafter \@firstoftwo
 \else \expandafter \@secondoftwo
 \fi
}%
\providecommand \@ifx [1]{%
 \ifx #1\expandafter \@firstoftwo
 \else \expandafter \@secondoftwo
 \fi
}%
\providecommand \natexlab [1]{#1}%
\providecommand \enquote  [1]{``#1''}%
\providecommand \bibnamefont  [1]{#1}%
\providecommand \bibfnamefont [1]{#1}%
\providecommand \citenamefont [1]{#1}%
\providecommand \href@noop [0]{\@secondoftwo}%
\providecommand \href [0]{\begingroup \@sanitize@url \@href}%
\providecommand \@href[1]{\@@startlink{#1}\@@href}%
\providecommand \@@href[1]{\endgroup#1\@@endlink}%
\providecommand \@sanitize@url [0]{\catcode `\\12\catcode `\$12\catcode
  `\&12\catcode `\#12\catcode `\^12\catcode `\_12\catcode `\%12\relax}%
\providecommand \@@startlink[1]{}%
\providecommand \@@endlink[0]{}%
\providecommand \url  [0]{\begingroup\@sanitize@url \@url }%
\providecommand \@url [1]{\endgroup\@href {#1}{\urlprefix }}%
\providecommand \urlprefix  [0]{URL }%
\providecommand \Eprint [0]{\href }%
\providecommand \doibase [0]{https://doi.org/}%
\providecommand \selectlanguage [0]{\@gobble}%
\providecommand \bibinfo  [0]{\@secondoftwo}%
\providecommand \bibfield  [0]{\@secondoftwo}%
\providecommand \translation [1]{[#1]}%
\providecommand \BibitemOpen [0]{}%
\providecommand \bibitemStop [0]{}%
\providecommand \bibitemNoStop [0]{.\EOS\space}%
\providecommand \EOS [0]{\spacefactor3000\relax}%
\providecommand \BibitemShut  [1]{\csname bibitem#1\endcsname}%
\let\auto@bib@innerbib\@empty
\bibitem [{\citenamefont {{Abbott}}\ \emph {et~al.}(2016)\citenamefont
  {{Abbott}} \emph {et~al.}}]{GW150914_TGR}%
  \BibitemOpen
  \bibfield  {author} {\bibinfo {author} {\bibfnamefont {B.~P.}\ \bibnamefont
  {{Abbott}}} \emph {et~al.} (\bibinfo {collaboration} {{LIGO Scientific
  Collaboration} and {Virgo Collaboration}}),\ }\bibfield  {title} {\bibinfo
  {title} {{Tests of General Relativity with GW150914}},\ }\href
  {https://doi.org/10.1103/PhysRevLett.116.221101} {\bibfield  {journal}
  {\bibinfo  {journal} {\prl}\ }\textbf {\bibinfo {volume} {116}},\ \bibinfo
  {eid} {221101} (\bibinfo {year} {2016})},\ \Eprint
  {https://arxiv.org/abs/1602.03841} {arXiv:1602.03841 [gr-qc]} \BibitemShut
  {NoStop}%
\bibitem [{\citenamefont {{Abbott}}\ \emph
  {et~al.}(2017{\natexlab{a}})\citenamefont {{Abbott}} \emph
  {et~al.}}]{GW170104_TGR}%
  \BibitemOpen
  \bibfield  {author} {\bibinfo {author} {\bibfnamefont {B.~P.}\ \bibnamefont
  {{Abbott}}} \emph {et~al.} (\bibinfo {collaboration} {{LIGO Scientific
  Collaboration} and {Virgo Collaboration}}),\ }\bibfield  {title} {\bibinfo
  {title} {{GW170104: Observation of a 50-Solar-Mass Binary Black Hole
  Coalescence at Redshift 0.2}},\ }\href
  {https://doi.org/10.1103/PhysRevLett.118.221101} {\bibfield  {journal}
  {\bibinfo  {journal} {\prl}\ }\textbf {\bibinfo {volume} {118}},\ \bibinfo
  {eid} {221101} (\bibinfo {year} {2017}{\natexlab{a}})},\ \Eprint
  {https://arxiv.org/abs/1706.01812} {arXiv:1706.01812 [gr-qc]} \BibitemShut
  {NoStop}%
\bibitem [{\citenamefont {{Abbott}}\ \emph
  {et~al.}(2017{\natexlab{b}})\citenamefont {{Abbott}} \emph
  {et~al.}}]{GW170814_TGR}%
  \BibitemOpen
  \bibfield  {author} {\bibinfo {author} {\bibfnamefont {B.~P.}\ \bibnamefont
  {{Abbott}}} \emph {et~al.} (\bibinfo {collaboration} {{LIGO Scientific
  Collaboration} and {Virgo Collaboration}}),\ }\bibfield  {title} {\bibinfo
  {title} {{GW170814: A Three-Detector Observation of Gravitational Waves from
  a Binary Black Hole Coalescence}},\ }\href
  {https://doi.org/10.1103/PhysRevLett.119.141101} {\bibfield  {journal}
  {\bibinfo  {journal} {\prl}\ }\textbf {\bibinfo {volume} {119}},\ \bibinfo
  {eid} {141101} (\bibinfo {year} {2017}{\natexlab{b}})},\ \Eprint
  {https://arxiv.org/abs/1709.09660} {arXiv:1709.09660 [gr-qc]} \BibitemShut
  {NoStop}%
\bibitem [{\citenamefont {{Abbott}}\ \emph
  {et~al.}(2019{\natexlab{a}})\citenamefont {{Abbott}} \emph
  {et~al.}}]{GW170817_TGR}%
  \BibitemOpen
  \bibfield  {author} {\bibinfo {author} {\bibfnamefont {B.~P.}\ \bibnamefont
  {{Abbott}}} \emph {et~al.} (\bibinfo {collaboration} {{LIGO Scientific
  Collaboration} and {Virgo Collaboration}}),\ }\bibfield  {title} {\bibinfo
  {title} {{Tests of General Relativity with GW170817}},\ }\href
  {https://doi.org/10.1103/PhysRevLett.123.011102} {\bibfield  {journal}
  {\bibinfo  {journal} {\prl}\ }\textbf {\bibinfo {volume} {123}},\ \bibinfo
  {eid} {011102} (\bibinfo {year} {2019}{\natexlab{a}})},\ \Eprint
  {https://arxiv.org/abs/1811.00364} {arXiv:1811.00364 [gr-qc]} \BibitemShut
  {NoStop}%
\bibitem [{\citenamefont {{Abbott}}\ \emph
  {et~al.}(2019{\natexlab{b}})\citenamefont {{Abbott}} \emph
  {et~al.}}]{GWTC1_TGR}%
  \BibitemOpen
  \bibfield  {author} {\bibinfo {author} {\bibfnamefont {B.~P.}\ \bibnamefont
  {{Abbott}}} \emph {et~al.} (\bibinfo {collaboration} {{LIGO Scientific
  Collaboration} and {Virgo Collaboration}}),\ }\bibfield  {title} {\bibinfo
  {title} {{Tests of general relativity with the binary black hole signals from
  the LIGO-Virgo catalog GWTC-1}},\ }\href
  {https://doi.org/10.1103/PhysRevD.100.104036} {\bibfield  {journal} {\bibinfo
   {journal} {\prd}\ }\textbf {\bibinfo {volume} {100}},\ \bibinfo {eid}
  {104036} (\bibinfo {year} {2019}{\natexlab{b}})},\ \Eprint
  {https://arxiv.org/abs/1903.04467} {arXiv:1903.04467 [gr-qc]} \BibitemShut
  {NoStop}%
\bibitem [{\citenamefont {{Abbott}}\ \emph
  {et~al.}(2021{\natexlab{a}})\citenamefont {{Abbott}} \emph
  {et~al.}}]{GWTC2_TGR}%
  \BibitemOpen
  \bibfield  {author} {\bibinfo {author} {\bibfnamefont {R.}~\bibnamefont
  {{Abbott}}} \emph {et~al.} (\bibinfo {collaboration} {{LIGO Scientific
  Collaboration} and {Virgo Collaboration}}),\ }\bibfield  {title} {\bibinfo
  {title} {{Tests of general relativity with binary black holes from the second
  LIGO-Virgo gravitational-wave transient catalog}},\ }\href
  {https://doi.org/10.1103/PhysRevD.103.122002} {\bibfield  {journal} {\bibinfo
   {journal} {\prd}\ }\textbf {\bibinfo {volume} {103}},\ \bibinfo {eid}
  {122002} (\bibinfo {year} {2021}{\natexlab{a}})},\ \Eprint
  {https://arxiv.org/abs/2010.14529} {arXiv:2010.14529 [gr-qc]} \BibitemShut
  {NoStop}%
\bibitem [{\citenamefont {{Abbott}}\ \emph
  {et~al.}(2021{\natexlab{b}})\citenamefont {{Abbott}} \emph
  {et~al.}}]{GWTC3_TGR}%
  \BibitemOpen
  \bibfield  {author} {\bibinfo {author} {\bibfnamefont {R.}~\bibnamefont
  {{Abbott}}} \emph {et~al.} (\bibinfo {collaboration} {{LIGO Scientific
  Collaboration} and {the Virgo Collaboration} and {the KAGRA
  Collaboration}}),\ }\bibfield  {title} {\bibinfo {title} {{Tests of General
  Relativity with GWTC-3}},\ }\href {https://doi.org/10.48550/arXiv.2112.06861}
  {\bibfield  {journal} {\bibinfo  {journal} {arXiv e-prints}\ ,\ \bibinfo
  {eid} {arXiv:2112.06861}} (\bibinfo {year} {2021}{\natexlab{b}})},\ \Eprint
  {https://arxiv.org/abs/2112.06861} {arXiv:2112.06861 [gr-qc]} \BibitemShut
  {NoStop}%
\bibitem [{\citenamefont {Aasi}\ \emph {et~al.}(2015)\citenamefont {Aasi} \emph
  {et~al.}}]{LIGO}%
  \BibitemOpen
  \bibfield  {author} {\bibinfo {author} {\bibfnamefont {J.}~\bibnamefont
  {Aasi}} \emph {et~al.} (\bibinfo {collaboration} {LIGO Scientific}),\
  }\bibfield  {title} {\bibinfo {title} {{Advanced LIGO}},\ }\href
  {https://doi.org/10.1088/0264-9381/32/7/074001} {\bibfield  {journal}
  {\bibinfo  {journal} {Class. Quant. Grav.}\ }\textbf {\bibinfo {volume}
  {32}},\ \bibinfo {pages} {074001} (\bibinfo {year} {2015})},\ \Eprint
  {https://arxiv.org/abs/1411.4547} {arXiv:1411.4547 [gr-qc]} \BibitemShut
  {NoStop}%
\bibitem [{\citenamefont {Acernese}\ \emph {et~al.}(2015)\citenamefont
  {Acernese} \emph {et~al.}}]{Virgo}%
  \BibitemOpen
  \bibfield  {author} {\bibinfo {author} {\bibfnamefont {F.}~\bibnamefont
  {Acernese}} \emph {et~al.} (\bibinfo {collaboration} {VIRGO}),\ }\bibfield
  {title} {\bibinfo {title} {{Advanced Virgo: a second-generation
  interferometric gravitational wave detector}},\ }\href
  {https://doi.org/10.1088/0264-9381/32/2/024001} {\bibfield  {journal}
  {\bibinfo  {journal} {Class. Quant. Grav.}\ }\textbf {\bibinfo {volume}
  {32}},\ \bibinfo {pages} {024001} (\bibinfo {year} {2015})},\ \Eprint
  {https://arxiv.org/abs/1408.3978} {arXiv:1408.3978 [gr-qc]} \BibitemShut
  {NoStop}%
\bibitem [{\citenamefont {Ghosh}\ \emph {et~al.}(2016)\citenamefont {Ghosh}
  \emph {et~al.}}]{Ghosh2016qgn}%
  \BibitemOpen
  \bibfield  {author} {\bibinfo {author} {\bibfnamefont {A.}~\bibnamefont
  {Ghosh}} \emph {et~al.},\ }\bibfield  {title} {\bibinfo {title} {{Testing
  general relativity using golden black-hole binaries}},\ }\href
  {https://doi.org/10.1103/PhysRevD.94.021101} {\bibfield  {journal} {\bibinfo
  {journal} {Phys. Rev. D}\ }\textbf {\bibinfo {volume} {94}},\ \bibinfo
  {pages} {021101} (\bibinfo {year} {2016})},\ \Eprint
  {https://arxiv.org/abs/1602.02453} {arXiv:1602.02453 [gr-qc]} \BibitemShut
  {NoStop}%
\bibitem [{\citenamefont {Ghosh}\ \emph {et~al.}(2018)\citenamefont {Ghosh},
  \citenamefont {Johnson-Mcdaniel}, \citenamefont {Ghosh}, \citenamefont
  {Mishra}, \citenamefont {Ajith}, \citenamefont {Del~Pozzo}, \citenamefont
  {Berry}, \citenamefont {Nielsen},\ and\ \citenamefont
  {London}}]{Ghosh2017gfp}%
  \BibitemOpen
  \bibfield  {author} {\bibinfo {author} {\bibfnamefont {A.}~\bibnamefont
  {Ghosh}}, \bibinfo {author} {\bibfnamefont {N.~K.}\ \bibnamefont
  {Johnson-Mcdaniel}}, \bibinfo {author} {\bibfnamefont {A.}~\bibnamefont
  {Ghosh}}, \bibinfo {author} {\bibfnamefont {C.~K.}\ \bibnamefont {Mishra}},
  \bibinfo {author} {\bibfnamefont {P.}~\bibnamefont {Ajith}}, \bibinfo
  {author} {\bibfnamefont {W.}~\bibnamefont {Del~Pozzo}}, \bibinfo {author}
  {\bibfnamefont {C.~P.~L.}\ \bibnamefont {Berry}}, \bibinfo {author}
  {\bibfnamefont {A.~B.}\ \bibnamefont {Nielsen}},\ and\ \bibinfo {author}
  {\bibfnamefont {L.}~\bibnamefont {London}},\ }\bibfield  {title} {\bibinfo
  {title} {{Testing general relativity using gravitational wave signals from
  the inspiral, merger and ringdown of binary black holes}},\ }\href
  {https://doi.org/10.1088/1361-6382/aa972e} {\bibfield  {journal} {\bibinfo
  {journal} {Class. Quant. Grav.}\ }\textbf {\bibinfo {volume} {35}},\ \bibinfo
  {pages} {014002} (\bibinfo {year} {2018})},\ \Eprint
  {https://arxiv.org/abs/1704.06784} {arXiv:1704.06784 [gr-qc]} \BibitemShut
  {NoStop}%
\bibitem [{\citenamefont {{Li}}\ \emph {et~al.}(2012)\citenamefont {{Li}},
  \citenamefont {{Del Pozzo}}, \citenamefont {{Vitale}}, \citenamefont {{Van
  Den Broeck}}, \citenamefont {{Agathos}}, \citenamefont {{Veitch}},
  \citenamefont {{Grover}}, \citenamefont {{Sidery}}, \citenamefont
  {{Sturani}},\ and\ \citenamefont {{Vecchio}}}]{Li2012PN}%
  \BibitemOpen
  \bibfield  {author} {\bibinfo {author} {\bibfnamefont {T.~G.~F.}\
  \bibnamefont {{Li}}}, \bibinfo {author} {\bibfnamefont {W.}~\bibnamefont
  {{Del Pozzo}}}, \bibinfo {author} {\bibfnamefont {S.}~\bibnamefont
  {{Vitale}}}, \bibinfo {author} {\bibfnamefont {C.}~\bibnamefont {{Van Den
  Broeck}}}, \bibinfo {author} {\bibfnamefont {M.}~\bibnamefont {{Agathos}}},
  \bibinfo {author} {\bibfnamefont {J.}~\bibnamefont {{Veitch}}}, \bibinfo
  {author} {\bibfnamefont {K.}~\bibnamefont {{Grover}}}, \bibinfo {author}
  {\bibfnamefont {T.}~\bibnamefont {{Sidery}}}, \bibinfo {author}
  {\bibfnamefont {R.}~\bibnamefont {{Sturani}}},\ and\ \bibinfo {author}
  {\bibfnamefont {A.}~\bibnamefont {{Vecchio}}},\ }\bibfield  {title} {\bibinfo
  {title} {{Towards a generic test of the strong field dynamics of general
  relativity using compact binary coalescence}},\ }\href
  {https://doi.org/10.1103/PhysRevD.85.082003} {\bibfield  {journal} {\bibinfo
  {journal} {\prd}\ }\textbf {\bibinfo {volume} {85}},\ \bibinfo {eid} {082003}
  (\bibinfo {year} {2012})},\ \Eprint {https://arxiv.org/abs/1110.0530}
  {arXiv:1110.0530 [gr-qc]} \BibitemShut {NoStop}%
\bibitem [{\citenamefont {{Agathos}}\ \emph {et~al.}(2014)\citenamefont
  {{Agathos}}, \citenamefont {{Del Pozzo}}, \citenamefont {{Li}}, \citenamefont
  {{Van Den Broeck}}, \citenamefont {{Veitch}},\ and\ \citenamefont
  {{Vitale}}}]{agathos2014TIGER}%
  \BibitemOpen
  \bibfield  {author} {\bibinfo {author} {\bibfnamefont {M.}~\bibnamefont
  {{Agathos}}}, \bibinfo {author} {\bibfnamefont {W.}~\bibnamefont {{Del
  Pozzo}}}, \bibinfo {author} {\bibfnamefont {T.~G.~F.}\ \bibnamefont {{Li}}},
  \bibinfo {author} {\bibfnamefont {C.}~\bibnamefont {{Van Den Broeck}}},
  \bibinfo {author} {\bibfnamefont {J.}~\bibnamefont {{Veitch}}},\ and\
  \bibinfo {author} {\bibfnamefont {S.}~\bibnamefont {{Vitale}}},\ }\bibfield
  {title} {\bibinfo {title} {{TIGER: A data analysis pipeline for testing the
  strong-field dynamics of general relativity with gravitational wave signals
  from coalescing compact binaries}},\ }\href
  {https://doi.org/10.1103/PhysRevD.89.082001} {\bibfield  {journal} {\bibinfo
  {journal} {\prd}\ }\textbf {\bibinfo {volume} {89}},\ \bibinfo {eid} {082001}
  (\bibinfo {year} {2014})},\ \Eprint {https://arxiv.org/abs/1311.0420}
  {arXiv:1311.0420 [gr-qc]} \BibitemShut {NoStop}%
\bibitem [{\citenamefont {Mehta}\ \emph {et~al.}(2023)\citenamefont {Mehta},
  \citenamefont {Buonanno}, \citenamefont {Cotesta}, \citenamefont {Ghosh},
  \citenamefont {Sennett},\ and\ \citenamefont {Steinhoff}}]{Mehta:2022pcn}%
  \BibitemOpen
  \bibfield  {author} {\bibinfo {author} {\bibfnamefont {A.~K.}\ \bibnamefont
  {Mehta}}, \bibinfo {author} {\bibfnamefont {A.}~\bibnamefont {Buonanno}},
  \bibinfo {author} {\bibfnamefont {R.}~\bibnamefont {Cotesta}}, \bibinfo
  {author} {\bibfnamefont {A.}~\bibnamefont {Ghosh}}, \bibinfo {author}
  {\bibfnamefont {N.}~\bibnamefont {Sennett}},\ and\ \bibinfo {author}
  {\bibfnamefont {J.}~\bibnamefont {Steinhoff}},\ }\bibfield  {title} {\bibinfo
  {title} {{Tests of general relativity with gravitational-wave observations
  using a flexible theory-independent method}},\ }\href
  {https://doi.org/10.1103/PhysRevD.107.044020} {\bibfield  {journal} {\bibinfo
   {journal} {Phys. Rev. D}\ }\textbf {\bibinfo {volume} {107}},\ \bibinfo
  {pages} {044020} (\bibinfo {year} {2023})},\ \Eprint
  {https://arxiv.org/abs/2203.13937} {arXiv:2203.13937 [gr-qc]} \BibitemShut
  {NoStop}%
\bibitem [{\citenamefont {Will}(1998)}]{Will1997}%
  \BibitemOpen
  \bibfield  {author} {\bibinfo {author} {\bibfnamefont {C.~M.}\ \bibnamefont
  {Will}},\ }\bibfield  {title} {\bibinfo {title} {{Bounding the mass of the
  graviton using gravitational wave observations of inspiralling compact
  binaries}},\ }\href {https://doi.org/10.1103/PhysRevD.57.2061} {\bibfield
  {journal} {\bibinfo  {journal} {Phys. Rev. D}\ }\textbf {\bibinfo {volume}
  {57}},\ \bibinfo {pages} {2061} (\bibinfo {year} {1998})},\ \Eprint
  {https://arxiv.org/abs/gr-qc/9709011} {arXiv:gr-qc/9709011} \BibitemShut
  {NoStop}%
\bibitem [{\citenamefont {Mirshekari}\ \emph {et~al.}(2012)\citenamefont
  {Mirshekari}, \citenamefont {Yunes},\ and\ \citenamefont
  {Will}}]{Mirshekari2011}%
  \BibitemOpen
  \bibfield  {author} {\bibinfo {author} {\bibfnamefont {S.}~\bibnamefont
  {Mirshekari}}, \bibinfo {author} {\bibfnamefont {N.}~\bibnamefont {Yunes}},\
  and\ \bibinfo {author} {\bibfnamefont {C.~M.}\ \bibnamefont {Will}},\
  }\bibfield  {title} {\bibinfo {title} {{Constraining Generic Lorentz
  Violation and the Speed of the Graviton with Gravitational Waves}},\ }\href
  {https://doi.org/10.1103/PhysRevD.85.024041} {\bibfield  {journal} {\bibinfo
  {journal} {Phys. Rev. D}\ }\textbf {\bibinfo {volume} {85}},\ \bibinfo
  {pages} {024041} (\bibinfo {year} {2012})},\ \Eprint
  {https://arxiv.org/abs/1110.2720} {arXiv:1110.2720 [gr-qc]} \BibitemShut
  {NoStop}%
\bibitem [{\citenamefont {Zhu}\ \emph {et~al.}(2023)\citenamefont {Zhu},
  \citenamefont {Zhao}, \citenamefont {Yan}, \citenamefont {Gong},\ and\
  \citenamefont {Wang}}]{Zhu:2023wci}%
  \BibitemOpen
  \bibfield  {author} {\bibinfo {author} {\bibfnamefont {T.}~\bibnamefont
  {Zhu}}, \bibinfo {author} {\bibfnamefont {W.}~\bibnamefont {Zhao}}, \bibinfo
  {author} {\bibfnamefont {J.-M.}\ \bibnamefont {Yan}}, \bibinfo {author}
  {\bibfnamefont {C.}~\bibnamefont {Gong}},\ and\ \bibinfo {author}
  {\bibfnamefont {A.}~\bibnamefont {Wang}},\ }\bibfield  {title} {\bibinfo
  {title} {{Tests of modified gravitational wave propagations with
  gravitational waves}},\ }\Eprint {https://arxiv.org/abs/2304.09025}
  {arXiv:2304.09025 [gr-qc]}  (\bibinfo {year} {2023})\BibitemShut {NoStop}%
\bibitem [{\citenamefont {Ng}\ \emph {et~al.}(2023)\citenamefont {Ng},
  \citenamefont {Isi}, \citenamefont {Wong},\ and\ \citenamefont
  {Farr}}]{Ng:2023jjt}%
  \BibitemOpen
  \bibfield  {author} {\bibinfo {author} {\bibfnamefont {T.~C.~K.}\
  \bibnamefont {Ng}}, \bibinfo {author} {\bibfnamefont {M.}~\bibnamefont
  {Isi}}, \bibinfo {author} {\bibfnamefont {K.~W.~K.}\ \bibnamefont {Wong}},\
  and\ \bibinfo {author} {\bibfnamefont {W.~M.}\ \bibnamefont {Farr}},\
  }\href@noop {} {\bibinfo {title} {{Constraining gravitational wave amplitude
  birefringence with GWTC-3}}} (\bibinfo {year} {2023}),\ \Eprint
  {https://arxiv.org/abs/2305.05844} {arXiv:2305.05844 [gr-qc]} \BibitemShut
  {NoStop}%
\bibitem [{\citenamefont {Eardley}\ \emph
  {et~al.}(1973{\natexlab{a}})\citenamefont {Eardley}, \citenamefont {Lee},
  \citenamefont {Lightman}, \citenamefont {Wagoner},\ and\ \citenamefont
  {Will}}]{Eardley1973}%
  \BibitemOpen
  \bibfield  {author} {\bibinfo {author} {\bibfnamefont {D.~M.}\ \bibnamefont
  {Eardley}}, \bibinfo {author} {\bibfnamefont {D.~L.}\ \bibnamefont {Lee}},
  \bibinfo {author} {\bibfnamefont {A.~P.}\ \bibnamefont {Lightman}}, \bibinfo
  {author} {\bibfnamefont {R.~V.}\ \bibnamefont {Wagoner}},\ and\ \bibinfo
  {author} {\bibfnamefont {C.~M.}\ \bibnamefont {Will}},\ }\bibfield  {title}
  {\bibinfo {title} {Gravitational-wave observations as a tool for testing
  relativistic gravity},\ }\href {https://doi.org/10.1103/PhysRevLett.30.884}
  {\bibfield  {journal} {\bibinfo  {journal} {Phys. Rev. Lett.}\ }\textbf
  {\bibinfo {volume} {30}},\ \bibinfo {pages} {884} (\bibinfo {year}
  {1973}{\natexlab{a}})}\BibitemShut {NoStop}%
\bibitem [{\citenamefont {Eardley}\ \emph
  {et~al.}(1973{\natexlab{b}})\citenamefont {Eardley}, \citenamefont {Lee},\
  and\ \citenamefont {Lightman}}]{Eardley1973b}%
  \BibitemOpen
  \bibfield  {author} {\bibinfo {author} {\bibfnamefont {D.~M.}\ \bibnamefont
  {Eardley}}, \bibinfo {author} {\bibfnamefont {D.~L.}\ \bibnamefont {Lee}},\
  and\ \bibinfo {author} {\bibfnamefont {A.~P.}\ \bibnamefont {Lightman}},\
  }\bibfield  {title} {\bibinfo {title} {Gravitational-wave observations as a
  tool for testing relativistic gravity},\ }\href
  {https://doi.org/10.1103/PhysRevD.8.3308} {\bibfield  {journal} {\bibinfo
  {journal} {Phys. Rev. D}\ }\textbf {\bibinfo {volume} {8}},\ \bibinfo {pages}
  {3308} (\bibinfo {year} {1973}{\natexlab{b}})}\BibitemShut {NoStop}%
\bibitem [{\citenamefont {{Isi}}\ and\ \citenamefont
  {{Weinstein}}(2017)}]{Isi:2017fbj}%
  \BibitemOpen
  \bibfield  {author} {\bibinfo {author} {\bibfnamefont {M.}~\bibnamefont
  {{Isi}}}\ and\ \bibinfo {author} {\bibfnamefont {A.~J.}\ \bibnamefont
  {{Weinstein}}},\ }\bibfield  {title} {\bibinfo {title} {{Probing
  gravitational wave polarizations with signals from compact binary
  coalescences}},\ }\href {https://doi.org/10.48550/arXiv.1710.03794}
  {\bibfield  {journal} {\bibinfo  {journal} {arXiv e-prints}\ ,\ \bibinfo
  {eid} {arXiv:1710.03794}} (\bibinfo {year} {2017})},\ \Eprint
  {https://arxiv.org/abs/1710.03794} {arXiv:1710.03794 [gr-qc]} \BibitemShut
  {NoStop}%
\bibitem [{\citenamefont {Pang}\ \emph {et~al.}(2020)\citenamefont {Pang},
  \citenamefont {Lo}, \citenamefont {Wong}, \citenamefont {Li},\ and\
  \citenamefont {Van Den~Broeck}}]{Pang2020pfz}%
  \BibitemOpen
  \bibfield  {author} {\bibinfo {author} {\bibfnamefont {P.~T.~H.}\
  \bibnamefont {Pang}}, \bibinfo {author} {\bibfnamefont {R.~K.~L.}\
  \bibnamefont {Lo}}, \bibinfo {author} {\bibfnamefont {I.~C.~F.}\ \bibnamefont
  {Wong}}, \bibinfo {author} {\bibfnamefont {T.~G.~F.}\ \bibnamefont {Li}},\
  and\ \bibinfo {author} {\bibfnamefont {C.}~\bibnamefont {Van Den~Broeck}},\
  }\bibfield  {title} {\bibinfo {title} {{Generic searches for alternative
  gravitational wave polarizations with networks of interferometric
  detectors}},\ }\href {https://doi.org/10.1103/PhysRevD.101.104055} {\bibfield
   {journal} {\bibinfo  {journal} {Phys. Rev. D}\ }\textbf {\bibinfo {volume}
  {101}},\ \bibinfo {pages} {104055} (\bibinfo {year} {2020})},\ \Eprint
  {https://arxiv.org/abs/2003.07375} {arXiv:2003.07375 [gr-qc]} \BibitemShut
  {NoStop}%
\bibitem [{\citenamefont {Chatziioannou}\ \emph {et~al.}(2021)\citenamefont
  {Chatziioannou}, \citenamefont {Isi}, \citenamefont {Haster},\ and\
  \citenamefont {Littenberg}}]{Chatziioannou2021mij}%
  \BibitemOpen
  \bibfield  {author} {\bibinfo {author} {\bibfnamefont {K.}~\bibnamefont
  {Chatziioannou}}, \bibinfo {author} {\bibfnamefont {M.}~\bibnamefont {Isi}},
  \bibinfo {author} {\bibfnamefont {C.-J.}\ \bibnamefont {Haster}},\ and\
  \bibinfo {author} {\bibfnamefont {T.~B.}\ \bibnamefont {Littenberg}},\
  }\bibfield  {title} {\bibinfo {title} {{Morphology-independent test of the
  mixed polarization content of transient gravitational wave signals}},\ }\href
  {https://doi.org/10.1103/PhysRevD.104.044005} {\bibfield  {journal} {\bibinfo
   {journal} {Phys. Rev. D}\ }\textbf {\bibinfo {volume} {104}},\ \bibinfo
  {pages} {044005} (\bibinfo {year} {2021})},\ \Eprint
  {https://arxiv.org/abs/2105.01521} {arXiv:2105.01521 [gr-qc]} \BibitemShut
  {NoStop}%
\bibitem [{\citenamefont {{Isi}}\ \emph {et~al.}(2019)\citenamefont {{Isi}},
  \citenamefont {{Chatziioannou}},\ and\ \citenamefont {{Farr}}}]{isi2019hier}%
  \BibitemOpen
  \bibfield  {author} {\bibinfo {author} {\bibfnamefont {M.}~\bibnamefont
  {{Isi}}}, \bibinfo {author} {\bibfnamefont {K.}~\bibnamefont
  {{Chatziioannou}}},\ and\ \bibinfo {author} {\bibfnamefont {W.~M.}\
  \bibnamefont {{Farr}}},\ }\bibfield  {title} {\bibinfo {title} {{Hierarchical
  Test of General Relativity with Gravitational Waves}},\ }\href
  {https://doi.org/10.1103/PhysRevLett.123.121101} {\bibfield  {journal}
  {\bibinfo  {journal} {\prl}\ }\textbf {\bibinfo {volume} {123}},\ \bibinfo
  {eid} {121101} (\bibinfo {year} {2019})},\ \Eprint
  {https://arxiv.org/abs/1904.08011} {arXiv:1904.08011 [gr-qc]} \BibitemShut
  {NoStop}%
\bibitem [{\citenamefont {{Saleem}}\ \emph {et~al.}(2022)\citenamefont
  {{Saleem}}, \citenamefont {{Krishnendu}}, \citenamefont {{Ghosh}},
  \citenamefont {{Gupta}}, \citenamefont {{Del Pozzo}}, \citenamefont
  {{Ghosh}},\ and\ \citenamefont {{Arun}}}]{saleem2022hier}%
  \BibitemOpen
  \bibfield  {author} {\bibinfo {author} {\bibfnamefont {M.}~\bibnamefont
  {{Saleem}}}, \bibinfo {author} {\bibfnamefont {N.~V.}\ \bibnamefont
  {{Krishnendu}}}, \bibinfo {author} {\bibfnamefont {A.}~\bibnamefont
  {{Ghosh}}}, \bibinfo {author} {\bibfnamefont {A.}~\bibnamefont {{Gupta}}},
  \bibinfo {author} {\bibfnamefont {W.}~\bibnamefont {{Del Pozzo}}}, \bibinfo
  {author} {\bibfnamefont {A.}~\bibnamefont {{Ghosh}}},\ and\ \bibinfo {author}
  {\bibfnamefont {K.~G.}\ \bibnamefont {{Arun}}},\ }\bibfield  {title}
  {\bibinfo {title} {{Population inference of spin-induced quadrupole moments
  as a probe for nonblack hole compact binaries}},\ }\href
  {https://doi.org/10.1103/PhysRevD.105.104066} {\bibfield  {journal} {\bibinfo
   {journal} {\prd}\ }\textbf {\bibinfo {volume} {105}},\ \bibinfo {eid}
  {104066} (\bibinfo {year} {2022})},\ \Eprint
  {https://arxiv.org/abs/2111.04135} {arXiv:2111.04135 [gr-qc]} \BibitemShut
  {NoStop}%
\bibitem [{\citenamefont {{Zimmerman}}\ \emph {et~al.}(2019)\citenamefont
  {{Zimmerman}}, \citenamefont {{Haster}},\ and\ \citenamefont
  {{Chatziioannou}}}]{zimmerman2019combining}%
  \BibitemOpen
  \bibfield  {author} {\bibinfo {author} {\bibfnamefont {A.}~\bibnamefont
  {{Zimmerman}}}, \bibinfo {author} {\bibfnamefont {C.-J.}\ \bibnamefont
  {{Haster}}},\ and\ \bibinfo {author} {\bibfnamefont {K.}~\bibnamefont
  {{Chatziioannou}}},\ }\bibfield  {title} {\bibinfo {title} {{On combining
  information from multiple gravitational wave sources}},\ }\href
  {https://doi.org/10.1103/PhysRevD.99.124044} {\bibfield  {journal} {\bibinfo
  {journal} {\prd}\ }\textbf {\bibinfo {volume} {99}},\ \bibinfo {eid} {124044}
  (\bibinfo {year} {2019})},\ \Eprint {https://arxiv.org/abs/1903.11008}
  {arXiv:1903.11008 [astro-ph.IM]} \BibitemShut {NoStop}%
\bibitem [{\citenamefont {Isi}\ \emph {et~al.}(2022)\citenamefont {Isi},
  \citenamefont {Farr},\ and\ \citenamefont {Chatziioannou}}]{Isi:2022cii}%
  \BibitemOpen
  \bibfield  {author} {\bibinfo {author} {\bibfnamefont {M.}~\bibnamefont
  {Isi}}, \bibinfo {author} {\bibfnamefont {W.~M.}\ \bibnamefont {Farr}},\ and\
  \bibinfo {author} {\bibfnamefont {K.}~\bibnamefont {Chatziioannou}},\
  }\bibfield  {title} {\bibinfo {title} {{Comparing Bayes factors and
  hierarchical inference for testing general relativity with gravitational
  waves}},\ }\href {https://doi.org/10.1103/PhysRevD.106.024048} {\bibfield
  {journal} {\bibinfo  {journal} {Phys. Rev. D}\ }\textbf {\bibinfo {volume}
  {106}},\ \bibinfo {pages} {024048} (\bibinfo {year} {2022})},\ \Eprint
  {https://arxiv.org/abs/2204.10742} {arXiv:2204.10742 [gr-qc]} \BibitemShut
  {NoStop}%
\bibitem [{\citenamefont {{Psaltis}}\ \emph {et~al.}(2021)\citenamefont
  {{Psaltis}}, \citenamefont {{Talbot}}, \citenamefont {{Payne}},\ and\
  \citenamefont {{Mandel}}}]{psaltis2021eht}%
  \BibitemOpen
  \bibfield  {author} {\bibinfo {author} {\bibfnamefont {D.}~\bibnamefont
  {{Psaltis}}}, \bibinfo {author} {\bibfnamefont {C.}~\bibnamefont {{Talbot}}},
  \bibinfo {author} {\bibfnamefont {E.}~\bibnamefont {{Payne}}},\ and\ \bibinfo
  {author} {\bibfnamefont {I.}~\bibnamefont {{Mandel}}},\ }\bibfield  {title}
  {\bibinfo {title} {{Probing the black hole metric: Black hole shadows and
  binary black-hole inspirals}},\ }\href
  {https://doi.org/10.1103/PhysRevD.103.104036} {\bibfield  {journal} {\bibinfo
   {journal} {\prd}\ }\textbf {\bibinfo {volume} {103}},\ \bibinfo {eid}
  {104036} (\bibinfo {year} {2021})},\ \Eprint
  {https://arxiv.org/abs/2012.02117} {arXiv:2012.02117 [gr-qc]} \BibitemShut
  {NoStop}%
\bibitem [{\citenamefont {{Wolfe}}\ \emph {et~al.}(2022)\citenamefont
  {{Wolfe}}, \citenamefont {{Talbot}},\ and\ \citenamefont
  {{Golomb}}}]{wolfe2022hybrid}%
  \BibitemOpen
  \bibfield  {author} {\bibinfo {author} {\bibfnamefont {N.~E.}\ \bibnamefont
  {{Wolfe}}}, \bibinfo {author} {\bibfnamefont {C.}~\bibnamefont {{Talbot}}},\
  and\ \bibinfo {author} {\bibfnamefont {J.}~\bibnamefont {{Golomb}}},\
  }\bibfield  {title} {\bibinfo {title} {{Accelerating Tests of General
  Relativity with Gravitational-Wave Signals using Hybrid Sampling}},\ }\href
  {https://doi.org/10.48550/arXiv.2208.12872} {\bibfield  {journal} {\bibinfo
  {journal} {arXiv e-prints}\ ,\ \bibinfo {eid} {arXiv:2208.12872}} (\bibinfo
  {year} {2022})},\ \Eprint {https://arxiv.org/abs/2208.12872}
  {arXiv:2208.12872 [gr-qc]} \BibitemShut {NoStop}%
\bibitem [{\citenamefont {{Abbott}}\ \emph
  {et~al.}(2021{\natexlab{c}})\citenamefont {{Abbott}} \emph
  {et~al.}}]{gwtc-3_pop}%
  \BibitemOpen
  \bibfield  {author} {\bibinfo {author} {\bibfnamefont {R.}~\bibnamefont
  {{Abbott}}} \emph {et~al.} (\bibinfo {collaboration} {{The LIGO Scientific
  Collaboration} and {the Virgo Collaboration} and {the KAGRA
  Collaboration}}),\ }\bibfield  {title} {\bibinfo {title} {{The population of
  merging compact binaries inferred using gravitational waves through
  GWTC-3}},\ }\href {https://doi.org/10.48550/arXiv.2111.03634} {\bibfield
  {journal} {\bibinfo  {journal} {arXiv e-prints}\ ,\ \bibinfo {eid}
  {arXiv:2111.03634}} (\bibinfo {year} {2021}{\natexlab{c}})},\ \Eprint
  {https://arxiv.org/abs/2111.03634} {arXiv:2111.03634 [astro-ph.HE]}
  \BibitemShut {NoStop}%
\bibitem [{\citenamefont {Abbott}\ \emph {et~al.}(2021)\citenamefont {Abbott}
  \emph {et~al.}}]{gwtc-2_pop}%
  \BibitemOpen
  \bibfield  {author} {\bibinfo {author} {\bibfnamefont {R.}~\bibnamefont
  {Abbott}} \emph {et~al.},\ }\bibfield  {title} {\bibinfo {title} {{Population
  Properties of Compact Objects from the Second LIGO-Virgo Gravitational-Wave
  Transient Catalog}},\ }\href@noop {} {\bibfield  {journal} {\bibinfo
  {journal} {Astrophys. J.}\ }\textbf {\bibinfo {volume} {913}},\ \bibinfo
  {pages} {L7} (\bibinfo {year} {2021})}\BibitemShut {NoStop}%
\bibitem [{\citenamefont {Yunes}\ and\ \citenamefont
  {Pretorius}(2009)}]{Yunes2009ke}%
  \BibitemOpen
  \bibfield  {author} {\bibinfo {author} {\bibfnamefont {N.}~\bibnamefont
  {Yunes}}\ and\ \bibinfo {author} {\bibfnamefont {F.}~\bibnamefont
  {Pretorius}},\ }\bibfield  {title} {\bibinfo {title} {{Fundamental
  Theoretical Bias in Gravitational Wave Astrophysics and the Parameterized
  Post-Einsteinian Framework}},\ }\href
  {https://doi.org/10.1103/PhysRevD.80.122003} {\bibfield  {journal} {\bibinfo
  {journal} {Phys. Rev. D}\ }\textbf {\bibinfo {volume} {80}},\ \bibinfo
  {pages} {122003} (\bibinfo {year} {2009})},\ \Eprint
  {https://arxiv.org/abs/0909.3328} {arXiv:0909.3328 [gr-qc]} \BibitemShut
  {NoStop}%
\bibitem [{\citenamefont {{Blanchet}}(2014)}]{Blanchet2014PN}%
  \BibitemOpen
  \bibfield  {author} {\bibinfo {author} {\bibfnamefont {L.}~\bibnamefont
  {{Blanchet}}},\ }\bibfield  {title} {\bibinfo {title} {{Gravitational
  Radiation from Post-Newtonian Sources and Inspiralling Compact Binaries}},\
  }\href {https://doi.org/10.12942/lrr-2014-2} {\bibfield  {journal} {\bibinfo
  {journal} {Living Reviews in Relativity}\ }\textbf {\bibinfo {volume} {17}},\
  \bibinfo {eid} {2} (\bibinfo {year} {2014})},\ \Eprint
  {https://arxiv.org/abs/1310.1528} {arXiv:1310.1528 [gr-qc]} \BibitemShut
  {NoStop}%
\bibitem [{\citenamefont {{Arun}}\ \emph {et~al.}(2005)\citenamefont {{Arun}},
  \citenamefont {{Iyer}}, \citenamefont {{Sathyaprakash}},\ and\ \citenamefont
  {{Sundararajan}}}]{arun2005PN}%
  \BibitemOpen
  \bibfield  {author} {\bibinfo {author} {\bibfnamefont {K.~G.}\ \bibnamefont
  {{Arun}}}, \bibinfo {author} {\bibfnamefont {B.~R.}\ \bibnamefont {{Iyer}}},
  \bibinfo {author} {\bibfnamefont {B.~S.}\ \bibnamefont {{Sathyaprakash}}},\
  and\ \bibinfo {author} {\bibfnamefont {P.~A.}\ \bibnamefont
  {{Sundararajan}}},\ }\bibfield  {title} {\bibinfo {title} {{Parameter
  estimation of inspiralling compact binaries using 3.5 post-Newtonian
  gravitational wave phasing: The nonspinning case}},\ }\href
  {https://doi.org/10.1103/PhysRevD.71.084008} {\bibfield  {journal} {\bibinfo
  {journal} {\prd}\ }\textbf {\bibinfo {volume} {71}},\ \bibinfo {eid} {084008}
  (\bibinfo {year} {2005})},\ \Eprint {https://arxiv.org/abs/gr-qc/0411146}
  {arXiv:gr-qc/0411146 [gr-qc]} \BibitemShut {NoStop}%
\bibitem [{\citenamefont {{Chatziioannou}}\ \emph {et~al.}(2017)\citenamefont
  {{Chatziioannou}}, \citenamefont {{Klein}}, \citenamefont {{Yunes}},\ and\
  \citenamefont {{Cornish}}}]{chatziioannou2017PN}%
  \BibitemOpen
  \bibfield  {author} {\bibinfo {author} {\bibfnamefont {K.}~\bibnamefont
  {{Chatziioannou}}}, \bibinfo {author} {\bibfnamefont {A.}~\bibnamefont
  {{Klein}}}, \bibinfo {author} {\bibfnamefont {N.}~\bibnamefont {{Yunes}}},\
  and\ \bibinfo {author} {\bibfnamefont {N.}~\bibnamefont {{Cornish}}},\
  }\bibfield  {title} {\bibinfo {title} {{Constructing gravitational waves from
  generic spin-precessing compact binary inspirals}},\ }\href
  {https://doi.org/10.1103/PhysRevD.95.104004} {\bibfield  {journal} {\bibinfo
  {journal} {\prd}\ }\textbf {\bibinfo {volume} {95}},\ \bibinfo {eid} {104004}
  (\bibinfo {year} {2017})},\ \Eprint {https://arxiv.org/abs/1703.03967}
  {arXiv:1703.03967 [gr-qc]} \BibitemShut {NoStop}%
\bibitem [{\citenamefont {Loutrel}\ \emph {et~al.}(2018)\citenamefont
  {Loutrel}, \citenamefont {Tanaka},\ and\ \citenamefont
  {Yunes}}]{Loutrel2018ydv}%
  \BibitemOpen
  \bibfield  {author} {\bibinfo {author} {\bibfnamefont {N.}~\bibnamefont
  {Loutrel}}, \bibinfo {author} {\bibfnamefont {T.}~\bibnamefont {Tanaka}},\
  and\ \bibinfo {author} {\bibfnamefont {N.}~\bibnamefont {Yunes}},\ }\bibfield
   {title} {\bibinfo {title} {{Spin-Precessing Black Hole Binaries in Dynamical
  Chern-Simons Gravity}},\ }\href {https://doi.org/10.1103/PhysRevD.98.064020}
  {\bibfield  {journal} {\bibinfo  {journal} {Phys. Rev. D}\ }\textbf {\bibinfo
  {volume} {98}},\ \bibinfo {pages} {064020} (\bibinfo {year} {2018})},\
  \Eprint {https://arxiv.org/abs/1806.07431} {arXiv:1806.07431 [gr-qc]}
  \BibitemShut {NoStop}%
\bibitem [{\citenamefont {Perkins}\ \emph {et~al.}(2021)\citenamefont
  {Perkins}, \citenamefont {Nair}, \citenamefont {Silva},\ and\ \citenamefont
  {Yunes}}]{Perkins2021mhb}%
  \BibitemOpen
  \bibfield  {author} {\bibinfo {author} {\bibfnamefont {S.~E.}\ \bibnamefont
  {Perkins}}, \bibinfo {author} {\bibfnamefont {R.}~\bibnamefont {Nair}},
  \bibinfo {author} {\bibfnamefont {H.~O.}\ \bibnamefont {Silva}},\ and\
  \bibinfo {author} {\bibfnamefont {N.}~\bibnamefont {Yunes}},\ }\bibfield
  {title} {\bibinfo {title} {{Improved gravitational-wave constraints on
  higher-order curvature theories of gravity}},\ }\href
  {https://doi.org/10.1103/PhysRevD.104.024060} {\bibfield  {journal} {\bibinfo
   {journal} {Phys. Rev. D}\ }\textbf {\bibinfo {volume} {104}},\ \bibinfo
  {pages} {024060} (\bibinfo {year} {2021})},\ \Eprint
  {https://arxiv.org/abs/2104.11189} {arXiv:2104.11189 [gr-qc]} \BibitemShut
  {NoStop}%
\bibitem [{\citenamefont {Loutrel}\ and\ \citenamefont
  {Yunes}(2022)}]{Loutrel2022tbk}%
  \BibitemOpen
  \bibfield  {author} {\bibinfo {author} {\bibfnamefont {N.}~\bibnamefont
  {Loutrel}}\ and\ \bibinfo {author} {\bibfnamefont {N.}~\bibnamefont
  {Yunes}},\ }\bibfield  {title} {\bibinfo {title} {{Parity violation in
  spin-precessing binaries: Gravitational waves from the inspiral of black
  holes in dynamical Chern-Simons gravity}},\ }\href
  {https://doi.org/10.1103/PhysRevD.106.064009} {\bibfield  {journal} {\bibinfo
   {journal} {Phys. Rev. D}\ }\textbf {\bibinfo {volume} {106}},\ \bibinfo
  {pages} {064009} (\bibinfo {year} {2022})},\ \Eprint
  {https://arxiv.org/abs/2205.02675} {arXiv:2205.02675 [gr-qc]} \BibitemShut
  {NoStop}%
\bibitem [{\citenamefont {Okounkova}\ \emph {et~al.}(2023)\citenamefont
  {Okounkova}, \citenamefont {Isi}, \citenamefont {Chatziioannou},\ and\
  \citenamefont {Farr}}]{Okounkova:2022grv}%
  \BibitemOpen
  \bibfield  {author} {\bibinfo {author} {\bibfnamefont {M.}~\bibnamefont
  {Okounkova}}, \bibinfo {author} {\bibfnamefont {M.}~\bibnamefont {Isi}},
  \bibinfo {author} {\bibfnamefont {K.}~\bibnamefont {Chatziioannou}},\ and\
  \bibinfo {author} {\bibfnamefont {W.~M.}\ \bibnamefont {Farr}},\ }\bibfield
  {title} {\bibinfo {title} {{Gravitational wave inference on a
  numerical-relativity simulation of a black hole merger beyond general
  relativity}},\ }\href {https://doi.org/10.1103/PhysRevD.107.024046}
  {\bibfield  {journal} {\bibinfo  {journal} {Phys. Rev. D}\ }\textbf {\bibinfo
  {volume} {107}},\ \bibinfo {pages} {024046} (\bibinfo {year} {2023})},\
  \Eprint {https://arxiv.org/abs/2208.02805} {arXiv:2208.02805 [gr-qc]}
  \BibitemShut {NoStop}%
\bibitem [{\citenamefont {{Abbott}}\ \emph {et~al.}(2021)\citenamefont
  {{Abbott}} \emph {et~al.}}]{gwtc2}%
  \BibitemOpen
  \bibfield  {author} {\bibinfo {author} {\bibfnamefont {R.}~\bibnamefont
  {{Abbott}}} \emph {et~al.} (\bibinfo {collaboration} {{LIGO Scientific
  Collaboration} and {Virgo Collaboration}}),\ }\bibfield  {title} {\bibinfo
  {title} {{GWTC-2: Compact Binary Coalescences Observed by LIGO and Virgo
  during the First Half of the Third Observing Run}},\ }\href
  {https://doi.org/10.1103/PhysRevX.11.021053} {\bibfield  {journal} {\bibinfo
  {journal} {Physical Review X}\ }\textbf {\bibinfo {volume} {11}},\ \bibinfo
  {eid} {021053} (\bibinfo {year} {2021})},\ \Eprint
  {https://arxiv.org/abs/2010.14527} {arXiv:2010.14527 [gr-qc]} \BibitemShut
  {NoStop}%
\bibitem [{\citenamefont {{Boh{\'e}}}\ \emph {et~al.}(2017)\citenamefont
  {{Boh{\'e}}}, \citenamefont {{Shao}}, \citenamefont {{Taracchini}},
  \citenamefont {{Buonanno}}, \citenamefont {{Babak}}, \citenamefont {{Harry}},
  \citenamefont {{Hinder}}, \citenamefont {{Ossokine}}, \citenamefont
  {{P{\"u}rrer}}, \citenamefont {{Raymond}}, \citenamefont {{Chu}},
  \citenamefont {{Fong}}, \citenamefont {{Kumar}}, \citenamefont {{Pfeiffer}},
  \citenamefont {{Boyle}}, \citenamefont {{Hemberger}}, \citenamefont
  {{Kidder}}, \citenamefont {{Lovelace}}, \citenamefont {{Scheel}},\ and\
  \citenamefont {{Szil{\'a}gyi}}}]{bohe2017}%
  \BibitemOpen
  \bibfield  {author} {\bibinfo {author} {\bibfnamefont {A.}~\bibnamefont
  {{Boh{\'e}}}}, \bibinfo {author} {\bibfnamefont {L.}~\bibnamefont {{Shao}}},
  \bibinfo {author} {\bibfnamefont {A.}~\bibnamefont {{Taracchini}}}, \bibinfo
  {author} {\bibfnamefont {A.}~\bibnamefont {{Buonanno}}}, \bibinfo {author}
  {\bibfnamefont {S.}~\bibnamefont {{Babak}}}, \bibinfo {author} {\bibfnamefont
  {I.~W.}\ \bibnamefont {{Harry}}}, \bibinfo {author} {\bibfnamefont
  {I.}~\bibnamefont {{Hinder}}}, \bibinfo {author} {\bibfnamefont
  {S.}~\bibnamefont {{Ossokine}}}, \bibinfo {author} {\bibfnamefont
  {M.}~\bibnamefont {{P{\"u}rrer}}}, \bibinfo {author} {\bibfnamefont
  {V.}~\bibnamefont {{Raymond}}}, \bibinfo {author} {\bibfnamefont
  {T.}~\bibnamefont {{Chu}}}, \bibinfo {author} {\bibfnamefont
  {H.}~\bibnamefont {{Fong}}}, \bibinfo {author} {\bibfnamefont
  {P.}~\bibnamefont {{Kumar}}}, \bibinfo {author} {\bibfnamefont {H.~P.}\
  \bibnamefont {{Pfeiffer}}}, \bibinfo {author} {\bibfnamefont
  {M.}~\bibnamefont {{Boyle}}}, \bibinfo {author} {\bibfnamefont {D.~A.}\
  \bibnamefont {{Hemberger}}}, \bibinfo {author} {\bibfnamefont {L.~E.}\
  \bibnamefont {{Kidder}}}, \bibinfo {author} {\bibfnamefont {G.}~\bibnamefont
  {{Lovelace}}}, \bibinfo {author} {\bibfnamefont {M.~A.}\ \bibnamefont
  {{Scheel}}},\ and\ \bibinfo {author} {\bibfnamefont {B.}~\bibnamefont
  {{Szil{\'a}gyi}}},\ }\bibfield  {title} {\bibinfo {title} {{Improved
  effective-one-body model of spinning, nonprecessing binary black holes for
  the era of gravitational-wave astrophysics with advanced detectors}},\ }\href
  {https://doi.org/10.1103/PhysRevD.95.044028} {\bibfield  {journal} {\bibinfo
  {journal} {\prd}\ }\textbf {\bibinfo {volume} {95}},\ \bibinfo {eid} {044028}
  (\bibinfo {year} {2017})},\ \Eprint {https://arxiv.org/abs/1611.03703}
  {arXiv:1611.03703 [gr-qc]} \BibitemShut {NoStop}%
\bibitem [{\citenamefont {{Cotesta}}\ \emph {et~al.}(2018)\citenamefont
  {{Cotesta}}, \citenamefont {{Buonanno}}, \citenamefont {{Boh{\'e}}},
  \citenamefont {{Taracchini}}, \citenamefont {{Hinder}},\ and\ \citenamefont
  {{Ossokine}}}]{Cotesta2018}%
  \BibitemOpen
  \bibfield  {author} {\bibinfo {author} {\bibfnamefont {R.}~\bibnamefont
  {{Cotesta}}}, \bibinfo {author} {\bibfnamefont {A.}~\bibnamefont
  {{Buonanno}}}, \bibinfo {author} {\bibfnamefont {A.}~\bibnamefont
  {{Boh{\'e}}}}, \bibinfo {author} {\bibfnamefont {A.}~\bibnamefont
  {{Taracchini}}}, \bibinfo {author} {\bibfnamefont {I.}~\bibnamefont
  {{Hinder}}},\ and\ \bibinfo {author} {\bibfnamefont {S.}~\bibnamefont
  {{Ossokine}}},\ }\bibfield  {title} {\bibinfo {title} {{Enriching the
  symphony of gravitational waves from binary black holes by tuning higher
  harmonics}},\ }\href {https://doi.org/10.1103/PhysRevD.98.084028} {\bibfield
  {journal} {\bibinfo  {journal} {\prd}\ }\textbf {\bibinfo {volume} {98}},\
  \bibinfo {eid} {084028} (\bibinfo {year} {2018})},\ \Eprint
  {https://arxiv.org/abs/1803.10701} {arXiv:1803.10701 [gr-qc]} \BibitemShut
  {NoStop}%
\bibitem [{\citenamefont {{Cotesta}}\ \emph {et~al.}(2020)\citenamefont
  {{Cotesta}}, \citenamefont {{Marsat}},\ and\ \citenamefont
  {{P{\"u}rrer}}}]{Cotesta2020}%
  \BibitemOpen
  \bibfield  {author} {\bibinfo {author} {\bibfnamefont {R.}~\bibnamefont
  {{Cotesta}}}, \bibinfo {author} {\bibfnamefont {S.}~\bibnamefont
  {{Marsat}}},\ and\ \bibinfo {author} {\bibfnamefont {M.}~\bibnamefont
  {{P{\"u}rrer}}},\ }\bibfield  {title} {\bibinfo {title} {{Frequency-domain
  reduced-order model of aligned-spin effective-one-body waveforms with
  higher-order modes}},\ }\href {https://doi.org/10.1103/PhysRevD.101.124040}
  {\bibfield  {journal} {\bibinfo  {journal} {\prd}\ }\textbf {\bibinfo
  {volume} {101}},\ \bibinfo {eid} {124040} (\bibinfo {year} {2020})},\ \Eprint
  {https://arxiv.org/abs/2003.12079} {arXiv:2003.12079 [gr-qc]} \BibitemShut
  {NoStop}%
\bibitem [{\citenamefont {{Brito}}\ \emph {et~al.}(2018)\citenamefont
  {{Brito}}, \citenamefont {{Buonanno}},\ and\ \citenamefont
  {{Raymond}}}]{Brito2018}%
  \BibitemOpen
  \bibfield  {author} {\bibinfo {author} {\bibfnamefont {R.}~\bibnamefont
  {{Brito}}}, \bibinfo {author} {\bibfnamefont {A.}~\bibnamefont
  {{Buonanno}}},\ and\ \bibinfo {author} {\bibfnamefont {V.}~\bibnamefont
  {{Raymond}}},\ }\bibfield  {title} {\bibinfo {title} {{Black-hole
  spectroscopy by making full use of gravitational-wave modeling}},\ }\href
  {https://doi.org/10.1103/PhysRevD.98.084038} {\bibfield  {journal} {\bibinfo
  {journal} {\prd}\ }\textbf {\bibinfo {volume} {98}},\ \bibinfo {eid} {084038}
  (\bibinfo {year} {2018})},\ \Eprint {https://arxiv.org/abs/1805.00293}
  {arXiv:1805.00293 [gr-qc]} \BibitemShut {NoStop}%
\bibitem [{\citenamefont {{LIGO Scientific Collaboration and Virgo
  Collaboration and KAGRA Collaboration}}(2022)}]{gwtc3_tgr_release}%
  \BibitemOpen
  \bibfield  {author} {\bibinfo {author} {\bibnamefont {{LIGO Scientific
  Collaboration and Virgo Collaboration and KAGRA Collaboration}}},\ }\href
  {https://doi.org/10.5281/zenodo.7007370} {\bibinfo {title} {Data release for
  tests of general relativity with gwtc-3}} (\bibinfo {year}
  {2022})\BibitemShut {NoStop}%
\bibitem [{\citenamefont {Abbott}\ \emph
  {et~al.}(2021{\natexlab{a}})\citenamefont {Abbott} \emph {et~al.}}]{gwtc-3}%
  \BibitemOpen
  \bibfield  {author} {\bibinfo {author} {\bibfnamefont {R.}~\bibnamefont
  {Abbott}} \emph {et~al.},\ }\bibfield  {title} {\bibinfo {title} {{GWTC-3:
  Compact Binary Coalescences Observed by LIGO and Virgo During the Second Part
  of the Third Observing Run}},\ }\href@noop {} {\  (\bibinfo {year}
  {2021}{\natexlab{a}})},\ \bibinfo {note} {arxiv/2111.03606}\BibitemShut
  {NoStop}%
\bibitem [{\citenamefont {Mandel}\ \emph {et~al.}(2019)\citenamefont {Mandel},
  \citenamefont {Farr},\ and\ \citenamefont {Gair}}]{Mandel2019}%
  \BibitemOpen
  \bibfield  {author} {\bibinfo {author} {\bibfnamefont {I.}~\bibnamefont
  {Mandel}}, \bibinfo {author} {\bibfnamefont {W.~M.}\ \bibnamefont {Farr}},\
  and\ \bibinfo {author} {\bibfnamefont {J.~R.}\ \bibnamefont {Gair}},\
  }\bibfield  {title} {\bibinfo {title} {Extracting distribution parameters
  from multiple uncertain observations with selection biases},\ }\href@noop {}
  {\bibfield  {journal} {\bibinfo  {journal} {Mon. Not. R. Ast. Soc.}\ }\textbf
  {\bibinfo {volume} {486}},\ \bibinfo {pages} {1086} (\bibinfo {year}
  {2019})}\BibitemShut {NoStop}%
\bibitem [{\citenamefont {Thrane}\ and\ \citenamefont {Talbot}(2019)}]{intro}%
  \BibitemOpen
  \bibfield  {author} {\bibinfo {author} {\bibfnamefont {E.}~\bibnamefont
  {Thrane}}\ and\ \bibinfo {author} {\bibfnamefont {C.}~\bibnamefont
  {Talbot}},\ }\bibfield  {title} {\bibinfo {title} {{An introduction to
  Bayesian inference in gravitational-wave astronomy: parameter estimation,
  model selection, and hierarchical models}},\ }\href@noop {} {\bibfield
  {journal} {\bibinfo  {journal} {Pub. Astron. Soc. Aust.}\ }\textbf {\bibinfo
  {volume} {36}},\ \bibinfo {pages} {E010} (\bibinfo {year}
  {2019})}\BibitemShut {NoStop}%
\bibitem [{\citenamefont {Vitale}\ \emph {et~al.}(2022)\citenamefont {Vitale},
  \citenamefont {Gerosa}, \citenamefont {Farr},\ and\ \citenamefont
  {Taylor}}]{Vitale2022}%
  \BibitemOpen
  \bibfield  {author} {\bibinfo {author} {\bibfnamefont {S.}~\bibnamefont
  {Vitale}}, \bibinfo {author} {\bibfnamefont {D.}~\bibnamefont {Gerosa}},
  \bibinfo {author} {\bibfnamefont {W.}~\bibnamefont {Farr}},\ and\ \bibinfo
  {author} {\bibfnamefont {S.}~\bibnamefont {Taylor}},\ }\href@noop {} {\emph
  {\bibinfo {title} {Inferring the Properties of a Population of Compact
  Binaries in Presence of Selection Effects}}},\ Handbook of Gravitational Wave
  Astronomy\ (\bibinfo  {publisher} {Springer, Singapore},\ \bibinfo {year}
  {2022})\BibitemShut {NoStop}%
\bibitem [{\citenamefont {Abbott}\ \emph
  {et~al.}(2019{\natexlab{a}})\citenamefont {Abbott} \emph
  {et~al.}}]{gwtc-1_pop}%
  \BibitemOpen
  \bibfield  {author} {\bibinfo {author} {\bibfnamefont {R.}~\bibnamefont
  {Abbott}} \emph {et~al.},\ }\bibfield  {title} {\bibinfo {title} {Binary
  black hole population properties inferred from the first and second observing
  runs of advanced {LIGO} and advanced virgo},\ }\href@noop {} {\bibfield
  {journal} {\bibinfo  {journal} {Astrophys. J.}\ }\textbf {\bibinfo {volume}
  {882}},\ \bibinfo {pages} {L24} (\bibinfo {year}
  {2019}{\natexlab{a}})}\BibitemShut {NoStop}%
\bibitem [{\citenamefont {Roulet}\ \emph {et~al.}(2021)\citenamefont {Roulet},
  \citenamefont {Chia}, \citenamefont {Olsen}, \citenamefont {Dai},
  \citenamefont {Venumadhav}, \citenamefont {Zackay},\ and\ \citenamefont
  {Zaldarriaga}}]{Roulet_2021}%
  \BibitemOpen
  \bibfield  {author} {\bibinfo {author} {\bibfnamefont {J.}~\bibnamefont
  {Roulet}}, \bibinfo {author} {\bibfnamefont {H.~S.}\ \bibnamefont {Chia}},
  \bibinfo {author} {\bibfnamefont {S.}~\bibnamefont {Olsen}}, \bibinfo
  {author} {\bibfnamefont {L.}~\bibnamefont {Dai}}, \bibinfo {author}
  {\bibfnamefont {T.}~\bibnamefont {Venumadhav}}, \bibinfo {author}
  {\bibfnamefont {B.}~\bibnamefont {Zackay}},\ and\ \bibinfo {author}
  {\bibfnamefont {M.}~\bibnamefont {Zaldarriaga}},\ }\bibfield  {title}
  {\bibinfo {title} {Distribution of effective spins and masses of binary black
  holes from the {LIGO} and virgo o1{\textendash}o3a observing runs},\
  }\href@noop {} {\bibfield  {journal} {\bibinfo  {journal} {Phys. Rev. D}\
  }\textbf {\bibinfo {volume} {104}},\ \bibinfo {pages} {083010} (\bibinfo
  {year} {2021})}\BibitemShut {NoStop}%
\bibitem [{\citenamefont {Farr}\ \emph {et~al.}(2017)\citenamefont {Farr},
  \citenamefont {Stevenson}, \citenamefont {Miller}, \citenamefont {Mandel},
  \citenamefont {Farr},\ and\ \citenamefont {Vecchio}}]{Farr_2017}%
  \BibitemOpen
  \bibfield  {author} {\bibinfo {author} {\bibfnamefont {W.~M.}\ \bibnamefont
  {Farr}}, \bibinfo {author} {\bibfnamefont {S.}~\bibnamefont {Stevenson}},
  \bibinfo {author} {\bibfnamefont {M.~C.}\ \bibnamefont {Miller}}, \bibinfo
  {author} {\bibfnamefont {I.}~\bibnamefont {Mandel}}, \bibinfo {author}
  {\bibfnamefont {B.}~\bibnamefont {Farr}},\ and\ \bibinfo {author}
  {\bibfnamefont {A.}~\bibnamefont {Vecchio}},\ }\bibfield  {title} {\bibinfo
  {title} {Distinguishing spin-aligned and isotropic black hole populations
  with gravitational waves},\ }\href@noop {} {\bibfield  {journal} {\bibinfo
  {journal} {Nature}\ }\textbf {\bibinfo {volume} {548}},\ \bibinfo {pages}
  {426} (\bibinfo {year} {2017})}\BibitemShut {NoStop}%
\bibitem [{\citenamefont {Talbot}\ and\ \citenamefont {Thrane}(2018)}]{mass}%
  \BibitemOpen
  \bibfield  {author} {\bibinfo {author} {\bibfnamefont {C.}~\bibnamefont
  {Talbot}}\ and\ \bibinfo {author} {\bibfnamefont {E.}~\bibnamefont
  {Thrane}},\ }\bibfield  {title} {\bibinfo {title} {Measuring the binary black
  hole mass spectrum with an astrophysically motivated parameterization},\
  }\href@noop {} {\bibfield  {journal} {\bibinfo  {journal} {Astrophys. J.}\
  }\textbf {\bibinfo {volume} {856}},\ \bibinfo {pages} {173} (\bibinfo {year}
  {2018})}\BibitemShut {NoStop}%
\bibitem [{\citenamefont {Talbot}\ and\ \citenamefont {Thrane}(2017)}]{spin}%
  \BibitemOpen
  \bibfield  {author} {\bibinfo {author} {\bibfnamefont {C.}~\bibnamefont
  {Talbot}}\ and\ \bibinfo {author} {\bibfnamefont {E.}~\bibnamefont
  {Thrane}},\ }\bibfield  {title} {\bibinfo {title} {Determining the population
  properties of spinning black holes},\ }\href@noop {} {\bibfield  {journal}
  {\bibinfo  {journal} {Phys. Rev. D}\ }\textbf {\bibinfo {volume} {96}},\
  \bibinfo {pages} {023012} (\bibinfo {year} {2017})}\BibitemShut {NoStop}%
\bibitem [{\citenamefont {Callister}\ \emph {et~al.}(2021)\citenamefont
  {Callister}, \citenamefont {Haster}, \citenamefont {Ng}, \citenamefont
  {Vitale},\ and\ \citenamefont {Farr}}]{Callister_2021}%
  \BibitemOpen
  \bibfield  {author} {\bibinfo {author} {\bibfnamefont {T.~A.}\ \bibnamefont
  {Callister}}, \bibinfo {author} {\bibfnamefont {C.}~\bibnamefont {Haster}},
  \bibinfo {author} {\bibfnamefont {K.~K.~Y.}\ \bibnamefont {Ng}}, \bibinfo
  {author} {\bibfnamefont {S.}~\bibnamefont {Vitale}},\ and\ \bibinfo {author}
  {\bibfnamefont {W.~M.}\ \bibnamefont {Farr}},\ }\bibfield  {title} {\bibinfo
  {title} {Who ordered that? unequal-mass binary black hole mergers have larger
  effective spins},\ }\href@noop {} {\bibfield  {journal} {\bibinfo  {journal}
  {Astrophys. J. Lett.}\ }\textbf {\bibinfo {volume} {922}},\ \bibinfo {pages}
  {L5} (\bibinfo {year} {2021})}\BibitemShut {NoStop}%
\bibitem [{\citenamefont {Fishbach}\ \emph {et~al.}(2022)\citenamefont
  {Fishbach}, \citenamefont {Kimball},\ and\ \citenamefont
  {Kalogera}}]{Fishbach_2022}%
  \BibitemOpen
  \bibfield  {author} {\bibinfo {author} {\bibfnamefont {M.}~\bibnamefont
  {Fishbach}}, \bibinfo {author} {\bibfnamefont {C.}~\bibnamefont {Kimball}},\
  and\ \bibinfo {author} {\bibfnamefont {V.}~\bibnamefont {Kalogera}},\
  }\bibfield  {title} {\bibinfo {title} {Limits on hierarchical black hole
  mergers from the most negative $\chi_\text{eff}$ systems},\ }\href@noop {}
  {\bibfield  {journal} {\bibinfo  {journal} {Astrophys. J. Lett.}\ }\textbf
  {\bibinfo {volume} {935}},\ \bibinfo {pages} {L26} (\bibinfo {year}
  {2022})}\BibitemShut {NoStop}%
\bibitem [{\citenamefont {Biscoveanu}\ \emph {et~al.}(2022)\citenamefont
  {Biscoveanu}, \citenamefont {Callister}, \citenamefont {Haster},
  \citenamefont {Ng}, \citenamefont {Vitale},\ and\ \citenamefont
  {Farr}}]{Biscoveanu_2022}%
  \BibitemOpen
  \bibfield  {author} {\bibinfo {author} {\bibfnamefont {S.}~\bibnamefont
  {Biscoveanu}}, \bibinfo {author} {\bibfnamefont {T.~A.}\ \bibnamefont
  {Callister}}, \bibinfo {author} {\bibfnamefont {C.}~\bibnamefont {Haster}},
  \bibinfo {author} {\bibfnamefont {K.~K.~Y.}\ \bibnamefont {Ng}}, \bibinfo
  {author} {\bibfnamefont {S.}~\bibnamefont {Vitale}},\ and\ \bibinfo {author}
  {\bibfnamefont {W.~M.}\ \bibnamefont {Farr}},\ }\bibfield  {title} {\bibinfo
  {title} {The binary black hole spin distribution likely broadens with
  redshift},\ }\href@noop {} {\bibfield  {journal} {\bibinfo  {journal}
  {Astrophys. J. Lett.}\ }\textbf {\bibinfo {volume} {932}},\ \bibinfo {pages}
  {L19} (\bibinfo {year} {2022})}\BibitemShut {NoStop}%
\bibitem [{\citenamefont {Vitale}\ \emph {et~al.}(2017)\citenamefont {Vitale},
  \citenamefont {Lynch}, \citenamefont {Sturani},\ and\ \citenamefont
  {Graff}}]{Vitale_2017}%
  \BibitemOpen
  \bibfield  {author} {\bibinfo {author} {\bibfnamefont {S.}~\bibnamefont
  {Vitale}}, \bibinfo {author} {\bibfnamefont {R.}~\bibnamefont {Lynch}},
  \bibinfo {author} {\bibfnamefont {R.}~\bibnamefont {Sturani}},\ and\ \bibinfo
  {author} {\bibfnamefont {P.}~\bibnamefont {Graff}},\ }\bibfield  {title}
  {\bibinfo {title} {Use of gravitational waves to probe the formation channels
  of compact binaries},\ }\href@noop {} {\bibfield  {journal} {\bibinfo
  {journal} {Class. Quantum Grav.}\ }\textbf {\bibinfo {volume} {34}},\
  \bibinfo {pages} {03LT01} (\bibinfo {year} {2017})}\BibitemShut {NoStop}%
\bibitem [{\citenamefont {Stevenson}\ \emph {et~al.}(2017)\citenamefont
  {Stevenson}, \citenamefont {Berry},\ and\ \citenamefont
  {Mandel}}]{Stevenson_2017}%
  \BibitemOpen
  \bibfield  {author} {\bibinfo {author} {\bibfnamefont {S.}~\bibnamefont
  {Stevenson}}, \bibinfo {author} {\bibfnamefont {C.~P.~L.}\ \bibnamefont
  {Berry}},\ and\ \bibinfo {author} {\bibfnamefont {I.}~\bibnamefont
  {Mandel}},\ }\bibfield  {title} {\bibinfo {title} {Hierarchical analysis of
  gravitational-wave measurements of binary black hole spin{\textendash}orbit
  misalignments},\ }\href@noop {} {\bibfield  {journal} {\bibinfo  {journal}
  {Mon. Not. R. Ast. Soc.}\ }\textbf {\bibinfo {volume} {471}},\ \bibinfo
  {pages} {2801} (\bibinfo {year} {2017})}\BibitemShut {NoStop}%
\bibitem [{\citenamefont {Miller}\ \emph {et~al.}(2020)\citenamefont {Miller},
  \citenamefont {Callister},\ and\ \citenamefont {Farr}}]{Miller2020}%
  \BibitemOpen
  \bibfield  {author} {\bibinfo {author} {\bibfnamefont {S.}~\bibnamefont
  {Miller}}, \bibinfo {author} {\bibfnamefont {T.~A.}\ \bibnamefont
  {Callister}},\ and\ \bibinfo {author} {\bibfnamefont {W.~M.}\ \bibnamefont
  {Farr}},\ }\bibfield  {title} {\bibinfo {title} {The low effective spin of
  binary black holes and implications for individual gravitational-wave
  events},\ }\href@noop {} {\bibfield  {journal} {\bibinfo  {journal}
  {Astrophys. J.}\ }\textbf {\bibinfo {volume} {895}},\ \bibinfo {pages} {128}
  (\bibinfo {year} {2020})}\BibitemShut {NoStop}%
\bibitem [{\citenamefont {Galaudage}\ \emph {et~al.}(2021)\citenamefont
  {Galaudage}, \citenamefont {Talbot}, \citenamefont {Nagar}, \citenamefont
  {Jain}, \citenamefont {Thrane},\ and\ \citenamefont {Mandel}}]{bbm}%
  \BibitemOpen
  \bibfield  {author} {\bibinfo {author} {\bibfnamefont {S.}~\bibnamefont
  {Galaudage}}, \bibinfo {author} {\bibfnamefont {C.}~\bibnamefont {Talbot}},
  \bibinfo {author} {\bibfnamefont {T.}~\bibnamefont {Nagar}}, \bibinfo
  {author} {\bibfnamefont {D.}~\bibnamefont {Jain}}, \bibinfo {author}
  {\bibfnamefont {E.}~\bibnamefont {Thrane}},\ and\ \bibinfo {author}
  {\bibfnamefont {I.}~\bibnamefont {Mandel}},\ }\bibfield  {title} {\bibinfo
  {title} {Building better spin models for merging binary black holes: Evidence
  for nonspinning and rapidly spinning nearly aligned subpopulations},\
  }\href@noop {} {\bibfield  {journal} {\bibinfo  {journal} {Astrophys. J.
  Lett.}\ }\textbf {\bibinfo {volume} {921}},\ \bibinfo {pages} {L15} (\bibinfo
  {year} {2021})}\BibitemShut {NoStop}%
\bibitem [{\citenamefont {Fishbach}\ \emph {et~al.}(2018)\citenamefont
  {Fishbach}, \citenamefont {Holz},\ and\ \citenamefont {Farr}}]{2018_maya}%
  \BibitemOpen
  \bibfield  {author} {\bibinfo {author} {\bibfnamefont {M.}~\bibnamefont
  {Fishbach}}, \bibinfo {author} {\bibfnamefont {D.~E.}\ \bibnamefont {Holz}},\
  and\ \bibinfo {author} {\bibfnamefont {W.~M.}\ \bibnamefont {Farr}},\
  }\bibfield  {title} {\bibinfo {title} {{Does the Black Hole Merger Rate
  Evolve with Redshift?}},\ }\href {https://doi.org/10.3847/2041-8213/aad800}
  {\bibfield  {journal} {\bibinfo  {journal} {Astrophys. J. Lett.}\ }\textbf
  {\bibinfo {volume} {863}},\ \bibinfo {pages} {L41} (\bibinfo {year}
  {2018})},\ \Eprint {https://arxiv.org/abs/1805.10270} {arXiv:1805.10270
  [astro-ph.HE]} \BibitemShut {NoStop}%
\bibitem [{\citenamefont {Edelman}\ \emph {et~al.}(2022)\citenamefont
  {Edelman}, \citenamefont {Doctor}, \citenamefont {Godfrey},\ and\
  \citenamefont {Farr}}]{Edelman_2022}%
  \BibitemOpen
  \bibfield  {author} {\bibinfo {author} {\bibfnamefont {B.}~\bibnamefont
  {Edelman}}, \bibinfo {author} {\bibfnamefont {Z.}~\bibnamefont {Doctor}},
  \bibinfo {author} {\bibfnamefont {J.}~\bibnamefont {Godfrey}},\ and\ \bibinfo
  {author} {\bibfnamefont {B.}~\bibnamefont {Farr}},\ }\bibfield  {title}
  {\bibinfo {title} {Ain't no mountain high enough: Semiparametric modeling of
  {LIGO}{\textendash}virgo's binary black hole mass distribution},\ }\href@noop
  {} {\bibfield  {journal} {\bibinfo  {journal} {Astrophys. J.}\ }\textbf
  {\bibinfo {volume} {924}},\ \bibinfo {pages} {101} (\bibinfo {year}
  {2022})}\BibitemShut {NoStop}%
\bibitem [{\citenamefont {{Edelman}}\ \emph {et~al.}(2023)\citenamefont
  {{Edelman}}, \citenamefont {{Farr}},\ and\ \citenamefont
  {{Doctor}}}]{Edelman_2023}%
  \BibitemOpen
  \bibfield  {author} {\bibinfo {author} {\bibfnamefont {B.}~\bibnamefont
  {{Edelman}}}, \bibinfo {author} {\bibfnamefont {B.}~\bibnamefont {{Farr}}},\
  and\ \bibinfo {author} {\bibfnamefont {Z.}~\bibnamefont {{Doctor}}},\
  }\bibfield  {title} {\bibinfo {title} {{Cover Your Basis: Comprehensive
  Data-driven Characterization of the Binary Black Hole Population}},\ }\href
  {https://doi.org/10.3847/1538-4357/acb5ed} {\bibfield  {journal} {\bibinfo
  {journal} {\apj}\ }\textbf {\bibinfo {volume} {946}},\ \bibinfo {eid} {16}
  (\bibinfo {year} {2023})},\ \Eprint {https://arxiv.org/abs/2210.12834}
  {arXiv:2210.12834 [astro-ph.HE]} \BibitemShut {NoStop}%
\bibitem [{\citenamefont {{Golomb}}\ and\ \citenamefont
  {{Talbot}}(2022)}]{golomb2022}%
  \BibitemOpen
  \bibfield  {author} {\bibinfo {author} {\bibfnamefont {J.}~\bibnamefont
  {{Golomb}}}\ and\ \bibinfo {author} {\bibfnamefont {C.}~\bibnamefont
  {{Talbot}}},\ }\bibfield  {title} {\bibinfo {title} {{Searching for structure
  in the binary black hole spin distribution}},\ }\href
  {https://doi.org/10.48550/arXiv.2210.12287} {\bibfield  {journal} {\bibinfo
  {journal} {arXiv e-prints}\ ,\ \bibinfo {eid} {arXiv:2210.12287}} (\bibinfo
  {year} {2022})},\ \Eprint {https://arxiv.org/abs/2210.12287}
  {arXiv:2210.12287 [astro-ph.HE]} \BibitemShut {NoStop}%
\bibitem [{\citenamefont {{Callister}}\ and\ \citenamefont
  {{Farr}}(2023)}]{Callister2023}%
  \BibitemOpen
  \bibfield  {author} {\bibinfo {author} {\bibfnamefont {T.~A.}\ \bibnamefont
  {{Callister}}}\ and\ \bibinfo {author} {\bibfnamefont {W.~M.}\ \bibnamefont
  {{Farr}}},\ }\bibfield  {title} {\bibinfo {title} {{A Parameter-Free Tour of
  the Binary Black Hole Population}},\ }\href
  {https://doi.org/10.48550/arXiv.2302.07289} {\bibfield  {journal} {\bibinfo
  {journal} {arXiv e-prints}\ ,\ \bibinfo {eid} {arXiv:2302.07289}} (\bibinfo
  {year} {2023})},\ \Eprint {https://arxiv.org/abs/2302.07289}
  {arXiv:2302.07289 [astro-ph.HE]} \BibitemShut {NoStop}%
\bibitem [{\citenamefont {{Farr}}(2019)}]{farr2019}%
  \BibitemOpen
  \bibfield  {author} {\bibinfo {author} {\bibfnamefont {W.~M.}\ \bibnamefont
  {{Farr}}},\ }\bibfield  {title} {\bibinfo {title} {{Accuracy Requirements for
  Empirically Measured Selection Functions}},\ }\href
  {https://doi.org/10.3847/2515-5172/ab1d5f} {\bibfield  {journal} {\bibinfo
  {journal} {Research Notes of the American Astronomical Society}\ }\textbf
  {\bibinfo {volume} {3}},\ \bibinfo {eid} {66} (\bibinfo {year} {2019})},\
  \Eprint {https://arxiv.org/abs/1904.10879} {arXiv:1904.10879 [astro-ph.IM]}
  \BibitemShut {NoStop}%
\bibitem [{\citenamefont {Magee}\ \emph {et~al.}(2023)\citenamefont {Magee}
  \emph {et~al.}}]{MageeInPrep}%
  \BibitemOpen
  \bibfield  {author} {\bibinfo {author} {\bibfnamefont {R.}~\bibnamefont
  {Magee}} \emph {et~al.},\ }\bibfield  {title} {\bibinfo {title} {Selection
  biases in tests of general relativity with gravitational waves}} (\bibinfo
  {year} {2023}),\ \bibinfo {note} {in prep.}\BibitemShut {Stop}%
\bibitem [{\citenamefont {Moore}\ and\ \citenamefont
  {Gerosa}(2021)}]{Moore2021xhn}%
  \BibitemOpen
  \bibfield  {author} {\bibinfo {author} {\bibfnamefont {C.~J.}\ \bibnamefont
  {Moore}}\ and\ \bibinfo {author} {\bibfnamefont {D.}~\bibnamefont {Gerosa}},\
  }\bibfield  {title} {\bibinfo {title} {{Population-informed priors in
  gravitational-wave astronomy}},\ }\href
  {https://doi.org/10.1103/PhysRevD.104.083008} {\bibfield  {journal} {\bibinfo
   {journal} {Phys. Rev. D}\ }\textbf {\bibinfo {volume} {104}},\ \bibinfo
  {pages} {083008} (\bibinfo {year} {2021})},\ \Eprint
  {https://arxiv.org/abs/2108.02462} {arXiv:2108.02462 [gr-qc]} \BibitemShut
  {NoStop}%
\bibitem [{\citenamefont {{Farr}}\ and\ \citenamefont
  {{Callister}}(2021)}]{farr_pop_informed}%
  \BibitemOpen
  \bibfield  {author} {\bibinfo {author} {\bibfnamefont {W.~M.}\ \bibnamefont
  {{Farr}}}\ and\ \bibinfo {author} {\bibfnamefont {T.~A.}\ \bibnamefont
  {{Callister}}},\ }\href
  {https://github.com/farr/Reweighting/blob/master-pdf/note/reweighting.pdf}
  {\emph {\bibinfo {title} {Re-Weighting Existing Samples to a Population
  Analysis}}},\ \bibinfo {type} {Tech. Rep.}\ (\bibinfo {year}
  {2021})\BibitemShut {NoStop}%
\bibitem [{\citenamefont {{Callister}}(2021)}]{callister_pop_informed}%
  \BibitemOpen
  \bibfield  {author} {\bibinfo {author} {\bibfnamefont {T.~A.}\ \bibnamefont
  {{Callister}}},\ }\href {https://dcc.ligo.org/LIGO-T2100301/public} {\emph
  {\bibinfo {title} {Reweighting Single Event Posteriors with Hyperparameter
  Marginalization}}},\ \bibinfo {type} {Tech. Rep.}\ \bibinfo {number}
  {LIGO-T2100301}\ (\bibinfo {year} {2021})\BibitemShut {NoStop}%
\bibitem [{\citenamefont {{Del Pozzo}}\ \emph {et~al.}(2011)\citenamefont {{Del
  Pozzo}}, \citenamefont {{Veitch}},\ and\ \citenamefont
  {{Vecchio}}}]{Pozzo2011}%
  \BibitemOpen
  \bibfield  {author} {\bibinfo {author} {\bibfnamefont {W.}~\bibnamefont {{Del
  Pozzo}}}, \bibinfo {author} {\bibfnamefont {J.}~\bibnamefont {{Veitch}}},\
  and\ \bibinfo {author} {\bibfnamefont {A.}~\bibnamefont {{Vecchio}}},\
  }\bibfield  {title} {\bibinfo {title} {{Testing general relativity using
  Bayesian model selection: Applications to observations of gravitational waves
  from compact binary systems}},\ }\href
  {https://doi.org/10.1103/PhysRevD.83.082002} {\bibfield  {journal} {\bibinfo
  {journal} {\prd}\ }\textbf {\bibinfo {volume} {83}},\ \bibinfo {eid} {082002}
  (\bibinfo {year} {2011})},\ \Eprint {https://arxiv.org/abs/1101.1391}
  {arXiv:1101.1391 [gr-qc]} \BibitemShut {NoStop}%
\bibitem [{\citenamefont {{Meidam}}\ \emph {et~al.}(2014)\citenamefont
  {{Meidam}}, \citenamefont {{Agathos}}, \citenamefont {{Van Den Broeck}},
  \citenamefont {{Veitch}},\ and\ \citenamefont
  {{Sathyaprakash}}}]{Meidam2014}%
  \BibitemOpen
  \bibfield  {author} {\bibinfo {author} {\bibfnamefont {J.}~\bibnamefont
  {{Meidam}}}, \bibinfo {author} {\bibfnamefont {M.}~\bibnamefont {{Agathos}}},
  \bibinfo {author} {\bibfnamefont {C.}~\bibnamefont {{Van Den Broeck}}},
  \bibinfo {author} {\bibfnamefont {J.}~\bibnamefont {{Veitch}}},\ and\
  \bibinfo {author} {\bibfnamefont {B.~S.}\ \bibnamefont {{Sathyaprakash}}},\
  }\bibfield  {title} {\bibinfo {title} {{Testing the no-hair theorem with
  black hole ringdowns using TIGER}},\ }\href
  {https://doi.org/10.1103/PhysRevD.90.064009} {\bibfield  {journal} {\bibinfo
  {journal} {\prd}\ }\textbf {\bibinfo {volume} {90}},\ \bibinfo {eid} {064009}
  (\bibinfo {year} {2014})},\ \Eprint {https://arxiv.org/abs/1406.3201}
  {arXiv:1406.3201 [gr-qc]} \BibitemShut {NoStop}%
\bibitem [{\citenamefont {{Ghosh}}\ \emph {et~al.}(2016)\citenamefont
  {{Ghosh}}, \citenamefont {{Ghosh}}, \citenamefont {{Johnson-McDaniel}},
  \citenamefont {{Mishra}}, \citenamefont {{Ajith}}, \citenamefont {{Del
  Pozzo}}, \citenamefont {{Nichols}}, \citenamefont {{Chen}}, \citenamefont
  {{Nielsen}}, \citenamefont {{Berry}},\ and\ \citenamefont
  {{London}}}]{Ghosh2016}%
  \BibitemOpen
  \bibfield  {author} {\bibinfo {author} {\bibfnamefont {A.}~\bibnamefont
  {{Ghosh}}}, \bibinfo {author} {\bibfnamefont {A.}~\bibnamefont {{Ghosh}}},
  \bibinfo {author} {\bibfnamefont {N.~K.}\ \bibnamefont {{Johnson-McDaniel}}},
  \bibinfo {author} {\bibfnamefont {C.~K.}\ \bibnamefont {{Mishra}}}, \bibinfo
  {author} {\bibfnamefont {P.}~\bibnamefont {{Ajith}}}, \bibinfo {author}
  {\bibfnamefont {W.}~\bibnamefont {{Del Pozzo}}}, \bibinfo {author}
  {\bibfnamefont {D.~A.}\ \bibnamefont {{Nichols}}}, \bibinfo {author}
  {\bibfnamefont {Y.}~\bibnamefont {{Chen}}}, \bibinfo {author} {\bibfnamefont
  {A.~B.}\ \bibnamefont {{Nielsen}}}, \bibinfo {author} {\bibfnamefont
  {C.~P.~L.}\ \bibnamefont {{Berry}}},\ and\ \bibinfo {author} {\bibfnamefont
  {L.}~\bibnamefont {{London}}},\ }\bibfield  {title} {\bibinfo {title}
  {{Testing general relativity using golden black-hole binaries}},\ }\href
  {https://doi.org/10.1103/PhysRevD.94.021101} {\bibfield  {journal} {\bibinfo
  {journal} {\prd}\ }\textbf {\bibinfo {volume} {94}},\ \bibinfo {eid} {021101}
  (\bibinfo {year} {2016})},\ \Eprint {https://arxiv.org/abs/1602.02453}
  {arXiv:1602.02453 [gr-qc]} \BibitemShut {NoStop}%
\bibitem [{\citenamefont {{Ghosh}}\ \emph {et~al.}(2018)\citenamefont
  {{Ghosh}}, \citenamefont {{Johnson-McDaniel}}, \citenamefont {{Ghosh}},
  \citenamefont {{Kant Mishra}}, \citenamefont {{Ajith}}, \citenamefont {{Del
  Pozzo}}, \citenamefont {{Berry}}, \citenamefont {{Nielsen}},\ and\
  \citenamefont {{London}}}]{Ghosh2018}%
  \BibitemOpen
  \bibfield  {author} {\bibinfo {author} {\bibfnamefont {A.}~\bibnamefont
  {{Ghosh}}}, \bibinfo {author} {\bibfnamefont {N.~K.}\ \bibnamefont
  {{Johnson-McDaniel}}}, \bibinfo {author} {\bibfnamefont {A.}~\bibnamefont
  {{Ghosh}}}, \bibinfo {author} {\bibfnamefont {C.}~\bibnamefont {{Kant
  Mishra}}}, \bibinfo {author} {\bibfnamefont {P.}~\bibnamefont {{Ajith}}},
  \bibinfo {author} {\bibfnamefont {W.}~\bibnamefont {{Del Pozzo}}}, \bibinfo
  {author} {\bibfnamefont {C.~P.~L.}\ \bibnamefont {{Berry}}}, \bibinfo
  {author} {\bibfnamefont {A.~B.}\ \bibnamefont {{Nielsen}}},\ and\ \bibinfo
  {author} {\bibfnamefont {L.}~\bibnamefont {{London}}},\ }\bibfield  {title}
  {\bibinfo {title} {{Testing general relativity using gravitational wave
  signals from the inspiral, merger and ringdown of binary black holes}},\
  }\href {https://doi.org/10.1088/1361-6382/aa972e} {\bibfield  {journal}
  {\bibinfo  {journal} {Classical and Quantum Gravity}\ }\textbf {\bibinfo
  {volume} {35}},\ \bibinfo {eid} {014002} (\bibinfo {year} {2018})},\ \Eprint
  {https://arxiv.org/abs/1704.06784} {arXiv:1704.06784 [gr-qc]} \BibitemShut
  {NoStop}%
\bibitem [{\citenamefont {{Meidam}}\ \emph {et~al.}(2018)\citenamefont
  {{Meidam}}, \citenamefont {{Tsang}}, \citenamefont {{Goldstein}},
  \citenamefont {{Agathos}}, \citenamefont {{Ghosh}}, \citenamefont {{Haster}},
  \citenamefont {{Raymond}}, \citenamefont {{Samajdar}}, \citenamefont
  {{Schmidt}}, \citenamefont {{Smith}}, \citenamefont {{Blackburn}},
  \citenamefont {{Del Pozzo}}, \citenamefont {{Field}}, \citenamefont {{Li}},
  \citenamefont {{P{\"u}rrer}}, \citenamefont {{Van Den Broeck}}, \citenamefont
  {{Veitch}},\ and\ \citenamefont {{Vitale}}}]{Meidam2018}%
  \BibitemOpen
  \bibfield  {author} {\bibinfo {author} {\bibfnamefont {J.}~\bibnamefont
  {{Meidam}}}, \bibinfo {author} {\bibfnamefont {K.~W.}\ \bibnamefont
  {{Tsang}}}, \bibinfo {author} {\bibfnamefont {J.}~\bibnamefont
  {{Goldstein}}}, \bibinfo {author} {\bibfnamefont {M.}~\bibnamefont
  {{Agathos}}}, \bibinfo {author} {\bibfnamefont {A.}~\bibnamefont {{Ghosh}}},
  \bibinfo {author} {\bibfnamefont {C.-J.}\ \bibnamefont {{Haster}}}, \bibinfo
  {author} {\bibfnamefont {V.}~\bibnamefont {{Raymond}}}, \bibinfo {author}
  {\bibfnamefont {A.}~\bibnamefont {{Samajdar}}}, \bibinfo {author}
  {\bibfnamefont {P.}~\bibnamefont {{Schmidt}}}, \bibinfo {author}
  {\bibfnamefont {R.}~\bibnamefont {{Smith}}}, \bibinfo {author} {\bibfnamefont
  {K.}~\bibnamefont {{Blackburn}}}, \bibinfo {author} {\bibfnamefont
  {W.}~\bibnamefont {{Del Pozzo}}}, \bibinfo {author} {\bibfnamefont {S.~E.}\
  \bibnamefont {{Field}}}, \bibinfo {author} {\bibfnamefont {T.}~\bibnamefont
  {{Li}}}, \bibinfo {author} {\bibfnamefont {M.}~\bibnamefont {{P{\"u}rrer}}},
  \bibinfo {author} {\bibfnamefont {C.}~\bibnamefont {{Van Den Broeck}}},
  \bibinfo {author} {\bibfnamefont {J.}~\bibnamefont {{Veitch}}},\ and\
  \bibinfo {author} {\bibfnamefont {S.}~\bibnamefont {{Vitale}}},\ }\bibfield
  {title} {\bibinfo {title} {{Parametrized tests of the strong-field dynamics
  of general relativity using gravitational wave signals from coalescing binary
  black holes: Fast likelihood calculations and sensitivity of the method}},\
  }\href {https://doi.org/10.1103/PhysRevD.97.044033} {\bibfield  {journal}
  {\bibinfo  {journal} {\prd}\ }\textbf {\bibinfo {volume} {97}},\ \bibinfo
  {eid} {044033} (\bibinfo {year} {2018})},\ \Eprint
  {https://arxiv.org/abs/1712.08772} {arXiv:1712.08772 [gr-qc]} \BibitemShut
  {NoStop}%
\bibitem [{\citenamefont {Wysocki}\ \emph {et~al.}(2019)\citenamefont
  {Wysocki}, \citenamefont {Lange},\ and\ \citenamefont
  {O’Shaughnessy}}]{Wysocki2019}%
  \BibitemOpen
  \bibfield  {author} {\bibinfo {author} {\bibfnamefont {D.}~\bibnamefont
  {Wysocki}}, \bibinfo {author} {\bibfnamefont {J.}~\bibnamefont {Lange}},\
  and\ \bibinfo {author} {\bibfnamefont {R.}~\bibnamefont {O’Shaughnessy}},\
  }\bibfield  {title} {\bibinfo {title} {Reconstructing phenomenological
  distributions of compact binaries via gravitational wave observations},\
  }\href@noop {} {\bibfield  {journal} {\bibinfo  {journal} {Phys. Rev. D}\
  }\textbf {\bibinfo {volume} {100}},\ \bibinfo {pages} {043012} (\bibinfo
  {year} {2019})}\BibitemShut {NoStop}%
\bibitem [{\citenamefont {Callister}\ \emph {et~al.}(2022)\citenamefont
  {Callister}, \citenamefont {Miller}, \citenamefont {Chatziioannou},\ and\
  \citenamefont {Farr}}]{Callister2022}%
  \BibitemOpen
  \bibfield  {author} {\bibinfo {author} {\bibfnamefont {T.~A.}\ \bibnamefont
  {Callister}}, \bibinfo {author} {\bibfnamefont {S.~J.}\ \bibnamefont
  {Miller}}, \bibinfo {author} {\bibfnamefont {K.}~\bibnamefont
  {Chatziioannou}},\ and\ \bibinfo {author} {\bibfnamefont {W.~M.}\
  \bibnamefont {Farr}},\ }\bibfield  {title} {\bibinfo {title} {No evidence
  that the majority of black holes in binaries have zero spin},\ }\href@noop {}
  {\  (\bibinfo {year} {2022})},\ \bibinfo {note}
  {arxiv/2205.08574}\BibitemShut {NoStop}%
\bibitem [{\citenamefont {{Khan}}\ \emph {et~al.}(2016)\citenamefont {{Khan}},
  \citenamefont {{Husa}}, \citenamefont {{Hannam}}, \citenamefont {{Ohme}},
  \citenamefont {{P{\"u}rrer}}, \citenamefont {{Forteza}},\ and\ \citenamefont
  {{Boh{\'e}}}}]{khan2016PN}%
  \BibitemOpen
  \bibfield  {author} {\bibinfo {author} {\bibfnamefont {S.}~\bibnamefont
  {{Khan}}}, \bibinfo {author} {\bibfnamefont {S.}~\bibnamefont {{Husa}}},
  \bibinfo {author} {\bibfnamefont {M.}~\bibnamefont {{Hannam}}}, \bibinfo
  {author} {\bibfnamefont {F.}~\bibnamefont {{Ohme}}}, \bibinfo {author}
  {\bibfnamefont {M.}~\bibnamefont {{P{\"u}rrer}}}, \bibinfo {author}
  {\bibfnamefont {X.~J.}\ \bibnamefont {{Forteza}}},\ and\ \bibinfo {author}
  {\bibfnamefont {A.}~\bibnamefont {{Boh{\'e}}}},\ }\bibfield  {title}
  {\bibinfo {title} {{Frequency-domain gravitational waves from nonprecessing
  black-hole binaries. II. A phenomenological model for the advanced detector
  era}},\ }\href {https://doi.org/10.1103/PhysRevD.93.044007} {\bibfield
  {journal} {\bibinfo  {journal} {\prd}\ }\textbf {\bibinfo {volume} {93}},\
  \bibinfo {eid} {044007} (\bibinfo {year} {2016})},\ \Eprint
  {https://arxiv.org/abs/1508.07253} {arXiv:1508.07253 [gr-qc]} \BibitemShut
  {NoStop}%
\bibitem [{\citenamefont {{LIGO Scientific Collaboration and Virgo
  Collaboration and KAGRA Collaboration}}(2021)}]{gwtc2_tgr_release}%
  \BibitemOpen
  \bibfield  {author} {\bibinfo {author} {\bibnamefont {{LIGO Scientific
  Collaboration and Virgo Collaboration and KAGRA Collaboration}}},\ }\href
  {https://doi.org/10.7935/903s-gx73} {\bibinfo {title} {Tests of general
  relativity with binary black holes from the second ligo–virgo
  gravitational-wave transient catalog - full posterior sample data release}}
  (\bibinfo {year} {2021})\BibitemShut {NoStop}%
\bibitem [{\citenamefont {Abbott}\ \emph {et~al.}(2020)\citenamefont {Abbott}
  \emph {et~al.}}]{gw190814}%
  \BibitemOpen
  \bibfield  {author} {\bibinfo {author} {\bibfnamefont {R.}~\bibnamefont
  {Abbott}} \emph {et~al.} (\bibinfo {collaboration} {LIGO Scientific,
  Virgo}),\ }\bibfield  {title} {\bibinfo {title} {{GW190814: Gravitational
  Waves from the Coalescence of a 23 Solar Mass Black Hole with a 2.6 Solar
  Mass Compact Object}},\ }\href {https://doi.org/10.3847/2041-8213/ab960f}
  {\bibfield  {journal} {\bibinfo  {journal} {Astrophys. J. Lett.}\ }\textbf
  {\bibinfo {volume} {896}},\ \bibinfo {pages} {L44} (\bibinfo {year}
  {2020})},\ \Eprint {https://arxiv.org/abs/2006.12611} {arXiv:2006.12611
  [astro-ph.HE]} \BibitemShut {NoStop}%
\bibitem [{\citenamefont {Abbott}\ \emph
  {et~al.}(2021{\natexlab{b}})\citenamefont {Abbott} \emph
  {et~al.}}]{gw200115}%
  \BibitemOpen
  \bibfield  {author} {\bibinfo {author} {\bibfnamefont {R.}~\bibnamefont
  {Abbott}} \emph {et~al.} (\bibinfo {collaboration} {LIGO Scientific, KAGRA,
  VIRGO}),\ }\bibfield  {title} {\bibinfo {title} {{Observation of
  Gravitational Waves from Two Neutron Star\textendash{}Black Hole
  Coalescences}},\ }\href {https://doi.org/10.3847/2041-8213/ac082e} {\bibfield
   {journal} {\bibinfo  {journal} {Astrophys. J. Lett.}\ }\textbf {\bibinfo
  {volume} {915}},\ \bibinfo {pages} {L5} (\bibinfo {year}
  {2021}{\natexlab{b}})},\ \Eprint {https://arxiv.org/abs/2106.15163}
  {arXiv:2106.15163 [astro-ph.HE]} \BibitemShut {NoStop}%
\bibitem [{\citenamefont {Abbott}\ \emph
  {et~al.}(2021{\natexlab{c}})\citenamefont {Abbott} \emph {et~al.}}]{gwtc-2}%
  \BibitemOpen
  \bibfield  {author} {\bibinfo {author} {\bibfnamefont {R.}~\bibnamefont
  {Abbott}} \emph {et~al.},\ }\bibfield  {title} {\bibinfo {title} {{GWTC-2:
  Compact Binary Coalescences Observed by LIGO and Virgo During the First Half
  of the Third Observing Run}},\ }\href@noop {} {\bibfield  {journal} {\bibinfo
   {journal} {Phys. Rev. X}\ }\textbf {\bibinfo {volume} {11}},\ \bibinfo
  {pages} {021053} (\bibinfo {year} {2021}{\natexlab{c}})}\BibitemShut
  {NoStop}%
\bibitem [{\citenamefont {Phan}\ \emph {et~al.}(2019)\citenamefont {Phan},
  \citenamefont {Pradhan},\ and\ \citenamefont
  {Jankowiak}}]{phan2019composable}%
  \BibitemOpen
  \bibfield  {author} {\bibinfo {author} {\bibfnamefont {D.}~\bibnamefont
  {Phan}}, \bibinfo {author} {\bibfnamefont {N.}~\bibnamefont {Pradhan}},\ and\
  \bibinfo {author} {\bibfnamefont {M.}~\bibnamefont {Jankowiak}},\ }\bibfield
  {title} {\bibinfo {title} {Composable effects for flexible and accelerated
  probabilistic programming in numpyro},\ }\href@noop {} {\bibfield  {journal}
  {\bibinfo  {journal} {arXiv preprint arXiv:1912.11554}\ } (\bibinfo {year}
  {2019})}\BibitemShut {NoStop}%
\bibitem [{\citenamefont {Bingham}\ \emph {et~al.}(2019)\citenamefont
  {Bingham}, \citenamefont {Chen}, \citenamefont {Jankowiak}, \citenamefont
  {Obermeyer}, \citenamefont {Pradhan}, \citenamefont {Karaletsos},
  \citenamefont {Singh}, \citenamefont {Szerlip}, \citenamefont {Horsfall},\
  and\ \citenamefont {Goodman}}]{bingham2019pyro}%
  \BibitemOpen
  \bibfield  {author} {\bibinfo {author} {\bibfnamefont {E.}~\bibnamefont
  {Bingham}}, \bibinfo {author} {\bibfnamefont {J.~P.}\ \bibnamefont {Chen}},
  \bibinfo {author} {\bibfnamefont {M.}~\bibnamefont {Jankowiak}}, \bibinfo
  {author} {\bibfnamefont {F.}~\bibnamefont {Obermeyer}}, \bibinfo {author}
  {\bibfnamefont {N.}~\bibnamefont {Pradhan}}, \bibinfo {author} {\bibfnamefont
  {T.}~\bibnamefont {Karaletsos}}, \bibinfo {author} {\bibfnamefont
  {R.}~\bibnamefont {Singh}}, \bibinfo {author} {\bibfnamefont {P.~A.}\
  \bibnamefont {Szerlip}}, \bibinfo {author} {\bibfnamefont {P.}~\bibnamefont
  {Horsfall}},\ and\ \bibinfo {author} {\bibfnamefont {N.~D.}\ \bibnamefont
  {Goodman}},\ }\bibfield  {title} {\bibinfo {title} {Pyro: Deep universal
  probabilistic programming},\ }\href {http://jmlr.org/papers/v20/18-403.html}
  {\bibfield  {journal} {\bibinfo  {journal} {J. Mach. Learn. Res.}\ }\textbf
  {\bibinfo {volume} {20}},\ \bibinfo {pages} {28:1} (\bibinfo {year}
  {2019})}\BibitemShut {NoStop}%
\bibitem [{\citenamefont {Bradbury}\ \emph {et~al.}(2018)\citenamefont
  {Bradbury}, \citenamefont {Frostig}, \citenamefont {Hawkins}, \citenamefont
  {Johnson}, \citenamefont {Leary}, \citenamefont {Maclaurin}, \citenamefont
  {Necula}, \citenamefont {Paszke}, \citenamefont {Vander{P}las}, \citenamefont
  {Wanderman-{M}ilne},\ and\ \citenamefont {Zhang}}]{jax2018github}%
  \BibitemOpen
  \bibfield  {author} {\bibinfo {author} {\bibfnamefont {J.}~\bibnamefont
  {Bradbury}}, \bibinfo {author} {\bibfnamefont {R.}~\bibnamefont {Frostig}},
  \bibinfo {author} {\bibfnamefont {P.}~\bibnamefont {Hawkins}}, \bibinfo
  {author} {\bibfnamefont {M.~J.}\ \bibnamefont {Johnson}}, \bibinfo {author}
  {\bibfnamefont {C.}~\bibnamefont {Leary}}, \bibinfo {author} {\bibfnamefont
  {D.}~\bibnamefont {Maclaurin}}, \bibinfo {author} {\bibfnamefont
  {G.}~\bibnamefont {Necula}}, \bibinfo {author} {\bibfnamefont
  {A.}~\bibnamefont {Paszke}}, \bibinfo {author} {\bibfnamefont
  {J.}~\bibnamefont {Vander{P}las}}, \bibinfo {author} {\bibfnamefont
  {S.}~\bibnamefont {Wanderman-{M}ilne}},\ and\ \bibinfo {author}
  {\bibfnamefont {Q.}~\bibnamefont {Zhang}},\ }\href
  {http://github.com/google/jax} {\bibinfo {title} {{JAX}: composable
  transformations of {P}ython+{N}um{P}y programs}} (\bibinfo {year}
  {2018})\BibitemShut {NoStop}%
\bibitem [{\citenamefont {{Robitaille}}\ \emph {et~al.}(2013)\citenamefont
  {{Robitaille}} \emph {et~al.}}]{astropy:2013}%
  \BibitemOpen
  \bibfield  {author} {\bibinfo {author} {\bibfnamefont {T.~P.}\ \bibnamefont
  {{Robitaille}}} \emph {et~al.},\ }\bibfield  {title} {\bibinfo {title}
  {{Astropy: A community Python package for astronomy}},\ }\href
  {https://doi.org/10.1051/0004-6361/201322068} {\bibfield  {journal} {\bibinfo
   {journal} {Astron. \& Astrophys.}\ }\textbf {\bibinfo {volume} {558}},\
  \bibinfo {eid} {A33} (\bibinfo {year} {2013})},\ \Eprint
  {https://arxiv.org/abs/1307.6212} {arXiv:1307.6212 [astro-ph.IM]}
  \BibitemShut {NoStop}%
\bibitem [{\citenamefont {{Price-Whelan}}\ \emph {et~al.}(2018)\citenamefont
  {{Price-Whelan}} \emph {et~al.}}]{astropy:2018}%
  \BibitemOpen
  \bibfield  {author} {\bibinfo {author} {\bibfnamefont {A.~M.}\ \bibnamefont
  {{Price-Whelan}}} \emph {et~al.},\ }\bibfield  {title} {\bibinfo {title}
  {{The Astropy Project: Building an Open-science Project and Status of the
  v2.0 Core Package}},\ }\href {https://doi.org/10.3847/1538-3881/aabc4f}
  {\bibfield  {journal} {\bibinfo  {journal} {Astron. J.}\ }\textbf {\bibinfo
  {volume} {156}},\ \bibinfo {eid} {123} (\bibinfo {year} {2018})},\ \Eprint
  {https://arxiv.org/abs/1801.02634} {arXiv:1801.02634 [astro-ph.IM]}
  \BibitemShut {NoStop}%
\bibitem [{\citenamefont {{Price-Whelan}}\ \emph {et~al.}(2022)\citenamefont
  {{Price-Whelan}} \emph {et~al.}}]{astropy:2022}%
  \BibitemOpen
  \bibfield  {author} {\bibinfo {author} {\bibfnamefont {A.~M.}\ \bibnamefont
  {{Price-Whelan}}} \emph {et~al.},\ }\bibfield  {title} {\bibinfo {title}
  {{The Astropy Project: Sustaining and Growing a Community-oriented
  Open-source Project and the Latest Major Release (v5.0) of the Core
  Package}},\ }\href {https://doi.org/10.3847/1538-4357/ac7c74} {\bibfield
  {journal} {\bibinfo  {journal} {apj}\ }\textbf {\bibinfo {volume} {935}},\
  \bibinfo {eid} {167} (\bibinfo {year} {2022})},\ \Eprint
  {https://arxiv.org/abs/2206.14220} {arXiv:2206.14220 [astro-ph.IM]}
  \BibitemShut {NoStop}%
\bibitem [{\citenamefont {Virtanen}\ \emph {et~al.}(2020)\citenamefont
  {Virtanen} \emph {et~al.}}]{2020SciPy-NMeth}%
  \BibitemOpen
  \bibfield  {author} {\bibinfo {author} {\bibfnamefont {P.}~\bibnamefont
  {Virtanen}} \emph {et~al.},\ }\bibfield  {title} {\bibinfo {title} {{{SciPy}
  1.0: Fundamental Algorithms for Scientific Computing in Python}},\ }\href
  {https://doi.org/10.1038/s41592-019-0686-2} {\bibfield  {journal} {\bibinfo
  {journal} {Nature Methods}\ }\textbf {\bibinfo {volume} {17}},\ \bibinfo
  {pages} {261} (\bibinfo {year} {2020})}\BibitemShut {NoStop}%
\bibitem [{\citenamefont {Hunter}(2007)}]{Hunter:2007}%
  \BibitemOpen
  \bibfield  {author} {\bibinfo {author} {\bibfnamefont {J.~D.}\ \bibnamefont
  {Hunter}},\ }\bibfield  {title} {\bibinfo {title} {Matplotlib: A 2d graphics
  environment},\ }\href {https://doi.org/10.1109/MCSE.2007.55} {\bibfield
  {journal} {\bibinfo  {journal} {Computing in Science \& Engineering}\
  }\textbf {\bibinfo {volume} {9}},\ \bibinfo {pages} {90} (\bibinfo {year}
  {2007})}\BibitemShut {NoStop}%
\bibitem [{\citenamefont {Kumar}\ \emph {et~al.}(2019)\citenamefont {Kumar},
  \citenamefont {Carroll}, \citenamefont {Hartikainen},\ and\ \citenamefont
  {Martin}}]{arviz_2019}%
  \BibitemOpen
  \bibfield  {author} {\bibinfo {author} {\bibfnamefont {R.}~\bibnamefont
  {Kumar}}, \bibinfo {author} {\bibfnamefont {C.}~\bibnamefont {Carroll}},
  \bibinfo {author} {\bibfnamefont {A.}~\bibnamefont {Hartikainen}},\ and\
  \bibinfo {author} {\bibfnamefont {O.}~\bibnamefont {Martin}},\ }\bibfield
  {title} {\bibinfo {title} {Arviz a unified library for exploratory analysis
  of bayesian models in python},\ }\href {https://doi.org/10.21105/joss.01143}
  {\bibfield  {journal} {\bibinfo  {journal} {Journal of Open Source Software}\
  }\textbf {\bibinfo {volume} {4}},\ \bibinfo {pages} {1143} (\bibinfo {year}
  {2019})}\BibitemShut {NoStop}%
\bibitem [{\citenamefont {Foreman-Mackey}(2016)}]{corner}%
  \BibitemOpen
  \bibfield  {author} {\bibinfo {author} {\bibfnamefont {D.}~\bibnamefont
  {Foreman-Mackey}},\ }\bibfield  {title} {\bibinfo {title} {corner.py:
  Scatterplot matrices in python},\ }\href
  {https://doi.org/10.21105/joss.00024} {\bibfield  {journal} {\bibinfo
  {journal} {The Journal of Open Source Software}\ }\textbf {\bibinfo {volume}
  {1}},\ \bibinfo {pages} {24} (\bibinfo {year} {2016})}\BibitemShut {NoStop}%
\bibitem [{\citenamefont {Payne}\ \emph {et~al.}(2023)\citenamefont {Payne},
  \citenamefont {Isi}, \citenamefont {Chatziioannou},\ and\ \citenamefont
  {Farr}}]{code_release}%
  \BibitemOpen
  \bibfield  {author} {\bibinfo {author} {\bibfnamefont {E.}~\bibnamefont
  {Payne}}, \bibinfo {author} {\bibfnamefont {M.}~\bibnamefont {Isi}}, \bibinfo
  {author} {\bibfnamefont {K.}~\bibnamefont {Chatziioannou}},\ and\ \bibinfo
  {author} {\bibfnamefont {W.~M.}\ \bibnamefont {Farr}},\ }\href
  {https://github.com/ethanpayne42/testingGR_astro.git} {\bibinfo {title} {Code
  release for ``fortifying gravitational-wave tests of general relativity
  against astrophysical assumption''}} (\bibinfo {year} {2023})\BibitemShut
  {NoStop}%
\bibitem [{\citenamefont {Abbott}\ \emph
  {et~al.}(2019{\natexlab{b}})\citenamefont {Abbott} \emph {et~al.}}]{gwtc1}%
  \BibitemOpen
  \bibfield  {author} {\bibinfo {author} {\bibfnamefont {B.~P.}\ \bibnamefont
  {Abbott}} \emph {et~al.} (\bibinfo {collaboration} {LIGO Scientific,
  Virgo}),\ }\bibfield  {title} {\bibinfo {title} {{GWTC-1: A
  Gravitational-Wave Transient Catalog of Compact Binary Mergers Observed by
  LIGO and Virgo during the First and Second Observing Runs}},\ }\href
  {https://doi.org/10.1103/PhysRevX.9.031040} {\bibfield  {journal} {\bibinfo
  {journal} {Phys. Rev. X}\ }\textbf {\bibinfo {volume} {9}},\ \bibinfo {pages}
  {031040} (\bibinfo {year} {2019}{\natexlab{b}})},\ \Eprint
  {https://arxiv.org/abs/1811.12907} {arXiv:1811.12907 [astro-ph.HE]}
  \BibitemShut {NoStop}%
\bibitem [{\citenamefont {Moore}\ \emph {et~al.}(2021)\citenamefont {Moore},
  \citenamefont {Finch}, \citenamefont {Buscicchio},\ and\ \citenamefont
  {Gerosa}}]{Moore2021}%
  \BibitemOpen
  \bibfield  {author} {\bibinfo {author} {\bibfnamefont {C.~J.}\ \bibnamefont
  {Moore}}, \bibinfo {author} {\bibfnamefont {E.}~\bibnamefont {Finch}},
  \bibinfo {author} {\bibfnamefont {R.}~\bibnamefont {Buscicchio}},\ and\
  \bibinfo {author} {\bibfnamefont {D.}~\bibnamefont {Gerosa}},\ }\bibfield
  {title} {\bibinfo {title} {Testing general relativity with gravitational-wave
  catalogs: The insidious nature of waveform systematics},\ }\href
  {https://doi.org/https://doi.org/10.1016/j.isci.2021.102577} {\bibfield
  {journal} {\bibinfo  {journal} {iScience}\ }\textbf {\bibinfo {volume}
  {24}},\ \bibinfo {pages} {102577} (\bibinfo {year} {2021})}\BibitemShut
  {NoStop}%
\bibitem [{\citenamefont {Hu}\ and\ \citenamefont {Veitch}(2023)}]{Hu2022bji}%
  \BibitemOpen
  \bibfield  {author} {\bibinfo {author} {\bibfnamefont {Q.}~\bibnamefont
  {Hu}}\ and\ \bibinfo {author} {\bibfnamefont {J.}~\bibnamefont {Veitch}},\
  }\bibfield  {title} {\bibinfo {title} {{Accumulating Errors in Tests of
  General Relativity with Gravitational Waves: Overlapping Signals and
  Inaccurate Waveforms}},\ }\href {https://doi.org/10.3847/1538-4357/acbc18}
  {\bibfield  {journal} {\bibinfo  {journal} {Astrophys. J.}\ }\textbf
  {\bibinfo {volume} {945}},\ \bibinfo {pages} {103} (\bibinfo {year}
  {2023})},\ \Eprint {https://arxiv.org/abs/2210.04769} {arXiv:2210.04769
  [gr-qc]} \BibitemShut {NoStop}%
\bibitem [{\citenamefont {Saini}\ \emph {et~al.}(2022)\citenamefont {Saini},
  \citenamefont {Favata},\ and\ \citenamefont {Arun}}]{Saini:2022igm}%
  \BibitemOpen
  \bibfield  {author} {\bibinfo {author} {\bibfnamefont {P.}~\bibnamefont
  {Saini}}, \bibinfo {author} {\bibfnamefont {M.}~\bibnamefont {Favata}},\ and\
  \bibinfo {author} {\bibfnamefont {K.~G.}\ \bibnamefont {Arun}},\ }\bibfield
  {title} {\bibinfo {title} {{Systematic bias on parametrized tests of general
  relativity due to neglect of orbital eccentricity}},\ }\href
  {https://doi.org/10.1103/PhysRevD.106.084031} {\bibfield  {journal} {\bibinfo
   {journal} {Phys. Rev. D}\ }\textbf {\bibinfo {volume} {106}},\ \bibinfo
  {pages} {084031} (\bibinfo {year} {2022})},\ \Eprint
  {https://arxiv.org/abs/2203.04634} {arXiv:2203.04634 [gr-qc]} \BibitemShut
  {NoStop}%
\bibitem [{\citenamefont {Bhat}\ \emph {et~al.}(2023)\citenamefont {Bhat},
  \citenamefont {Saini}, \citenamefont {Favata},\ and\ \citenamefont
  {Arun}}]{Bhat:2022amc}%
  \BibitemOpen
  \bibfield  {author} {\bibinfo {author} {\bibfnamefont {S.~A.}\ \bibnamefont
  {Bhat}}, \bibinfo {author} {\bibfnamefont {P.}~\bibnamefont {Saini}},
  \bibinfo {author} {\bibfnamefont {M.}~\bibnamefont {Favata}},\ and\ \bibinfo
  {author} {\bibfnamefont {K.~G.}\ \bibnamefont {Arun}},\ }\bibfield  {title}
  {\bibinfo {title} {{Systematic bias on the inspiral-merger-ringdown
  consistency test due to neglect of orbital eccentricity}},\ }\href
  {https://doi.org/10.1103/PhysRevD.107.024009} {\bibfield  {journal} {\bibinfo
   {journal} {Phys. Rev. D}\ }\textbf {\bibinfo {volume} {107}},\ \bibinfo
  {pages} {024009} (\bibinfo {year} {2023})},\ \Eprint
  {https://arxiv.org/abs/2207.13761} {arXiv:2207.13761 [gr-qc]} \BibitemShut
  {NoStop}%
\bibitem [{\citenamefont {Abbott}\ \emph {et~al.}(2023)\citenamefont {Abbott}
  \emph {et~al.}}]{LIGOScientific2021aug}%
  \BibitemOpen
  \bibfield  {author} {\bibinfo {author} {\bibfnamefont {R.}~\bibnamefont
  {Abbott}} \emph {et~al.} (\bibinfo {collaboration} {LIGO Scientific, Virgo,,
  KAGRA, VIRGO}),\ }\bibfield  {title} {\bibinfo {title} {{Constraints on the
  Cosmic Expansion History from GWTC\textendash{}3}},\ }\href
  {https://doi.org/10.3847/1538-4357/ac74bb} {\bibfield  {journal} {\bibinfo
  {journal} {Astrophys. J.}\ }\textbf {\bibinfo {volume} {949}},\ \bibinfo
  {pages} {76} (\bibinfo {year} {2023})},\ \Eprint
  {https://arxiv.org/abs/2111.03604} {arXiv:2111.03604 [astro-ph.CO]}
  \BibitemShut {NoStop}%
\bibitem [{\citenamefont {Wysocki}\ \emph {et~al.}(2020)\citenamefont
  {Wysocki}, \citenamefont {O'Shaughnessy}, \citenamefont {Wade},\ and\
  \citenamefont {Lange}}]{Wysocki:2020myz}%
  \BibitemOpen
  \bibfield  {author} {\bibinfo {author} {\bibfnamefont {D.}~\bibnamefont
  {Wysocki}}, \bibinfo {author} {\bibfnamefont {R.}~\bibnamefont
  {O'Shaughnessy}}, \bibinfo {author} {\bibfnamefont {L.}~\bibnamefont
  {Wade}},\ and\ \bibinfo {author} {\bibfnamefont {J.}~\bibnamefont {Lange}},\
  }\bibfield  {title} {\bibinfo {title} {{Inferring the neutron star equation
  of state simultaneously with the population of merging neutron stars}},\
  }\href@noop {} {\  (\bibinfo {year} {2020})},\ \Eprint
  {https://arxiv.org/abs/2001.01747} {arXiv:2001.01747 [gr-qc]} \BibitemShut
  {NoStop}%
\bibitem [{\citenamefont {Romero-Shaw}\ \emph {et~al.}(2022)\citenamefont
  {Romero-Shaw}, \citenamefont {Thrane},\ and\ \citenamefont {Lasky}}]{wmf}%
  \BibitemOpen
  \bibfield  {author} {\bibinfo {author} {\bibfnamefont {I.~M.}\ \bibnamefont
  {Romero-Shaw}}, \bibinfo {author} {\bibfnamefont {E.}~\bibnamefont
  {Thrane}},\ and\ \bibinfo {author} {\bibfnamefont {P.~D.}\ \bibnamefont
  {Lasky}},\ }\bibfield  {title} {\bibinfo {title} {When models fail: an
  introduction to posterior predictive checks and model misspecification in
  gravitational-wave astronomy},\ }\href@noop {} {\bibfield  {journal}
  {\bibinfo  {journal} {Pub. Astron. Soc. Aust.}\ }\textbf {\bibinfo {volume}
  {39}},\ \bibinfo {pages} {E025} (\bibinfo {year} {2022})}\BibitemShut
  {NoStop}%
\bibitem [{\citenamefont {Gelman}\ \emph {et~al.}(2013)\citenamefont {Gelman},
  \citenamefont {Carlin}, \citenamefont {Stern}, \citenamefont {Dunson},
  \citenamefont {Vehtari},\ and\ \citenamefont {Rubin}}]{Gelman}%
  \BibitemOpen
  \bibfield  {author} {\bibinfo {author} {\bibfnamefont {A.}~\bibnamefont
  {Gelman}}, \bibinfo {author} {\bibfnamefont {J.~B.}\ \bibnamefont {Carlin}},
  \bibinfo {author} {\bibfnamefont {H.~S.}\ \bibnamefont {Stern}}, \bibinfo
  {author} {\bibfnamefont {D.~B.}\ \bibnamefont {Dunson}}, \bibinfo {author}
  {\bibfnamefont {A.}~\bibnamefont {Vehtari}},\ and\ \bibinfo {author}
  {\bibfnamefont {D.~B.}\ \bibnamefont {Rubin}},\ }\href@noop {} {\emph
  {\bibinfo {title} {{Bayesian Data Analysis, Third Edition}}}},\ Chapman {\&}
  Hall/CRC Texts in Statistical Science\ (\bibinfo  {publisher} {Taylor {\&}
  Francis},\ \bibinfo {year} {2013})\BibitemShut {NoStop}%
\bibitem [{\citenamefont {Payne}\ and\ \citenamefont
  {Thrane}(2023)}]{Payne2023}%
  \BibitemOpen
  \bibfield  {author} {\bibinfo {author} {\bibfnamefont {E.}~\bibnamefont
  {Payne}}\ and\ \bibinfo {author} {\bibfnamefont {E.}~\bibnamefont {Thrane}},\
  }\bibfield  {title} {\bibinfo {title} {Model exploration in
  gravitational-wave astronomy with the maximum population likelihood},\ }\href
  {https://doi.org/10.1103/PhysRevResearch.5.023013} {\bibfield  {journal}
  {\bibinfo  {journal} {Phys. Rev. Res.}\ }\textbf {\bibinfo {volume} {5}},\
  \bibinfo {pages} {023013} (\bibinfo {year} {2023})}\BibitemShut {NoStop}%
\bibitem [{\citenamefont {Gupta}\ \emph {et~al.}(2020)\citenamefont {Gupta},
  \citenamefont {Datta}, \citenamefont {Kastha}, \citenamefont {Borhanian},
  \citenamefont {Arun},\ and\ \citenamefont {Sathyaprakash}}]{Gupta:2020lxa}%
  \BibitemOpen
  \bibfield  {author} {\bibinfo {author} {\bibfnamefont {A.}~\bibnamefont
  {Gupta}}, \bibinfo {author} {\bibfnamefont {S.}~\bibnamefont {Datta}},
  \bibinfo {author} {\bibfnamefont {S.}~\bibnamefont {Kastha}}, \bibinfo
  {author} {\bibfnamefont {S.}~\bibnamefont {Borhanian}}, \bibinfo {author}
  {\bibfnamefont {K.~G.}\ \bibnamefont {Arun}},\ and\ \bibinfo {author}
  {\bibfnamefont {B.~S.}\ \bibnamefont {Sathyaprakash}},\ }\bibfield  {title}
  {\bibinfo {title} {{Multiparameter tests of general relativity using
  multiband gravitational-wave observations}},\ }\href
  {https://doi.org/10.1103/PhysRevLett.125.201101} {\bibfield  {journal}
  {\bibinfo  {journal} {Phys. Rev. Lett.}\ }\textbf {\bibinfo {volume} {125}},\
  \bibinfo {pages} {201101} (\bibinfo {year} {2020})},\ \Eprint
  {https://arxiv.org/abs/2005.09607} {arXiv:2005.09607 [gr-qc]} \BibitemShut
  {NoStop}%
\bibitem [{\citenamefont {Datta}\ \emph {et~al.}(2021)\citenamefont {Datta},
  \citenamefont {Gupta}, \citenamefont {Kastha}, \citenamefont {Arun},\ and\
  \citenamefont {Sathyaprakash}}]{Datta:2020vcj}%
  \BibitemOpen
  \bibfield  {author} {\bibinfo {author} {\bibfnamefont {S.}~\bibnamefont
  {Datta}}, \bibinfo {author} {\bibfnamefont {A.}~\bibnamefont {Gupta}},
  \bibinfo {author} {\bibfnamefont {S.}~\bibnamefont {Kastha}}, \bibinfo
  {author} {\bibfnamefont {K.~G.}\ \bibnamefont {Arun}},\ and\ \bibinfo
  {author} {\bibfnamefont {B.~S.}\ \bibnamefont {Sathyaprakash}},\ }\bibfield
  {title} {\bibinfo {title} {{Tests of general relativity using multiband
  observations of intermediate mass binary black hole mergers}},\ }\href
  {https://doi.org/10.1103/PhysRevD.103.024036} {\bibfield  {journal} {\bibinfo
   {journal} {Phys. Rev. D}\ }\textbf {\bibinfo {volume} {103}},\ \bibinfo
  {pages} {024036} (\bibinfo {year} {2021})},\ \Eprint
  {https://arxiv.org/abs/2006.12137} {arXiv:2006.12137 [gr-qc]} \BibitemShut
  {NoStop}%
\bibitem [{\citenamefont {{Shoom}}\ \emph {et~al.}(2023)\citenamefont
  {{Shoom}}, \citenamefont {{Gupta}}, \citenamefont {{Krishnan}}, \citenamefont
  {{Nielsen}},\ and\ \citenamefont {{Capano}}}]{Shoom:2021mdj}%
  \BibitemOpen
  \bibfield  {author} {\bibinfo {author} {\bibfnamefont {A.~A.}\ \bibnamefont
  {{Shoom}}}, \bibinfo {author} {\bibfnamefont {P.~K.}\ \bibnamefont
  {{Gupta}}}, \bibinfo {author} {\bibfnamefont {B.}~\bibnamefont {{Krishnan}}},
  \bibinfo {author} {\bibfnamefont {A.~B.}\ \bibnamefont {{Nielsen}}},\ and\
  \bibinfo {author} {\bibfnamefont {C.~D.}\ \bibnamefont {{Capano}}},\
  }\bibfield  {title} {\bibinfo {title} {{Testing the post-Newtonian expansion
  with GW170817}},\ }\href {https://doi.org/10.1007/s10714-023-03100-z}
  {\bibfield  {journal} {\bibinfo  {journal} {General Relativity and
  Gravitation}\ }\textbf {\bibinfo {volume} {55}},\ \bibinfo {eid} {55}
  (\bibinfo {year} {2023})},\ \Eprint {https://arxiv.org/abs/2105.02191}
  {arXiv:2105.02191 [gr-qc]} \BibitemShut {NoStop}%
\bibitem [{\citenamefont {Perkins}\ and\ \citenamefont
  {Yunes}(2022)}]{Perkins:2022fhr}%
  \BibitemOpen
  \bibfield  {author} {\bibinfo {author} {\bibfnamefont {S.}~\bibnamefont
  {Perkins}}\ and\ \bibinfo {author} {\bibfnamefont {N.}~\bibnamefont
  {Yunes}},\ }\bibfield  {title} {\bibinfo {title} {{Are parametrized tests of
  general relativity with gravitational waves robust to unknown higher
  post-Newtonian order effects?}},\ }\href
  {https://doi.org/10.1103/PhysRevD.105.124047} {\bibfield  {journal} {\bibinfo
   {journal} {Phys. Rev. D}\ }\textbf {\bibinfo {volume} {105}},\ \bibinfo
  {pages} {124047} (\bibinfo {year} {2022})},\ \Eprint
  {https://arxiv.org/abs/2201.02542} {arXiv:2201.02542 [gr-qc]} \BibitemShut
  {NoStop}%
\bibitem [{\citenamefont {{Datta}}\ \emph {et~al.}(2022)\citenamefont
  {{Datta}}, \citenamefont {{Saleem}}, \citenamefont {{Arun}},\ and\
  \citenamefont {{Sathyaprakash}}}]{Datta:2022izc}%
  \BibitemOpen
  \bibfield  {author} {\bibinfo {author} {\bibfnamefont {S.}~\bibnamefont
  {{Datta}}}, \bibinfo {author} {\bibfnamefont {M.}~\bibnamefont {{Saleem}}},
  \bibinfo {author} {\bibfnamefont {K.~G.}\ \bibnamefont {{Arun}}},\ and\
  \bibinfo {author} {\bibfnamefont {B.~S.}\ \bibnamefont {{Sathyaprakash}}},\
  }\bibfield  {title} {\bibinfo {title} {{Multiparameter tests of general
  relativity using principal component analysis with next-generation
  gravitational wave detectors}},\ }\href
  {https://doi.org/10.48550/arXiv.2208.07757} {\bibfield  {journal} {\bibinfo
  {journal} {arXiv e-prints}\ ,\ \bibinfo {eid} {arXiv:2208.07757}} (\bibinfo
  {year} {2022})},\ \Eprint {https://arxiv.org/abs/2208.07757}
  {arXiv:2208.07757 [gr-qc]} \BibitemShut {NoStop}%
\bibitem [{\citenamefont {Ade}\ \emph {et~al.}(2016)\citenamefont {Ade} \emph
  {et~al.}}]{Planck:2015fie}%
  \BibitemOpen
  \bibfield  {author} {\bibinfo {author} {\bibfnamefont {P.~A.~R.}\
  \bibnamefont {Ade}} \emph {et~al.} (\bibinfo {collaboration} {Planck}),\
  }\bibfield  {title} {\bibinfo {title} {{Planck 2015 results. XIII.
  Cosmological parameters}},\ }\href
  {https://doi.org/10.1051/0004-6361/201525830} {\bibfield  {journal} {\bibinfo
   {journal} {Astron. Astrophys.}\ }\textbf {\bibinfo {volume} {594}},\
  \bibinfo {pages} {A13} (\bibinfo {year} {2016})},\ \Eprint
  {https://arxiv.org/abs/1502.01589} {arXiv:1502.01589 [astro-ph.CO]}
  \BibitemShut {NoStop}%
\bibitem [{\citenamefont {{Sathyaprakash}}\ and\ \citenamefont
  {{Dhurandhar}}(1991)}]{sathyaprakash1991PN}%
  \BibitemOpen
  \bibfield  {author} {\bibinfo {author} {\bibfnamefont {B.~S.}\ \bibnamefont
  {{Sathyaprakash}}}\ and\ \bibinfo {author} {\bibfnamefont {S.~V.}\
  \bibnamefont {{Dhurandhar}}},\ }\bibfield  {title} {\bibinfo {title} {{Choice
  of filters for the detection of gravitational waves from coalescing
  binaries}},\ }\href {https://doi.org/10.1103/PhysRevD.44.3819} {\bibfield
  {journal} {\bibinfo  {journal} {\prd}\ }\textbf {\bibinfo {volume} {44}},\
  \bibinfo {pages} {3819} (\bibinfo {year} {1991})}\BibitemShut {NoStop}%
\bibitem [{\citenamefont {{Blanchet}}\ and\ \citenamefont
  {{Schafer}}(1993)}]{Blanchet1993tails}%
  \BibitemOpen
  \bibfield  {author} {\bibinfo {author} {\bibfnamefont {L.}~\bibnamefont
  {{Blanchet}}}\ and\ \bibinfo {author} {\bibfnamefont {G.}~\bibnamefont
  {{Schafer}}},\ }\bibfield  {title} {\bibinfo {title} {{Gravitational wave
  tails and binary star systems}},\ }\href
  {https://doi.org/10.1088/0264-9381/10/12/026} {\bibfield  {journal} {\bibinfo
   {journal} {Classical and Quantum Gravity}\ }\textbf {\bibinfo {volume}
  {10}},\ \bibinfo {pages} {2699} (\bibinfo {year} {1993})}\BibitemShut
  {NoStop}%
\bibitem [{\citenamefont {{Blanchet}}\ and\ \citenamefont
  {{Sathyaprakash}}(1994)}]{Blanchet1994tails}%
  \BibitemOpen
  \bibfield  {author} {\bibinfo {author} {\bibfnamefont {L.}~\bibnamefont
  {{Blanchet}}}\ and\ \bibinfo {author} {\bibfnamefont {B.~S.}\ \bibnamefont
  {{Sathyaprakash}}},\ }\bibfield  {title} {\bibinfo {title} {{Signal analysis
  of gravitational wave tails}},\ }\href
  {https://doi.org/10.1088/0264-9381/11/11/020} {\bibfield  {journal} {\bibinfo
   {journal} {Classical and Quantum Gravity}\ }\textbf {\bibinfo {volume}
  {11}},\ \bibinfo {pages} {2807} (\bibinfo {year} {1994})}\BibitemShut
  {NoStop}%
\bibitem [{\citenamefont {{Will}}(2014)}]{Will2014confront}%
  \BibitemOpen
  \bibfield  {author} {\bibinfo {author} {\bibfnamefont {C.~M.}\ \bibnamefont
  {{Will}}},\ }\bibfield  {title} {\bibinfo {title} {{The Confrontation between
  General Relativity and Experiment}},\ }\href
  {https://doi.org/10.12942/lrr-2014-4} {\bibfield  {journal} {\bibinfo
  {journal} {Living Reviews in Relativity}\ }\textbf {\bibinfo {volume} {17}},\
  \bibinfo {eid} {4} (\bibinfo {year} {2014})},\ \Eprint
  {https://arxiv.org/abs/1403.7377} {arXiv:1403.7377 [gr-qc]} \BibitemShut
  {NoStop}%
\bibitem [{\citenamefont {Cutler}\ and\ \citenamefont
  {Flanagan}(1994)}]{Cutler1994}%
  \BibitemOpen
  \bibfield  {author} {\bibinfo {author} {\bibfnamefont {C.}~\bibnamefont
  {Cutler}}\ and\ \bibinfo {author} {\bibfnamefont {E.~E.}\ \bibnamefont
  {Flanagan}},\ }\bibfield  {title} {\bibinfo {title} {{Gravitational waves
  from merging compact binaries: How accurately can one extract the binary's
  parameters from the inspiral wave form?}},\ }\href
  {https://doi.org/10.1103/PhysRevD.49.2658} {\bibfield  {journal} {\bibinfo
  {journal} {Phys. Rev. D}\ }\textbf {\bibinfo {volume} {49}},\ \bibinfo
  {pages} {2658} (\bibinfo {year} {1994})},\ \Eprint
  {https://arxiv.org/abs/gr-qc/9402014} {arXiv:gr-qc/9402014} \BibitemShut
  {NoStop}%
\bibitem [{\citenamefont {Campanelli}\ \emph {et~al.}(2006)\citenamefont
  {Campanelli}, \citenamefont {Lousto},\ and\ \citenamefont
  {Zlochower}}]{Campanelli2006uy}%
  \BibitemOpen
  \bibfield  {author} {\bibinfo {author} {\bibfnamefont {M.}~\bibnamefont
  {Campanelli}}, \bibinfo {author} {\bibfnamefont {C.~O.}\ \bibnamefont
  {Lousto}},\ and\ \bibinfo {author} {\bibfnamefont {Y.}~\bibnamefont
  {Zlochower}},\ }\bibfield  {title} {\bibinfo {title} {{Spinning-black-hole
  binaries: The orbital hang up}},\ }\href
  {https://doi.org/10.1103/PhysRevD.74.041501} {\bibfield  {journal} {\bibinfo
  {journal} {Phys. Rev. D}\ }\textbf {\bibinfo {volume} {74}},\ \bibinfo
  {pages} {041501} (\bibinfo {year} {2006})},\ \Eprint
  {https://arxiv.org/abs/gr-qc/0604012} {arXiv:gr-qc/0604012} \BibitemShut
  {NoStop}%
\bibitem [{\citenamefont {{Essick}}\ and\ \citenamefont
  {{Farr}}(2022)}]{essick2022}%
  \BibitemOpen
  \bibfield  {author} {\bibinfo {author} {\bibfnamefont {R.}~\bibnamefont
  {{Essick}}}\ and\ \bibinfo {author} {\bibfnamefont {W.}~\bibnamefont
  {{Farr}}},\ }\bibfield  {title} {\bibinfo {title} {{Precision Requirements
  for Monte Carlo Sums within Hierarchical Bayesian Inference}},\ }\href
  {https://doi.org/10.48550/arXiv.2204.00461} {\bibfield  {journal} {\bibinfo
  {journal} {arXiv e-prints}\ ,\ \bibinfo {eid} {arXiv:2204.00461}} (\bibinfo
  {year} {2022})},\ \Eprint {https://arxiv.org/abs/2204.00461}
  {arXiv:2204.00461 [astro-ph.IM]} \BibitemShut {NoStop}%
\bibitem [{\citenamefont {{Talbot}}\ and\ \citenamefont
  {{Golomb}}(2023)}]{talbot2023}%
  \BibitemOpen
  \bibfield  {author} {\bibinfo {author} {\bibfnamefont {C.}~\bibnamefont
  {{Talbot}}}\ and\ \bibinfo {author} {\bibfnamefont {J.}~\bibnamefont
  {{Golomb}}},\ }\bibfield  {title} {\bibinfo {title} {{Growing Pains:
  Understanding the Impact of Likelihood Uncertainty on Hierarchical Bayesian
  Inference for Gravitational-Wave Astronomy}},\ }\href
  {https://doi.org/10.48550/arXiv.2304.06138} {\bibfield  {journal} {\bibinfo
  {journal} {arXiv e-prints}\ ,\ \bibinfo {eid} {arXiv:2304.06138}} (\bibinfo
  {year} {2023})},\ \Eprint {https://arxiv.org/abs/2304.06138}
  {arXiv:2304.06138 [astro-ph.IM]} \BibitemShut {NoStop}%
\bibitem [{\citenamefont {Scott}(1979)}]{scott1979optimal}%
  \BibitemOpen
  \bibfield  {author} {\bibinfo {author} {\bibfnamefont {D.~W.}\ \bibnamefont
  {Scott}},\ }\bibfield  {title} {\bibinfo {title} {On optimal and data-based
  histograms},\ }\href@noop {} {\bibfield  {journal} {\bibinfo  {journal}
  {Biometrika}\ }\textbf {\bibinfo {volume} {66}},\ \bibinfo {pages} {605}
  (\bibinfo {year} {1979})}\BibitemShut {NoStop}%
\bibitem [{\citenamefont {Kish}(1995)}]{Kish}%
  \BibitemOpen
  \bibfield  {author} {\bibinfo {author} {\bibfnamefont {L.}~\bibnamefont
  {Kish}},\ }\href@noop {} {\emph {\bibinfo {title} {Survey Sampling}}},\
  \bibinfo {edition} {3rd}\ ed.\ (\bibinfo  {publisher} {Wiley-Interscience},\
  \bibinfo {address} {Oxford, England},\ \bibinfo {year} {1995})\BibitemShut
  {NoStop}%
\bibitem [{\citenamefont {{Elvira}}\ \emph {et~al.}(2018)\citenamefont
  {{Elvira}}, \citenamefont {{Martino}},\ and\ \citenamefont
  {{Robert}}}]{Elvira}%
  \BibitemOpen
  \bibfield  {author} {\bibinfo {author} {\bibfnamefont {V.}~\bibnamefont
  {{Elvira}}}, \bibinfo {author} {\bibfnamefont {L.}~\bibnamefont
  {{Martino}}},\ and\ \bibinfo {author} {\bibfnamefont {C.~P.}\ \bibnamefont
  {{Robert}}},\ }\bibfield  {title} {\bibinfo {title} {{Rethinking the
  Effective Sample Size}},\ }\href {https://doi.org/10.48550/arXiv.1809.04129}
  {\bibfield  {journal} {\bibinfo  {journal} {arXiv e-prints}\ ,\ \bibinfo
  {eid} {arXiv:1809.04129}} (\bibinfo {year} {2018})},\ \Eprint
  {https://arxiv.org/abs/1809.04129} {arXiv:1809.04129 [stat.CO]} \BibitemShut
  {NoStop}%
\bibitem [{\citenamefont {{Hogg}}\ \emph {et~al.}(2020)\citenamefont {{Hogg}},
  \citenamefont {{Price-Whelan}},\ and\ \citenamefont {{Leistedt}}}]{Hogg2020}%
  \BibitemOpen
  \bibfield  {author} {\bibinfo {author} {\bibfnamefont {D.~W.}\ \bibnamefont
  {{Hogg}}}, \bibinfo {author} {\bibfnamefont {A.~M.}\ \bibnamefont
  {{Price-Whelan}}},\ and\ \bibinfo {author} {\bibfnamefont {B.}~\bibnamefont
  {{Leistedt}}},\ }\bibfield  {title} {\bibinfo {title} {{Data Analysis
  Recipes: Products of multivariate Gaussians in Bayesian inferences}},\ }\href
  {https://doi.org/10.48550/arXiv.2005.14199} {\bibfield  {journal} {\bibinfo
  {journal} {arXiv e-prints}\ ,\ \bibinfo {eid} {arXiv:2005.14199}} (\bibinfo
  {year} {2020})},\ \Eprint {https://arxiv.org/abs/2005.14199}
  {arXiv:2005.14199 [stat.CO]} \BibitemShut {NoStop}%
\end{thebibliography}%

\end{document}